\documentclass[12pt]{article}
\usepackage[margin=0.7in]{geometry}
  
\usepackage{hyperref}
\hypersetup{
	colorlinks=true,
	linkcolor=blue,
	filecolor=blue,      
	urlcolor=red,
	citecolor=blue,
}

\usepackage{amsfonts}
\usepackage{amsgen,amsmath,amstext,amsbsy,amsopn,amssymb,subfigure, amsthm}
\usepackage[dvips]{graphicx}
\usepackage{comment}
\usepackage{enumitem}
\usepackage{natbib}
\usepackage{booktabs}
\usepackage{blkarray}
\usepackage{algorithm}
\usepackage{algpseudocode}
\usepackage{url}
\usepackage{longtable}
\usepackage{mleftright}

\usepackage[usenames]{color}

\allowdisplaybreaks

\newtheorem{manualtheoreminner}{Theorem}
\newenvironment{manualtheorem}[1]{%
	\renewcommand\themanualtheoreminner{#1}%
	\manualtheoreminner
}{\endmanualtheoreminner}

\newtheorem{manualassumptioninner}{Assumption}
\newenvironment{manualassumption}[1]{%
	\renewcommand\themanualassumptioninner{#1}%
	\manualassumptioninner
}{\endmanualassumptioninner}

\newtheorem{Notation}{Notation}  
\newtheorem{Example}{Example}  
\newtheorem{Definition}{Definition}
\newtheorem{Theorem}{Theorem}
\newtheorem*{Theorem*}{Theorem}
\newtheorem{Lemma}{Lemma}
\newtheorem{Remark}{Remark}
\newtheorem{Corollary}{Corollary}
\newtheorem{Assumption}{Assumption}
\newtheorem*{Assumption*}{Assumption}
\newtheorem{Proposition}{Proposition}

\newcommand{\be}{\begin{equation}}
	\newcommand{\ee}{\end{equation}}
\newcommand{\bea}{\begin{eqnarray}}
	\newcommand{\eea}{\end{eqnarray}}
\newcommand{\beas}{\begin{eqnarray*}}
	\newcommand{\eeas}{\end{eqnarray*}}

\newcommand{\T}{^{\top}}

\newcommand{\EE}{\mathbb{E}}

\newcommand{\br}{{\mathbf{r}}}
\newcommand{\bp}{{\mathbf{p}}}

\newcommand{\X}{{\mathbf{X}}}
\newcommand{\B}{{\mathbf{B}}}

\newcommand{\I}{{\mathbf{I}}}
\newcommand{\M}{{\mathbf{M}}}

\renewcommand{\S}{{\mathbf{S}}}

\newcommand{\Var}{{\rm Var}}

\newcommand{\Cov}{{\rm Cov}}

\newcommand{\A}{{\mathbf{A}}}
\newcommand{\Y}{{\mathbf{Y}}}
\newcommand{\Z}{{\mathbf{Z}}}

\newcommand{\Span}{\operatorname{span}}

\newcommand{\rank}{{\rm rank}}
\newcommand{\var}{{\rm var}}
\newcommand{\SSigma}{\boldsymbol{\Sigma}}
\newcommand{\tr}{{\rm tr}}

\newcommand{\diag}{{\rm diag}}

\newcommand{\SVD}{{\rm SVD}}
\newcommand{\Eigen}{{\rm Eigen}}

\newcommand{\argmin}{\mathop{\rm arg\min}}
\newcommand{\argmax}{\mathop{\rm arg\max}}
\newcommand{\RR}{\mathbb{R}}
\newcommand{\PP}{\mathbb{P}}
\newcommand{\MM}{\mathcal{M}}
\newcommand{\vect}{{\operatorname{vec}}}

\allowdisplaybreaks

\begin{document}

\title{Mode-wise Principal Subspace Pursuit and Matrix Spiked Covariance Model}
    \author{Runshi Tang\footnote{Department of Statistics, University of Wisconsin-Madison}, ~ Ming Yuan\footnote{Department of Statistics, Columbia University}, ~ and ~ Anru R. Zhang\footnote{Departments of Biostatistics \& Bioinformatics and Computer Science, Duke University}}
\date{(\today)}
\maketitle

\bigskip

\begin{sloppypar}

\abstract{This paper introduces a novel framework called \underline{Mo}de-wise \underline{P}rincipal S\underline{u}bspace \underline{P}ursuit (\texttt{MOP-UP}) to extract hidden variations in both the row and column dimensions for matrix data. To enhance the understanding of the framework, we introduce a class of matrix-variate spiked covariance models that serve as inspiration for the development of the \texttt{MOP-UP} algorithm. The \texttt{MOP-UP} algorithm consists of two steps: Average Subspace Capture (\texttt{ASC}) and Alternating Projection (\texttt{AP}). These steps are specifically designed to capture the row-wise and column-wise dimension-reduced subspaces which contain the most informative features of the data. \texttt{ASC} utilizes a novel average projection operator as initialization and achieves exact recovery in the noiseless setting. We analyze the convergence and non-asymptotic error bounds of \texttt{MOP-UP}, introducing a blockwise matrix eigenvalue perturbation bound that proves the desired bound, where classic perturbation bounds fail. The effectiveness and practical merits of the proposed framework are demonstrated through experiments on both simulated and real datasets. Lastly, we discuss generalizations of our approach to higher-order data.}

\maketitle

%%%%%%%%%%%%%%%%%%%%%%%%
\section{Introduction}\label{sec:intro}
%%%%%%%%%%%%%%%%%%%%%%%%

In modern scientific applications, data are often observed in the form of multiple matrices or tensors that pertain to different subjects from a certain population. For instance, longitudinal gene expression data consist of a matrix of gene expression levels across time for each subject \citep{liu2017characterizing}; MRI imaging data contain one order-3 tensor image for each patient \citep{zhou2013tensor}; multilayer network can be represented by an order-3 tensor, where each layer (i.e., a matrix) represents one network \citep{jing2021community}; $m$-uniform hypergraph is typically viewed as an order-$m$ tensor, whose entries denote all hyper-edges \citep{zhen2022community}; atomic-resolution 4D scanning transmission electron microscopy data can be expressed as an order-3 tensor with two models denoting scan location and the other denoting the convergent beam electron diffraction pattern \citep{zhang2020denoising}. Combining information from all subjects results in a high-order tensor with subject independence along one mode and some covariance structure along the other modes that represent the relationship among the measured covariates.

Principal Component Analysis (PCA) is a widely accepted method for analyzing data consisting of vectors associated with individual subjects. Its primary objective is to identify a lower-dimensional subspace within the feature domain that captures the majority of data variance \citep{pearson1901liii}. PCA is a reliable technique for reducing the dimensionality of data. Singular Value Decomposition (SVD) is an efficient approach commonly used to compute PCA. However, when the dataset is in the form of a series of matrices, PCA encounters challenges.

In the literature, the tensor SVD framework (also known as tensor PCA in the machine learning and information theory community) is discussed \citep{richard2014statistical,zhang2018tensor,wang2020learning,zhou2022optimal,han2022optimal}. This framework revolves around a signal-plus-noise model: $\Y = \X + \Z$, where $\X$ represents a mean tensor with certain low-complexity structures (e.g., CP, Tucker, tubal, tensor-train low-rank, etc.), and $\Z$ denotes mean-zero random observational noise. The goal of tensor SVD (or tensor PCA) is to efficiently extract $\X$ from $\Y$. However, this approach is not suitable for analyzing high-order covariance structures of tensor data due to several reasons. First, most mean-based SVD methodologies assume that the dataset has some tensor low rankness, but this assumption may not always hold true. Second, tensor SVD or low-rank tensor factorization primarily focuses on the mean structure of the data tensor, simplifying the problem to a significantly lower number of parameters compared to the covariance structure.  To fix ideas, consider for instance repeated observations of matrix data. While $n$ i.i.d. copies of $p$-by-$p$ matrix result in a data tensor with $np^2$ entries, the associated covariance tensor includes $p^4$ entries. Most importantly, tensor SVD or low-rank tensor factorization \citep{kolda2009tensor,zhang2018tensor} may not fit for treating the data tensor as information obtained from independent replicates of a certain population. Consequently, achieving good performance in covariance tensor statistical inference using mean-based models cannot be expected.

Since the direct analysis of the covariance tensor of $p$-by-$p$ observational data matrices involves $p^4$ parameters and is typically difficult in high-dimensional settings, a number of simplified covariance tensor structures were introduced, including the (approximate) Kronecker product distribution (see, e.g.,  \cite{dawid1981some,dutilleul1999mle,yin2012model,tsiligkaridis2013covariance,zhou2014gemini,chen2015statistical,hoff2015multilinear,ding2018matrix,hoff2022core}):
\begin{equation*}
    \SSigma = \Sigma_1\otimes_K \Sigma_2, \quad \text{i.e.,}\quad \SSigma_{ijkl} = (\Sigma_1)_{ik} \cdot (\Sigma_2)_{jl},
\end{equation*}
and Kronecker sum distribution \citep{greenewald2013kronecker,greenewald2017tensor}: 
\begin{equation*}
    \Sigma = \Sigma_1 \oplus_K \Sigma_2 := I_{p_1} \otimes_K \Sigma_2 +  \Sigma_1 \otimes_K I_{p_2}, \quad \text{i.e.,}\quad \SSigma_{ijkl} = (\Sigma_1)_{ik}1_{\{j=l\}} + (\Sigma_2)_{jl}1_{\{i=k\}}, i,j,k,l =1,\ldots, p, 
\end{equation*}
where $\otimes_K$ denotes the Kronecker product: $(A \otimes_K B)_{p_3(r-1)+v, p_4(s-1)+w}=A_{r s} B_{v w}$ for matrices $A\in\RR^{p_1, p_2}$ and $B\in\RR^{p_3, p_4}$.
These models simplify the entire covariance tensor into two matrices $\Sigma_1$ and $\Sigma_2$, which greatly streamline subsequent analysis. Nevertheless, these tensor-to-matrix simplifications can impose certain limitations. Additionally, the simplified covariance tensor fails to discern the direction of covariates with higher variances, unlike the vector-based principal component analysis (PCA) technique. As a consequence, the existing literature does not provide a direct equivalent of PCA specifically designed for tensor data. Therefore, there is a disparity in the current research.

To address this disparity, this paper aims to introduce a novel framework for dimension reduction in a series of matrix data, referred to as \underline{Mo}de-wise \underline{P}rincipal S\underline{u}bspace \underline{P}ursuit (\texttt{MOP-UP}). The primary objective of \texttt{MOP-UP} is to extract concealed variations in both the row and column dimensions of data matrices. Specifically, for a collection of matrix data with a shared dimension, denoted as $X_1,\ldots, X_n\in \mathbb{R}^{p_1\times p_2}$, we aim to identify the common column and row subspaces represented by semi-orthogonal matrices\footnote{A semi-orthogonal matrix is defined as a matrix with orthonormal columns.}, $U\in \mathbb{R}^{p_1\times r_1}$ and $V\in \mathbb{R}^{p_2\times r_2}$, respectively. The objective is to approximate the following decomposition for each matrix $X_i$:
\begin{equation}\label{eq:X_k-approxiate-decomposition}
X_i \approx M + UA_i+B_i V^\top,
\end{equation}
where $i$ ranges from 1 to $n$ and $A_i$ and $B_i$ are score matrices that vary across the indices. Intuitively, the decomposition \eqref{eq:X_k-approxiate-decomposition} captures the row-wise and column-wise dimension-reduced subspaces, denoted by $U$ and $V$ respectively, which encompass the majority of the informative features present in $X_i$. 

%%%%%%%%%%
\subsection{Matrix Spiked Covariance Models and Higher-order Generalizations}\label{sec:high-order-PCA}
%%%%%%%%%%

To establish a statistical foundation for the \texttt{MOP-UP} framework and to serve as a source of inspiration for algorithmic and theoretical development, it is beneficial to review the conventional probabilistic PCA model \citep{tipping1999probabilistic} before delving deeper. Suppose $x_1,\ldots, x_n$ are a series of $p$-dimensional i.i.d. observations with mean vector $\mu$ and covariance matrix $\Sigma$. The goal of PCA is to seek a few loading vectors that explain most of the variance in data through the following decomposition,
\begin{equation}\label{eq:PCA-regular-decomposition}
	x_i = \mu + Ua_i^\top + z_i =\mu + \sum_{j=1}^r u_j a_{ij} + z_i.
\end{equation}
Here $U = [u_1,\ldots, u_r]\in \mathbb{R}^{p\times r}$ is a set of fixed and uniform orthogonal vectors for all observations, $a_{i1}, \ldots, a_{ir}$ are random values, $z_i$ represents the noise. Particularly, $U$ and $a$ are often referred to as ``loading" and ``principal component (PC) scores" in the literature. To theoretically analyze the performance of PCA, the following spiked covariance model was introduced and widely studied \citep{johnstone2001distribution,paul2007asymptotics,cai2013sparse,donoho2018optimal,cai2015optimal},
\begin{equation*}
    \Sigma = \sigma^2 I + U\Lambda U^\top = \sigma^2I + \sum_{i=1}^r\lambda_iu_iu_i^\top, \quad  U \in \mathbb{O}_{p, r}.
\end{equation*}
An equivalent form of this model can be obtained by algebraic calculation as
\begin{equation}\label{eq:Sigma-spiked-property}
    (\Sigma - \sigma^2I)U_{\perp}=0,\quad \text{$U_{\perp}$ is the orthogonal complement of $U$.}
\end{equation}
In the noiseless setting (i.e., $\sigma^2=0$), the low-rank property of the data is equivalent to the low-rank property of its covariance matrix, as illustrated by the correspondence between \eqref{eq:PCA-regular-decomposition} and \eqref{eq:Sigma-spiked-property}. We aim to extend this connection to the matrix-variate scenario.
Suppose $\X = [X_1,\ldots, X_n]$ is an order-3 dataset, where $X_1,\ldots, X_n$ are i.i.d. matrix observations with mean matrix $M$. 
Now we still seek a low-dimensional row subspace $U$ and a low-dimensional column subspace $V$ that can together explain most of the variance in $\X$. In analogy to the matrix PCA of \eqref{eq:PCA-regular-decomposition} and \eqref{eq:Sigma-spiked-property}, we consider the following two models 
\begin{equation}\label{eq:X_k-decomposition-highorder}
    X_i = M + UA_i^\top + B_i V^\top + Z_i, \quad i=1,\ldots, n,
\end{equation}
\begin{equation}\label{eq:matrix-covariate-spiked-model-kronecker}
    (V_\perp \otimes_K U_\perp)\T (\Cov(\vect(X)) - \sigma^2 I_{p_1p_2}) = 0,\quad \text{for some fixed semi-orthogonal matrices $U_{\perp}$ and $V_{\perp}$}, 
\end{equation}
where $\vect(X)$ denotes the vectorization of the matrix $X$, formed by stacking the columns of $X$ into a single column vector. \eqref{eq:matrix-covariate-spiked-model-kronecker} can be equivalently written as 
\begin{equation}\label{eq:matrix-covariate-spiked-model}
    (\Cov(X) - \sigma^2 {\I_{(p_1\times p_2)_2}}) \times_1 U_{\perp} \times_2 V_{\perp} = 0,\quad \text{for some fixed semi-orthogonal matrices $U_{\perp}$ and $V_{\perp}$}.
\end{equation}
Here, $\Cov(X) = \EE ((X - \EE X)\otimes (X - \EE X))$ denote the covariance tensor, $\otimes$ denotes the tensor product, and $\times_1$ and $\times_2$ represent the tensor-matrix product, which will be introduced in Section \ref{sec:notation}. The matrices $U$ and $V$ are analogous to $U$ in the regular PCA \eqref{eq:PCA-regular-decomposition} and can be referred to as the column and row loading matrices, respectively. The matrices $A_i$ and $B_i$ are random matrices that correspond to the scores $a_k$ in PCA \eqref{eq:PCA-regular-decomposition}, and can be referred to as score matrices. Additionally, $Z_i$ represents the noise involved in the process.

\begin{figure}[!ht]
    \centering
    \includegraphics[width=0.5\textwidth]{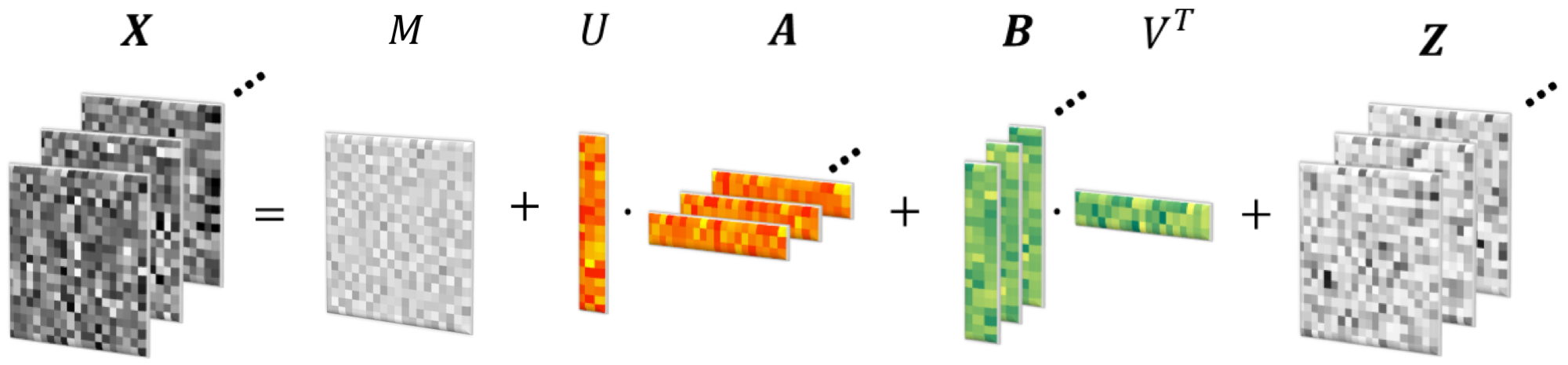}
    \caption{Illustration of a matrix spiked covariance model in a decomposition form}
    \label{fig:illustration-pca-model}
\end{figure}

Model \eqref{eq:X_k-decomposition-highorder} provides a rigorous statistical interpretation for the \texttt{MOP-UP} framework. Additionally, Models \eqref{eq:X_k-decomposition-highorder} and \eqref{eq:matrix-covariate-spiked-model-kronecker} (or \eqref{eq:matrix-covariate-spiked-model}) correspond to \eqref{eq:PCA-regular-decomposition} and \eqref{eq:Sigma-spiked-property} respectively, which are part of the classical spiked covariance model. Formulations \eqref{eq:X_k-decomposition-highorder} and \eqref{eq:matrix-covariate-spiked-model-kronecker} (or \eqref{eq:matrix-covariate-spiked-model}) are proven to be equivalent in the upcoming Theorem \ref{theorem_equivalence}. Based on this equivalence, this paper introduces and studies the class of {\bf matrix spiked covariance models} that satisfies either decomposition \eqref{eq:X_k-decomposition-highorder} or condition \eqref{eq:matrix-covariate-spiked-model-kronecker} (or \eqref{eq:matrix-covariate-spiked-model}). See Figure \ref{fig:illustration-pca-model} for an illustration of matrix spiked covariance model. 

%The definition of the matrix spiked covariance model can be generalized for the order-$d$ tensor variate. 
Furthermore, we say $\X\in \mathbb{R}^{\bp}$ has a {\bf rank-$\br$ high-order spiked covariance} if 
\begin{equation}\label{eq_tensor_spiked_cov_decomp_sec1}
    \X = \M + \sum_{k=1}^d \A_k \times_k U_k + \Z,
\end{equation}
or equivalently
\be \label{eq_tensor_spiked_cov_sec1}
\Cov(\X) = \SSigma_0 + \sigma^2 {\I_{\bp}},\ \SSigma_0 \in \mathbb{R}^{\bp \times \bp},\ \SSigma_0 \times_{k = 1}^d U_{k\perp} = 0.
\ee
Here, $\bp = p_1\times \cdots \times p_d$, $\M\in \mathbb{R}^{\bp}$ is a fixed mean tensor, $U_k\in \mathbb{O}_{p_k, r_k}$ are fixed semi-orthogonal matrices, $\A_k \in \mathbb{R}^{p_1\times \cdots \times p_{k-1}\times r_k\times p_{k+1}\times \cdots \times p_d}$ are random tensors with mean zero, and $\Z \in \mathbb{R}^{\bp}$ is a noise tensor, where all entries of $\Z$ has mean zero, covariance $\sigma^2 \I_{\bp}$, and is uncorrelated with random tensors $\A_1,\ldots, \A_d$. $\I_{\mathbf{p}}$ is the order-$(2d)$ tensor in $\mathbb{R}^{\mathbf{p} \times \mathbf{p}}$ with entries $(\I_{\mathbf{p}})_{\mathbf{q}, \mathbf{q}} = 1$ and $\mathbf{q} = (q_1, q_2, \cdots, q_d)$, $q_k \in \{1, \cdots, p_k\}$, and 0 elsewhere. \eqref{eq_tensor_spiked_cov_decomp_sec1} and \eqref{eq_tensor_spiked_cov_sec1} can be viewed as generalization of \eqref{eq:X_k-decomposition-highorder} and \eqref{eq:matrix-covariate-spiked-model-kronecker} (or \eqref{eq:matrix-covariate-spiked-model}), respectively, and their equivalence will be discussed in Theorem \ref{theorem_equivalence_tensor}. 

In summary, the proposed matrix and higher-order spiked covariance models relax the restrictive assumptions (such as the Kronecker product and sum) while still allowing a large number of free variables in the covariance tensor $\SSigma$. 

%%%%%%%%%%
\subsection{Our Contributions}\label{sec:contribution}
%%%%%%%%%%

We present the Mode-wise Principal Subspace Pursuit (\texttt{MOP-UP}) framework, designed to uncover concealed variations in both the row and column dimensions of data matrices. \texttt{MOP-UP} is supported by a novel class of matrix-variate spiked covariance models, representing a significant generalization beyond the traditional vector-case spiked covariance model. The decomposition formula \eqref{eq:X_k-decomposition-highorder} we introduce offers enhanced flexibility compared to existing dimension reduction formulations in the literature, enabling effective decomposition of a series of matrices. Our framework also extends the spiked covariance model to accommodate matrix and higher-order tensor samples, broadening its applicability from a statistical perspective.

To address dimension reduction for data matrices adhering to the \texttt{POP-UP} framework and the matrix spiked covariance model, we propose two novel methods: Average Subspace Capture (\texttt{ASC}) and Alternating Projection (\texttt{AP}). The \texttt{ASC} method introduces a new average projector estimator, distinct from the commonly used spectral initialization method found in existing literature. We highlight the geometric interpretations of \texttt{ASC} and provide theoretical guarantees that it achieves precise recovery of singular spaces almost sure in the noiseless scenarios. In contrast, our \texttt{AP} iteration procedure significantly deviates from the prevailing class of power iteration algorithms seen in the literature. We establish that \texttt{AP} essentially performs alternating minimization for an objective function that can be readily interpreted. Furthermore, we derive a statistical upper bound on the  estimation error for \texttt{ASC}, \texttt{AP}, as well as their combined usage, providing valuable insights into their performance.

We also study the methods and theory for higher-order spiked covariance models. Our investigation reveals notable differences in the algorithmic procedures for the spiked covariance model across various cases, including vector-variate, matrix-variate, and higher-order-variate scenarios. To provide a comprehensive overview, we summarize a comparison of the decomposition procedures for these different variate cases in Table \ref{table_comparison}.
\begin{table*}[ht]
\centering
\caption{Comparison of procedures for spiked models in different variate cases}\label{table_comparison}
    \begin{tabular}{c|ccc}
    \hline
    & Vector case & Matrix case & Higher-order tensor case\\ \hline
Initialization & SVD & Average Subspace Capture (\texttt{ASC}) & \texttt{HOSVD} \\
Followup iteration? & No & \begin{tabular}{c}
No (noiseless case)\\
Yes (noisy case)\\
\end{tabular} & Yes \\\hline
\end{tabular}
\end{table*}

To validate the efficacy of our model, we conduct data experiments on both synthetic and real-world datasets. First, we do simulation studies to show the tightness of our error bounds. Second, we apply the \texttt{MOP-UP} method to preprocess the MNIST dataset, reducing the dimensionality of the digit images before training a classifier. This approach yields interpretable dimension-reduced image features and demonstrated accurate prediction accuracy in the testing set when compared to traditional tensor methods. Third, we utilize the \texttt{MOP-UP} method on a human brain fMRI dataset obtained from a clinical study on cocaine use. Our results clearly demonstrate the effectiveness of our framework in preprocessing the data for the classification of cocaine and non-cocaine users, as well as for clustering region of interest (ROI) tasks. In both cases, our method showcases notable advantages in terms of the best prediction measurement and robustness across different input hyperparameters. 

Furthermore, we introduce a new technical tool of a matrix perturbation bound, which greatly aids in the technical analysis of the proposed \texttt{MOP-UP}. Our innovative methodology focuses on deriving a blockwise eigenspace perturbation bound, enabling us to establish our primary result with precision. This approach holds substantial value not only in situations where classical perturbation bounds, such as Davis-Kahan's theorem, may fall short in accurately assessing errors but also in other scenarios. Its applicability extends beyond the immediate context of our proposed \texttt{MOP-UP}, making it of independent interest. 

%%%%%%%%%%
\subsection{Literature Review}\label{sec:literature-review}
%%%%%%%%%%

In this section, we provide a brief overview of the related literature in the field. Principal Component Analysis (PCA) is one of the most well-established dimensionality reduction techniques, and numerous variations and related methods have been extensively studied. Textbooks such as \cite{jolliffe2005principal, abdi2010principal} offer comprehensive coverage of PCA and its variants, including factor analysis, independent component analysis, and projection pursuit. Several studies have investigated the distribution of eigenvalues in PCA under various assumptions. For example, \cite{johnstone2001distribution} examined the distribution of the largest eigenvalue in PCA when the covariance matrix is an identity matrix under Gaussianity. \cite{paul2007asymptotics} analyzed the eigenvalue distribution assuming Gaussianity and a specific covariance matrix structure. Shrinkage methods for eigenvalue regularization were studied by \cite{donoho2018optimal} under more general settings. Asymptotic properties of eigenvalues and eigenvectors were explored by \cite{bao2022statistical}. Extensions of PCA to matrices and images have also been investigated. Matrix PCA or 2-D PCA methods were developed to analyze matrix objects and images \citep{ye2004gpca, 10.1145/1015330.1015347}. \cite{1261097} considered applying linear transformations to the right side of observed matrices, while \cite{ye2004gpca} proposed an algorithm that incorporated spatial correlation of image pixels and applied linear transformations to both the left and right sides of observed matrices. \cite{he2005tensor} introduced the tensor subspace analysis algorithm, which treats input images as matrices residing in a tensor space and detects local geometric structures within that space. Furthermore, studies by \cite{koltchinskii2016asymptotics, koltchinskii2017concentration, koltchinskii2020efficient} have focused on the spectral distribution of sample covariance matrices. \cite{zhang2022heteroskedastic} proposed HeteroPCA, a variation of PCA that accounts for heteroskedasticity in the data.
\cite{efron2009set} considered a matrix $X$ whose rows are possibly correlated and aimed to test the hypothesis that the columns are independent of each other. He found that the row and column correlations of $X$ interact with each other in a way that complicates test procedures, essentially by reducing the accuracy of the relevant estimators. In contrast, our paper explores distinct problems, focusing on reducing multiple data matrices to dimension-reduced row and column subspaces. 

PCA relies on the mathematical tool of Singular Value Decomposition (SVD), which is a widely used matrix decomposition method. In recent years, SVD has been extended to tensor objects, leading to various generalizations such as Canonical Polyadic (CP) decomposition \citep{anandkumar2014guaranteed, ouyang2023multiway}, tensor train \citep{zhou2022optimal}, and Tucker decomposition \citep{hitchcock1927expression}. To find the best low Tucker rank approximation of a given tensor,  \cite{de2000multilinear} introduced Higher Order Singular Value Decomposition (\texttt{HOSVD}), and \cite{de2000best} introduced an alternating least squares algorithm known as High Order Orthogonal Iteration (\texttt{HOOI}). \texttt{HOOI} iteratively projects the tensor into a lower-dimensional space along each mode. The statistical modeling and performance analysis of \texttt{HOOI} were explored in \cite{zhang2018tensor}. However, these previous works focused on decomposing a single tensor without considering multiple samples from different subjects. The most relevant paper to our work is \cite{mpca}, which addressed this limitation by generalizing \texttt{HOOI} to handle multiple tensor observations. Their method, called Multilinear Principal Component Analysis (\texttt{MPCA}), extended the framework to incorporate multiple tensors. Several variations of \texttt{MPCA} have been proposed, including a TTP-based MSL algorithm \citep{tao2008bayesian}, robust \texttt{MPCA} \citep{inoue2009robust}, nonnegative \texttt{MPCA} \citep{panagakis2009non}, and others. A survey by \cite{LU20111540} provides a comprehensive summary of methods in this field, including these variations and techniques. 

These developments in PCA, SVD, and tensor decomposition methods have partly inspired the framework and algorithms proposed in our work.

%%%%%%%%%%
\subsection{Organization}\label{sec:organization}
%%%%%%%%%%

The remainder of this paper is organized as follows. In Section \ref{sec:model}, we provide notation, preliminaries, and a detailed discussion of the matrix spiked covariance model. We then introduce our algorithm in Section \ref{section_algorithm} and discuss its interpretation in Section \ref{section_geom_interpretaion}. We compare our model and algorithm to other methods in Section \ref{section_difference_comparison}. The theoretical properties of the algorithms are developed in Section \ref{section_theo_analysis}. Specifically in Section \ref{section_tech_lemma}, we introduce a technical lemma, a blockwise eigenspace perturbation bound, which plays a key role in our analysis. Furthermore, we present real data experiments in Section \ref{section_real_data}. Finally, we discuss the generalization to higher-order tensor cases and summarise our results in Section \ref{sec:extend_to_tensor}. Simulation studies, additional real data analyses, and all technical proofs are collected in the Supplementary Materials.

%%%%%%%%%%%%%%%%%%%%%%%%
\section{Models}\label{sec:model}
%%%%%%%%%%%%%%%%%%%%%%%%

%%%%%%%%%%
\subsection{Notation and Preliminaries}\label{sec:notation}
%%%%%%%%%%

In this work, lowercase letters ($u,v,\mu$, etc.) represent scalars or vectors; uppercase letters ($A, B, U$, etc.) represent matrices; and bold uppercase letters ($\X, \Z$, etc.) represent tensors. 
For variables $x$ and $y$, $x \lesssim y$ represents that there exists some constant $c$ that does not depend on $x$ or $y$ such that $x \leq cy$. 
For a vector $a$, $\|a\|$ denotes its $l_2$ norm. 
Let $I$ be the identity matrix with an appropriate dimension based on the context. 
For a matrix $A\in \RR^{p\times q}$, 
$\vect(A) \in \RR^{pq}$ denotes the vectorization of the matrix $A$, formed by stacking the columns of $A$ into a single vector.
$\sigma_i(A)$ represents the $i$th singular value of $A$, and all the singular values are ordered by its magnitude: $\sigma_1(A)\geq \sigma_{2}(A) \geq \cdots \geq \sigma_{\min\{p,q\}}\geq 0$; 
$\SVD_{r}(A) = [u_1, \cdots, u_r]$ represents the matrix consisting of the top $r$ left singular vectors of $A$, where $u_i$ is the singular vector of matrix $A$ corresponding to the singular value $\sigma_i(A)$;
$P_A = A(A^\top A)^\dagger A$ denotes an orthogonal projection matrix onto its column space, where $(\cdot)^\dagger$ is the matrix pseudo-inverse; 
$\|A\|$ is the spectral norm of $A$, which is equal to its largest singular value, $\sigma_1(A)$; 
$\|A\|_F = \sqrt{\tr(AA^\top)}$ is the Frobenius norm of $A$. 

The kernel (null space) of $A$ is denoted as $\text{ker}(A) = \{v : Av = 0\}$. The linear space spanned by all columns of $A$ is denoted as $\text{Span}(A) = \{v = Aw : w \in \mathbb{R}^p\}$. The sum of two linear spaces $\mathcal{V}$ and $\mathcal{W}$ is represented as $\mathcal{V} + \mathcal{W} = \{u = v + w : v \in \mathcal{V}, w \in \mathcal{W}\}$. We define $A\mathcal{V} = \{Av : v \in \mathcal{V}\}$ as the range of $A$ constrained to $\mathcal{V}$. When $A$ is symmetric with dimensions $p = q$, $\lambda_{r}(A)$ represents its $r$th eigenvalue, ordered such that $\lambda_1 \geq \cdots \geq \lambda_p$, and $\Eigen_r(A) = [u_1, \cdots, u_2]$ represents the matrix consisting of the top $r$ eigenvectors of $A$. Notice when $A$ is positive semi-definite, we have $\Eigen_r(A) = \SVD_r(A)$. For matrices $A\in\RR^{p_1, p_2}$ and $B\in\RR^{p_3, p_4}$, $A \otimes_K B \in \RR^{p_1p_3 \times p_2p_4}$ denotes their Kronecker product, which is defined element-wise as $(A \otimes_K B)_{p_3(r-1)+v, p_4(s-1)+w}=A_{r s} B_{v w}$ for $r \in \{1, \cdots, p_1\}, s \in \{1, \cdots, p_2\}, v \in \{1, \cdots, p_3\}$ and $w \in \{1, \cdots, p_4\}$. We denote $\mathbb{O}_{p, r}:=\left\{{U} \in \mathbb{R}^{p \times r}: {U}^{\top} {U}={I}\right\}$ as the set of all $p$-by-$r$ semi-orthonormal matrices, i.e., matrices with orthonormal columns. For $U \in \mathbb{O}_{p, r}$, $U_{\perp}$ represents a matrix in $\mathbb{O}_{p, p-r}$ whose columns are orthogonal to the columns of $U$. %To measure the difference between two subspaces, several popular distance metrics are discussed in the literature, including Distance with optimal rotation, Distance between projection matrices, and Geometric construction via principal angles. 
In this work, we employ the $\sin \Theta$ distance to characterize the distance between subspaces. For any $U, V \in \mathbb{O}_{p, r}$, we define $\|\sin \Theta(U, V)\| = \|U_\perp^\top V\| = \|UU^\top - VV^\top\|$.

An order-$d$ tensor $\A\in \RR^{p_1\times \cdots \times p_d}$ can be viewed as a multidimensional array, where $(i_1, \cdots, i_d)$ maps to $\A_{i_1, \cdots, i_d} \in \RR$. For convenience, we define $\bp = p_1\times \cdots \times p_d$. For a matrix $B \in \mathbb{R}^{p_k \times r_k}$, the mode-$k$ product of tensor $\A$ by matrix $B$ is denoted as $\A \times_k B \in \mathbb{R}^{p_1 \times \cdots \times p_{k-1} \times r_k \times p_{k+1} \times \cdots \times p_d}$ and defined as $(\A \times_k B)_{i_1, \cdots, i_d} = \sum_{j=1}^{p_k} \A_{i_1 \cdots i_{k-1} j i_{k+1} \cdots i_d} B_{i_k j}$. The mode-$k$ unfolding of tensor $\A$ is denoted as $\mathcal{M}_k(\A) \in \mathbb{R}^{p_k \times p_{-k}}$ and defined as $(\mathcal{M}_k(\A))_{i_k, h} = \A_{i_1 \cdots i_d}$, where $h = i_1 + p_1(i_2 - 1) + \cdots + \prod_{j=1}^{k-1} p_j (i_{k+1}-1) + p_{k+1} \prod_{j=1}^{k-1} p_j (i_{k+2}-1) + \cdots + \prod_{j \neq k, j \leq d-1} p_j (i_d - 1)$. When referring to a random tensor $\X$, ${\X_i}$ denotes its i.i.d. samples. If the random tensor already has a sub-index (e.g., $\A_k$), a comma is used to separate the sample index and its original sub-index (e.g., $\A_{i,k}$). 
For two tensors $\A \in \RR^{p_1\times \cdots \times p_d}$ and $\B \in \RR^{q_1\times \cdots \times q_k}$, the operation $\A \otimes \B \in \RR^{\bp \times \bold{q}}$ denotes the tensor product and $(\A \otimes \B)_{(i_1, \cdots, i_d, j_1, \cdots, j_k)} = \A_{(i_1, \cdots, i_d)} \B_{(j_1, \cdots, j_k)}$. The tensor product ``$\otimes$" should not be confused with the Kronecker product ``$\otimes_K$", which was defined earlier.
The covariance tensor $\Cov(\X)$ of random tensor $\X \in \RR^{p_1\times \cdots \times p_d}$ is defined as $\Cov(\X) = \EE ((\X - \EE \X) \otimes (\X - \EE \X))$, i.e., $\Cov(\X)_{(i_1, \cdots, i_d, j_1, \cdots, j_d)} =\EE [(\X - \EE \X)_{(i_1, \cdots, i_d)}(\X - \EE \X)_{(j_1, \cdots, j_d)}] $.
When $x$ is a random vector, $\Cov(x)$ is the covariance matrix. 
The symbol $\I_{\mathbf{p}^d}$ represents an order-$(2d)$ tensor in $\mathbb{R}^{\mathbf{p} \times \mathbf{p}}$ with entries $(\I_{\mathbf{p}_d})_{\mathbf{q}, \mathbf{q}} = 1$, where $\mathbf{q} = (q_1, q_2, \cdots, q_d)$, $q_k \in \{1, \cdots, p_k\}$, and 0 elsewhere. The symbol $\I_{(p_1\times p_2)_2}$ represents an order-$4$ tensor in $\mathbb{R}^{p_1 \times p_2 \times p_1 \times p_2}$ with entries $(\I_{(p_1\times p_2)_2})_{\mathbf{q}, \mathbf{q}} = 1$, where $\mathbf{q} = (q_1, q_2)$, $q_1 \in \{1, \cdots, p_1\}, q_2 \in \{1, \cdots, p_2\}$, and 0 elsewhere. 

We summarise notations in Table \ref{table_notation}, and any additional notation will be introduced and defined when they are first used. 
\begin{table}[ht]
    \centering
    \begin{tabular}{ll}
         \hline
         Notation&  \\
         \hline
         $\vect(A)$ & Vectorization of matrix $A$ by stacking the columns\\
         $\|A\|$ & Operator norm of matrix $A$  \\
         $\|A\|_F$ & Frobenius norm of matrix $A$  \\
         $\sigma_i(A)$ & $i$th singular value of matrix $A$ \\
         $\lambda_i(A)$ & $i$th eigenvalue of symmetric matrix $A$ \\
         $\SVD_i(A)$ & Matrix of top $i$ left singular vectors of matrix $A$ \\
         $\Eigen_i(A)$ & Matrix of top $i$ eigenvectors of symmetric matrix $A$ \\
         $\Span(A)$ & Linear span (range) of matrix $A$ \\
         $\ker(A)$ & Kernel (null space) of matrix $A$ \\
         $P_A$ & Orthogonal projection matrix onto column space of matrix $A$\\
         $U_\perp$ & Orthonormal complement to semi-orthonormal matrix $U$ \\
         $\|\sin \Theta(U, V)\|$ & Sine theta distance between semi-orthonormal matrices $U$ and $V$ \\
         $A \otimes_K B$ & Kronecker product of matrix $A$ and matrix $B$ \\
         $A \oplus_K B$ & Kronecker sum of matrix $A$ and matrix $B$ \\
         $A \mathcal{V}$ & Range of matrix A constrained to linear space $\mathcal{V}$ \\
         $\mathcal{V} + \mathcal{W}$ & Sum of linear space $\mathcal{V}$ and linear space $\mathcal{W}$ \\
         $\mathbb{O}_{p, r}$ &  Space of $p$-by-$r$ semi-orthonormal matrices \\
         $\X \otimes \Y$ & Tensor product of tensor $\X$ and tensor $\Y$ \\
         $\A \times_k B$ & mode-k product of tensor $\A$ by matrix $B$\\
         $\MM_k (\A)$ & mode-$k$ unfolding of tensor $\A$\\
         $\Cov (\X)$ & Covariance tensor of random tensor $\X$\\
         $\I_{\mathbf{p}_d}$ & A tensor with entries $(\I_{\mathbf{p}_d})_{\mathbf{q}, \mathbf{q}} = 1$, where $\mathbf{q} = (q_1, q_2, \cdots, q_d)$, and 0 elsewhere \\
         \hline
    \end{tabular}
    \caption{Notations. See detailed explanation in Section \ref{sec:notation}. }
    \label{table_notation}
\end{table}
%%%%%%%%%%
\subsection{Matrix Spiked Covariance Model}
%%%%%%%%%%

We formally introduce the following matrix spiked covariance model as follows.
\begin{Definition}[High-order Spiked Covariance Model Matrix Variate Case]\rm \label{definition_spiked_covariance}
    Suppose $X \in \mathbb{R}^{p_1 \times p_2}$ is a random matrix. We say $X$ has a \emph{rank-$(r_1 , r_2)$ high-order spiked covariance}, if there exists $\sigma^2>0$, $U\in \mathbb{O}_{p_1, r_1}$, and $V\in \mathbb{O}_{p_2, r_2}$, such that
    \[
    \Cov(\vect(X)) = \EE (\vect(X) - \EE \vect(X))\T (\vect(X) - \EE \vect(X)) = \Sigma_0 + \sigma^2 I_{p_1p_2},\ \Sigma_0 \in \mathbb{R}^{p_1p_2 \times  p_1p_2},
    \]
    \begin{equation}
        (V_\perp \otimes_K U_\perp)\T \Sigma_0  = 0; \notag
    \end{equation}    
     or equivalently, with tensor notations,
    $$
    \Cov(X) = \EE \left((X - \EE X) \otimes (X - \EE X)\right)  = \SSigma_0 + \sigma^2 {\I_{(p_1\times p_2)_2}},\ \SSigma_0 \in \mathbb{R}^{p_1 \times p_2 \times p_1 \times p_2}, \quad         \SSigma_0 \times_1 U_\perp \times_2 V_\perp = 0.$$
\end{Definition}
To ensure the existence of $U_\perp$ and $V_\perp$, we always assume $p_i > r_i$ for all $i$ in this work. The following theorem shows that the high-order spiked covariance model can be equivalently written as a decomposition form \eqref{eq:decomposition-equivalence-matrix} as depicted in Figure \ref{fig:illustration-pca-model}. 

\begin{Theorem}[Equivalent Definitions for High-order Spiked Covariance]\label{theorem_equivalence}
    $X\in \mathbb{R}^{p_1\times p_2}$ satisfies the high-order spiked covariance model if and only if there exist a deterministic matrix $M$, random matrices $B\in \mathbb{R}^{p_1\times r_2}$ and $A\in \mathbb{R}^{r_1\times p_2}$ with mean 0 such that
    \begin{equation}\label{eq:decomposition-equivalence-matrix}
        X = M + U A + BV\T + Z.
    \end{equation}
    Here, $U\in \mathbb{O}_{p_1, r_1}, V\in \mathbb{O}_{p_2, r_2}$ are fixed semi-orthogonal matrices, $Z\in \mathbb{R}^{p_1\times p_2}$ is a random matrix, where all entries of $Z$ are independent with mean zero and covariance $\sigma^2$, and are uncorrelated with $A,B$.
\end{Theorem}

The question of identifiability is particularly important: if a population covariance tensor $\text{Cov}(X)$ satisfies a high-order spiked covariance model (i.e., \eqref{eq:decomposition-equivalence-matrix} holds), when can the subspaces $\Span(U)$ and $\Span(V)$ be uniquely identified based on $X$? The following theorem provides a mild sufficient condition for identifiability.
\begin{Theorem}[Identifiability Condition for Matrix Spiked Covariance Model]\label{theorem_identifiability}
    Suppose $Y = UA + BV\T$, where $U\in \mathbb{O}_{p_1, r_1}, V\in \mathbb{O}_{p_2, r_2}$ are deterministic matrices and $A \in \mathbb{R}^{r_1 \times p_2}, B \in \mathbb{R}^{p_1 \times r_2}$ are random matrices. Suppose for any nonzero $v\in \mathbb{R}^{p_2}$ and any affine subspace (In this work, affine subspace refers to $\{v+e_1 u_1 + \cdots + e_ru_r: e_1,\ldots, e_r\in \mathbb{R}\}$, where $v, u_1,\ldots, u_r$ are all vectors of the same dimension.) $\mathcal{W} \subseteq \RR^{p_1}$, either $\PP\left(U A v \in \mathcal{W}| B \right) < 1$ or $\Span(U)\subseteq \mathcal{W}$. 
    Then, $U$ is identifiable in the sense that for any fixed $U^\prime \in \mathbb{O}_{p_1, r_1}$, if $\|\sin \Theta (U, U^\prime)\| \neq 0$, then $\SSigma \times_1 P_{U_{\perp}^\prime} \times_2 P_{V_{\perp}^\prime} \neq 0$ for any fixed $V^\prime \in \mathbb{O}_{p_2, r_2}$, where $\SSigma$ is the covariance tensor of $Y$.
\end{Theorem}

\begin{Remark}
    The condition on $A$ is guaranteed if, for any fixed vector $v_1 \in \mathbb{R}^{r_1}\setminus \{0\}$, the random vector $A v_1$ has a conditional density given $B$. When $d = 1$, this condition reduces to for a random variable $A$, $P(A = r| B) = 0$ for all $r\in \mathbb{R}$.
\end{Remark}

\begin{Example}[An Identifiable Example of Matrix Spiked Covariance Model]
    Let all entries of $A$ be i.i.d. Gaussian and independent of $B$. Note that for any given nonzero vector $v\in \mathbb{R}^{p_2}$, entries of $Av$ are also i.i.d. Gaussian. So, we have $\rank(\Cov(UAv)) = \rank(\EE(UAv v\T A\T U\T )) = \rank(UU\T) = r_1$. Thus, $\PP(UAv \in \mathcal{W} | B ) = \PP(UAv \in \mathcal{W} ) = 0$ for any $\mathcal{W}$ affine subspace such that $\Span(U)\subsetneq \mathcal{W}$, which implies $U$ is identifiable by Theorem \ref{theorem_identifiability}. 
\end{Example}

%\begin{Example}[Unidentifiable Example of Matrix Spiked Covariance Model]
%    Assume that $A$ is independent of $B$, and that the column vector $a_j, i = 1, \ldots, p_2$ of $A$ are i.i.d. with some distribution such that for some nontrivial subspace $\mathcal{W} \subsetneq \mathbb{R}^{r_1}$ with dimention $d_\mathcal{W}$ and some constant $\delta > 0$, we have $\PP(\exists \alpha_1 \in \RR: a_1 \in \mathcal{W}) > \delta$. Then, $\PP(\Span(UA) \subseteq U \mathcal{W}) > \delta^{p_2}$, where $U \mathcal{W}$ is the image of $U$ restricted on $\mathcal{W}$. Note that the dimension of $U \mathcal{W}$ is $d_\mathcal{W} < r_1$. So, for any subspace $\mathcal{V} \in \RR^{p_1}$ with dimension $r_1 - d_\mathcal{W}$, let $U'$ be the projector to $\mathcal{V} \oplus U \mathcal{W}$, then we have $\PP(P_{U'_\perp} X P{V_\perp} = 0) \geq \PP(P_{U'_\perp} (UA + BV\T) P{V_\perp} = 0 | \Span(UA) \subseteq U \mathcal{W}) \PP(\Span(UA) \subseteq U \mathcal{W}) = \delta^{p_2}$. Thus, $U$ is not identifiable. 
%\end{Example}

\begin{Example}[An Unidentifiable Example of Matrix Spiked Covariance Model]\label{example_unidentifiable}
    Assume that $A$ is independent of $B$, and that the column vectors $a_j$, for $j = 1, \ldots, p_2$, of $A$ are i.i.d. with some distribution. Suppose there exists a fixed subspace $\mathcal{W} \subsetneq \mathbb{R}^{r_1}$ with dimension $1\leq \dim(\mathcal{W}) \leq r_1-1$ such that $\PP(a_j \in \mathcal{W}) = 1$.

    In this construction, $U$ is not identifiable. This is because $\PP(\Span(UA) \subseteq U \mathcal{W}) = 1$, where $U \mathcal{W} = \{Uw: w\in \mathcal{W}\}$ is the image of map $U$ with the input $\mathcal{W}$. Note that $\text{dim}(U \mathcal{W}) < r_1$. So, for any subspace $\mathcal{V} \in \RR^{p_1}$ with dimension $r_1 - \dim(\mathcal{W})$, let $U'$ be the projector to $\mathcal{V} + U \mathcal{W}$, then we have $\PP(P_{U'_\perp} X P_{V_\perp} = 0) \geq \PP(P_{U'_\perp} (UA + BV\T) P_{V_\perp} = 0 | \Span(UA) \subseteq U \mathcal{W}) \PP(\Span(UA) \subseteq U \mathcal{W}) = 1$. In this case, $\Cov X \times_1 U'_\perp \times_2 V_\perp = (\EE X \otimes X)\times_1 U'_\perp \times_2 V_\perp = \EE [(X\times_1 U'_\perp \times_2 V_\perp) \otimes X] = 0$. Thus, $X$ also satisfies the spiked covariance model with $(U', V)$ by definition. Meanwhile, the condition ``Suppose for any nonzero $v\in \mathbb{R}^{p_2}$ and any affine subspace $\mathcal{W} \subseteq \RR^{p_1}$, either $\PP\left(U A v \in \mathcal{W}| B \right) < 1$ or $\Span(U)\subseteq \mathcal{W}$" also fails, because $\mathcal{W} \subsetneq \Span(U)$ and $1 = \PP\left(U A v \in \mathcal{W}| B \right) = \PP\left(U A v \in \mathcal{W}\right)$ for $\forall v$. 
    %Meanwhile, Condition ``Suppose for any nonzero $v\in \mathbb{R}^{p_2}$ and any affine subspace $\mathcal{W} \subseteq \RR^{p_1}$, either $\PP\left(U A v \in \mathcal{W}| B \right) < 1$ or $\Span(U)\subseteq \mathcal{W}$. '' also fails.
\end{Example}

%%%%%%%%%%%%%%%%%%%%%%%%
\section{Algorithm: MOP-UP}\label{section_algorithm}
%%%%%%%%%%%%%%%%%%%%%%%%

In this section, we focus on the following key question of \texttt{MOU-UP}: given observations $\{X_i\}_{i=1}^n \in \mathbb{R}^{p_1 \times p_2}$ with the high-order spiked covariance, how we can achieve a sufficient dimension reduction by recovering the loading matrices $U$ and $V$.

\subsection{Algorithm}
The overall algorithm includes two steps: initialization and iterative update, which are described below. The algorithms will be interpreted in Section \ref{section_geom_interpretaion}.

\subsubsection*{Initialization via Average Subspace Capture (ASC).}

We first centralize $\{X_i\}$ by subtracting their mean matrix $\bar{X} = \frac{1}{n}\sum_{i=1}^n X_i$. Then we introduce an initialization method as summarized in Algorithm \ref{algorithm_average_projection}. 
Assume $p_1 \geq p_2$, then if $r_1+ r_2 < p_2$, the time complexity of \texttt{ASC} is $O(n(p_1p_2^2 + p_1^2(r_1+ r_2)))$. 
The initialization method builds upon the geometric analysis to be presented in Section \ref{section_geom_interpretaion}.

\begin{algorithm}
    \caption{Initialization: Average Subspace Capture (\texttt{ASC})}
    \label{algorithm_average_projection}
    \algrenewcommand\algorithmicensure{\textbf{Output:}}
        \algrenewcommand\algorithmicrequire{\textbf{Input:}}
    \begin{algorithmic} 		
    \Require Data matrices $\{X_i\}_{i = 1}^n \in \mathbb{R}^{ p_1\times p_2}$, target rank $(r_1, r_2)$
        \Ensure {Estimation $\hat U, \hat V$}
        \State Centralization: $X_i \gets X_i - \bar{X}$

        \If{$r_1 + r_2 < p_1$}
        \State $\hat{U} \gets \Eigen_{r_1}\left( \frac{1}{n}\sum_{i = 1}^n \SVD_{r_1 + r_2}\left( X_i \right)\cdot {\SVD_{r_1 + r_2}\left(  X_i \right)}\T\right)$
        \Else
        \State  $\hat{U} \gets I_{p_1}$
        \EndIf
        \If{$r_1 + r_2 < p_2$}
        \State  $\hat{V} \gets \Eigen_{r_2}\left( \frac{1}{n}\sum_{i = 1}^n \SVD_{r_1 + r_2}\left( X_i\T \right)\cdot {\SVD_{r_1 + r_2}\left( X_i\T  \right)}\T\right)$
        \Else
        \State $\hat{V} \gets I_{p_2}$
        \EndIf\\
        \Return $\hat U, \hat V$ 
    \end{algorithmic}
\end{algorithm}

\subsubsection*{Update via Alternating Projection (AP).}

Next, starting from the initialization $\{\hat{U}_j^{(0)}\}_{j=1}^d$ obtained above, we perform the following iterative steps, summarized in Algorithm \ref{algorithm_iterative_projection}:

\begin{algorithm}[t]
\caption{Alternating Projection (\texttt{AP})}
    \label{algorithm_iterative_projection}
        \algrenewcommand\algorithmicensure{\textbf{Output:}}
        \algrenewcommand\algorithmicrequire{\textbf{Input:}}
    \begin{algorithmic}[t]
    \Require Data matrices $\{ X_i\}_{i = 1}^n \in \mathbb{R}^{ p_1\times p_2}$, target rank $(r_1, r_2)$, initialization $\hat U^{(0)}, \hat V^{(0)}$, maximal number of iteration $t_0$.
        \Ensure{Estimation $\hat U^{(t)}, \hat V^{(t)}$}
        \State Centralization: $X_i \gets X_i - \bar{X}$
        \For{$t$ in $1:t_0$}
        \State $\hat{V}^{(t)} \gets \Eigen_{r_2}\left(\sum_{i = 1}^{n} X_i \T \hat{U}_{\perp}^{(t-1)} \hat{U}_{\perp}^{(t-1)\top}  X_i \right)$
        \State $\hat{U}^{(t)} \gets \Eigen_{r_1}\left(\sum_{i = 1}^{n} X_i \hat{V}_{\perp}^{(t-1)} \hat{V}_{\perp}^{(t-1)\top}  X_i\T\right)$
        \State Break the for loop if converged or maximum number of iteration $t_0$ reached
        \EndFor\\
        \Return $\hat U^{(t)}, \hat V^{(t)}$
    \end{algorithmic}
\end{algorithm}

\begin{enumerate}
\item Multiply each centralized sample $(X_i - \bar{X})$ by $\hat{U}_{\perp}^{(t-1)}$ on its left or $\hat{V}_{\perp}^{(t-1)}$ on its right, and then multiply the transpose of the resulting matrix: $X_i \T \hat{U}_{\perp}^{(t-1)} \hat{U}_{\perp}^{(t-1)\top}  X_i$ and $X_i \hat{V}_{\perp}^{(t-1)} \hat{V}_{\perp}^{(t-1)\top}  X_i\T$.
\item Define $\hat{U}^{(t)}$ and $\hat{V}^{(t)}$ as the matrix consisting of the first $r_1$ and $r_2$ eigenvectors of the sum of the matrices obtained from the previous step:
\begin{align*}
\hat{V}^{(t)} = \Eigen_{r_2}\left(\sum_{i = 1}^{n} X_i \T \hat{U}_{\perp}^{(t-1)} \hat{U}_{\perp}^{(t-1)\top}  X_i \right),\\
\hat{U}^{(t)} = \Eigen_{r_1}\left(\sum_{i = 1}^{n} X_i \hat{V}_{\perp}^{(t-1)} \hat{V}_{\perp}^{(t-1)\top}  X_i\T\right).
\end{align*}
\end{enumerate}

We repeat these steps until convergence or a maximum number of iterations is reached. By iterating this procedure, we obtain estimates $\hat{U}^{(t)}$ and $ \hat{V}^{(t)}$ that capture the loading matrices $U$ and $V$ in the high-order spiked covariance model. 
Assume $p_1 \geq p_2$, then the time complexity of each iteration in \texttt{AP} is $O(n(p_1^2(p_2-r_2) + p_2^2(p_1 - r_1)) + p_1^3)$. 
Our algorithm is inspired by alternating minimization, where a detailed explanation is given in Section \ref{section_geom_interpretaion}. 

We further consider how to denoise each matrix observation, i.e., to estimate $X_i - Z_i = UA_i + B_i V^\top$. First, matrices $A_i, B_i$ are not identifiable from $X_i$ even if $U$ and $V$ are known exactly because there are multiple equivalent decompositions of $UA_i+B_iV^\top$:
$$UA_i + B_i V^\top = U(A_i + U^\top B_i V^\top) + U_\perp U_\perp^\top B_i V^\top = UA_i V_\perp V_\perp^\top + (UA_i V + B_i)V^\top.$$
So, it is infeasible to apply the plugin estimates of $A_i, B_i$ to estimate $UA_i + B_iV^\top$. On the other hand, $UA_i + B_iV^\top$ is in the subspace $\mathcal{P}(U, V) = \{H \in \mathbb{R}^{p_1\times p_2}: P_{U}HP_V = 0\}$. Thus, it is natural to apply the projection operator to estimate the signal part $UA_i + B_i V\T$ of the observation matrix $X_i$:
\begin{equation}\label{eq:hat-X_i}
\hat{X}_i = P_{\mathcal{P}(\hat{U}, \hat{V})}(X_i) = X_i - \hat{U}_\perp \hat{U}_\perp^\top (X_i - \bar{X})\hat{V}_\perp \hat{V}_\perp^\top.
\end{equation}

\subsubsection*{Rank Selection.} \label{section_rank_selection}
The target rank can be determined through two approaches. Suppose $\hat{U}^{(r_1, r_2)}, \hat{V}^{(r_1, r_2)}$ 
are the output of \texttt{MOP-UP} with the input rank $(r_1, r_2)$. Firstly, a scree plot of the loss $\sum_{i = 1}^n\|P_{\hat U_{\perp}^{(r_1, r_2)}} (X_i - \bar{X}) P_{\hat V_{\perp}^{(r_1, r_2)}}\|_F^2$
%$\sum_{i = 1}^n \|\hat X_i^{(r_1, r_2)} - X_i\|_F^2$ 
can be utilized. Alternatively, a BIC-type criterion can be employed. Note that for a $p$-by-$r$ matrix with orthogonal columns, the number of free parameters is given by $(p-1) + (p-2) + \cdots + (p-r) = (2p-r-1)\times r/2$. Hence, in our model, the total number of parameters is $(r_1(2p_1-r_1-1) + r_2(2p_2-r_2-1))/2$. Consequently, the penalization term in BIC is defined as $\log\left(np_1 p_2 \right)(r_1(2p_1-r_1-1) + r_2(2p_2-r_2-1))/2$, and the rank $r_1, r_2$ can be determined by
\begin{equation*}
    \begin{split}
        (\texttt{BIC}) \quad (\hat{r}_1, \hat{r}_2) = \argmin_{r_1,r_2} & \log\left( \sum_{i = 1}^n\|P_{\hat U_{\perp}^{(r_1, r_2)}} (X_i - \bar{X}) P_{\hat V_{\perp}^{(r_1, r_2)}}\|_F^2\right)  \\
        & + \frac{\log\left(np_1 p_2 \right)}{2np_1p_2}(r_1(2p_1-r_1-1) + r_2(2p_2-r_2-1)).
    \end{split}
\end{equation*}

%%%%%%%%%%%%
\subsection{Interpretations}\label{section_geom_interpretaion}
%%%%%%%%%%%%

In this section, we provide interpretations for both the proposed \texttt{ASC} and \texttt{AP} algorithms. 

\subsubsection*{Interpretation of \texttt{ASC}.} We introduce the following key observation. 

\begin{Theorem}\label{theorem_span_sum}
Suppose $U\in \mathbb{O}_{p_1, r_1}, V \in \mathbb{O}_{p_2, r_2}$ are semi-orthogonal matrices, $A$ and $B$ are some random matrices with densities in $\RR^{p_2r_1}$ and $\RR^{p_1r_2}$ respectively, the population matrix satisfies $X = UA + BV\T$, and $\{X_i\}_{i=1}^n$ are i.i.d. copies of $X$. If $p_2 \geq r_1 + r_2$ and $nr_2 \leq  (n-1)(p_1- r_1)$, then $\Span(U)$ equals the common subspace of column spaces of all $X_i$, $\Span(U) {=} \bigcap_{i = 1}^n \Span(X_i)$, almost surely.
\end{Theorem}

Theorem \ref{theorem_span_sum} reveals that finding $U$ can be reduced to finding the intersection space of all $\Span(X_i)$ in the noiseless matrix spiked covariance model. Note that $\hat{P}_i := \SVD_{r_1+r_2}\left( X_i \right)\cdot {\SVD_{r_1+r_2}\left( X_i \right)}\T$ is a projection matrix and we have $\|\sum_{i = 1}^{n} \hat{P}_i/n\| \leq \sum_{i=1}^n \|\hat{P}_i/n\| = 1$. Suppose $\lambda_j$ and $e_j$ are the $j$-th eigenvalue and eigenvector of $\sum_{i = 1}^{n} \hat{P}_i/n$, respectively. Then $\lambda_j=1$ if and only if $e_j \in \bigcap_{i = 1}^n \Span(\hat{P}_i)=\bigcap_{i = 1}^n \Span(X_i)$. By Theorem \ref{theorem_span_sum}, we have $\Span(U) = \bigcap_{i = 1}^n \Span(X_i)$ and hence for $\forall u \in \RR^{p_1}$, $u \in \Span(U)$ is equivalent to that $u$ is an eigenvector of $\sum_{i = 1}^{n} \hat{P}_i/n$ corresponding to the eigenvalue 1. This leads to the following Corollary \ref{corollary_matrix_initialization}, which shows that \texttt{ASC} exactly recovers $U$ almost surely in the noiseless case under mild conditions.
\begin{Corollary} \label{corollary_matrix_initialization}
    Under the same condition as in Theorem \ref{theorem_span_sum}, Algorithm \ref{algorithm_average_projection} (\texttt{ASC}) exactly recovers $\Span(U)$ almost surely in the sense that $\hat U = UO$ for some orthogonal matrix $O \in \mathbb{O}_{r_1}$ almost surely. 
\end{Corollary}

On the contrary, the classical high-order singular value decomposition (\texttt{HOSVD}) \citep{de2000multilinear}, denoted as $\hat U = \SVD_{r_1}([X_1 ~ X_2 ~ \cdots ~ X_n]),$
has often been employed for initialization in various tensor problems \citep{zhang2018tensor,han2022exact}. However, it fails to exactly recover $U$. This limitation arises from the fact that $\Span(U)$ does not necessarily correspond to the singular subspace of $[X_1 ~ X_2 ~ \cdots ~ X_n]$. This discrepancy can even be observed in a simple scenario when $r_1 = r_2 = 1$, i.e., $X_i = u a_i\T + b_i v\T$. If $b_i \neq u$ and $b_i\T u \neq 0$, $u$ is not the left singular vector of $X_i$. 

\subsubsection*{Interpretation of \texttt{AP}.}

Given the nature of the high-order spiked covariance model from Definition \ref{definition_spiked_covariance}, it is logical to explore the minimization of the following objective function:
\begin{equation}\label{eq:objective-function}
    \begin{split}
        \min_{\substack{U\in \mathbb{O}_{p_1, r_1}\\ V\in \mathbb{O}_{p_2, r_2}}} \sum_{i=1}^n \left\|U_{\perp}^\top (X_i-\bar{X}) V_{\perp}\right\|_F^2.
    \end{split}
\end{equation}
However, the objective function \eqref{eq:objective-function} poses a significant challenge as it is highly non-convex and, in general, evaluating it can be NP-hard. To address this computational difficulty, the proposed \texttt{AP} (Algorithm \ref{algorithm_iterative_projection}) offers a solution that leverages the insights presented in the following proposition: Algorithm \ref{algorithm_iterative_projection} (\texttt{AP}) can be viewed as an alternative minimization scheme involving $U^{(t)}$ and $V^{(t)}$.
\begin{Proposition}\label{proposition_minimizer}
For any given matrices $X_i, i = 1, \cdots n$ and $V^\prime \in \mathbb{O}_{p_2, r_2}$,
we have
\[\begin{aligned}
        &\argmin_{U \in \mathbb{O}_{p_1, r_1}} \sum_{i = 1}^{n} \left\|U_{\perp}^\top (X_i-\bar{X}) V^\prime_\perp \right\|_F^2
        =& \left\{\Eigen_{r_1}\left( \sum_{i = 1}^{n}  (X_i-\bar{X}) V^\prime_\perp V^{\prime \top}_\perp (X_i-\bar{X})  \T \right) O: \forall O\in \mathbb{O}_{r_k}\right\}.
\end{aligned}\]
A similar result holds symmetrically for minimization over $V$. 
\end{Proposition}

%%%%%%%%%%%%%%%%%%%%%%%%
\subsection{Matrix Spiked Covariance Model versus Existing Models}\label{section_difference_comparison}
%%%%%%%%%%%%%%%%%%%%%%%%
Next, we briefly compare the proposed procedure with the conventional methods in the existing literature. 

\subsubsection*{Classic Spiked Covariance Model and PCA}

As mentioned in the introduction, the matrix and higher-order spiked covariance model can be viewed as a generalization of the classic spiked covariance model discussed in previous studies \citep{johnstone2001distribution, donoho2018optimal, paul2007asymptotics} and our \texttt{MOP-UP} framework can be viewed as a generalization of the regular PCA. In the classic spiked covariance model, we consider a scenario where $x_1,\ldots, x_n$ are independent and identically distributed (i.i.d.) instances of a $p$-dimensional random vector $x$, satisfying the condition:
\begin{equation}
    \EE x = \mu, \quad \Var(x) = \Sigma_0 + \sigma^2 I, \quad \Sigma_0 = \sum_{i=1}^r\lambda_i u_iu_i^\top, \notag
\end{equation}
where $\lambda_1\geq \cdots \geq \lambda_r \geq 0$ are the eigenvalues, $\{u_1,\ldots, u_r\}$ are orthonormal eigenvectors. Denote $U = [u_1,\ldots, u_r]$, and $U_{\perp}\in \mathbb{O}_{p, p-r}$ as the orthogonal complement of $U$. Then we have $\Sigma_0U_{\perp} = 0.$ 

Meanwhile, the proposed \texttt{AP} (Algorithm \ref{algorithm_iterative_projection}) in vector-variate case reduces to the regular PCA estimator:
$$\hat{U} = \Eigen_r\left(\frac{1}{n}\sum_{i = 1}^n(x_i - \bar{x})(x_i-\bar{x})\T\right).$$
There is no need to include any initialization step in this vector-variate case.

\subsubsection*{Mean-Based Methods in Matrix Denoising}

The decomposition $X=UA+BV\T +Z$ within our \texttt{MOP-UP} framework can be perceived as a ``signal-plus-noise" model, specifically falling under the category of the matrix perturbation problems. This problem has been extensively explored in the literature, with significant contributions documented in works such as \cite{cai2015optimal,cai2018rate,gavish2014optimal,koltchinskii2016asymptotics}, among others. In the context of these studies, the typical data format is $X = M + Z$, where $M$ represents a deterministic low-rank matrix, and $Z$ accounts for random noise. In such scenarios, a single observation often yields theoretically guaranteed estimations of both $M$ and singular subspaces. When dealing with multiple observations, \texttt{MPCA} \cite{mpca} offers a solution, which will be discussed later. However, in our specific case, even in the absence of noise ($X=UA+BV\T$), it is impossible to recover both $U$ and $V$ simultaneously from a single observation. As highlighted in the matrix perturbation literature, when recovering $U$, $BV\T$ essentially acts as noise, necessitating that $BV\T$ be bounded to satisfy certain signal-to-noise ratio conditions \cite{cai2018rate}, and vice versa. Therefore, our models require multiple observations, which distinguishes them significantly from the existing literature on matrix signal-plus-noise models.

%From the perspective of data denoising, the decomposition $X=UA+BV\T +Z$ within our \texttt{MOP-UP} framework can be perceived as a signal-plus-noise model, specifically falling under the category of the matrix perturbation problem. This problem has been extensively explored in the literature, with significant contributions documented in works such as \cite{gavish2014optimal, cai2015optimal, koltchinskii2016asymptotics, cai2018rate}, among others.

%In the context of these studies, the typical data format is $X = M + Z$, where $M$ represents a deterministic low-rank matrix, and $Z$ accounts for random noise. In such scenarios, a single observation often yields theoretically guaranteed estimations of both $M$ and singular spaces. When dealing with multiple observations, \texttt{MPCA} \citep{mpca} offers a solution, which we will discuss later.

%However, in our specific case, even in the absence of noise ($X=UA+BV\T$), attempting to recover both $U$ and $V$ simultaneously from a single observation is futile. As highlighted in the matrix perturbation literature, when recovering $U$, $BV\T$ acts as noise, necessitating that $BV\T$ be sufficiently small to satisfy certain signal-to-noise ratio conditions \citep{cai2018rate}, and vice versa. Therefore, multiple observations become imperative for our framework.

\subsubsection*{MPCA (2D-PCA) and HOOI}

The proposed matrix-variate high-order spiked covariance model is also related to the matrix case of \texttt{MPCA} \citep{mpca} (also known as \texttt{2D-PCA} \cite{ye2004generalized}), and both fit into the signal-plus-noise dimension reduction framework. \texttt{MPCA} aims to decompose the observation matrices to
\be \label{eq_mpca}
X_i=U S_i V\T + Z_i, \quad i=1,\ldots, n, 
\ee
where $S_i \in \RR^{r_1 \times r_2}$ is the core matrix representing individual unique signal and $Z_i$ is the noise. 
By decomposing $X_i$ into four blocks, we have:
$$X_i = P_U X_i P_V + P_U X_i P_{V_\perp} + P_{U_\perp} X_i P_V + P_{U_\perp} X_i P_{V_\perp}.$$
While \texttt{MPCA} focuses on extracting $P_U X_i P_V$ and treating the other three parts as residuals, our high-order spiked covariance model captures $P_U X_i P_V$, $P_U X_i P_{V_\perp}$, and $P_{U_\perp} X_i P_V$, while reducing the contribution of the fourth block $P_{U_\perp} X_i P_{V_{\perp}}$. As a result, the proposed \texttt{MOP-UP} outperforms \texttt{MPCA} when the columns and rows of the data contain important information that is not solely derived from their common space $P_U X_i P_V$.

\texttt{MPCA} can be solved using a variant of high-order orthogonal iteration \citep[\texttt{HOOI};][]{de2000best}, a broader class of algorithms widely employed in Tucker low-rank tensor decomposition. See \cite{mpca}. In the case of \texttt{MPCA}, $\hat{U}^{(t)}$ is computed at each iteration by projecting $X_i\T$ onto $\Span(\hat{V}^{(t-1)})$. In contrast, Algorithm \ref{algorithm_iterative_projection} in our approach projects $X_i\T$ onto the orthogonal complement of $\Span(\hat{V}^{(t-1)})$, denoted as $\Span(\hat{V}^{(t-1)})^\perp$:
\begin{equation*}
    \begin{split}
        \text{\texttt{HOOI}:}\quad & \hat{U}^{(t)} = \Eigen_{r_1}\left( \sum_{i = 1}^{n} \left(X_i  \hat{V}^{(t-1)} \hat{V}^{(t-1) \top} X_i\T \right)\right),\\
        \text{\texttt{AP} (Algorithm \ref{algorithm_iterative_projection}):} \quad & \hat{U}^{(t)} = \Eigen_{r_1}\left( \sum_{i = 1}^{n} \left(X_i  \hat{V}_\perp^{(t-1)} \hat{V}_\perp^{(t-1)\top} X_i\T \right)\right).
    \end{split}
\end{equation*}
This distinction arises from the fact that the matrix spiked covariance model considers only $U_\perp\T X_i V_\perp$ as the decomposition residual, whereas \texttt{MPCA} treats $U\T X_i V_\perp$, $U_\perp\T X_i V$, and $U_\perp\T X_i V_\perp$ as the decomposition residuals.

\subsubsection*{Kronecker Product and Kronecker Sum Models}

%Similar to the vector-variate case, 
The low-rankness of the covariance tensor serves as a model for reducing the covariance's number of free parameters. In the literature, the Kronecker product \citep{tsiligkaridis2013convergence, zhou2014gemini} and Kronecker sum \citep{banerjee2008model, greenewald2019tensor} structures are other well-studied models of the covariance, which were discussed in Section \ref{sec:intro}. %In the matrix-variate case, the covariance of a random matrix $X \in \RR^{p_1 \times p_2}$ is parameterized by $\Cov(\vect{X}) = \Sigma_1\otimes_K \Sigma_2$ (Kronecker product model) or $\Cov(\vect{X}) = \Sigma_1 \oplus_K \Sigma_2$ (Kronecker sum model), where $\Sigma_1 \in \RR^{p_1 \times p_1}$ and $\Sigma_2 \in \RR^{p_2 \times p_2}$. 
The covariance matrices of the Kronecker product and Kronecker sum models are full rank, and the number of free parameters is $p_1(p_1+1)/2 + p_2(p_2+1)/2 - 1$. The Kronecker product model admits the parameterization of the data matrix $X = M + \Sigma_1^{1/2}Z\Sigma_2^{1/2}$, where $M$ is a fixed matrix and all entries of $Z$ are i.i.d. standard normal. %whereas the Kronecker sum model does not have a simple decomposition of $X$. 
Furthermore, error bounds and convergence rates for the algorithms have been established to estimate the covariance matrix under Gaussianity or sub-Gaussianity assumptions. Examples include the Kronecker Graphical Lasso \citep{tsiligkaridis2013convergence}, Gemini \citep{zhou2014gemini}, and TeraLasso \citep{greenewald2019tensor}. 

    In comparison, the covariance structure considered in our framework is given by $(V_\perp \otimes_K U_\perp)\T (\Cov(\vect(X)) - \sigma^2 I_{p_1p_2}) = 0$, as described by Theorem 1. The number of free parameters is $(p_2r_1 + p_1r_2 - r_1r_2)(p_2r_1 + p_1r_2 - r_1r_2 + 1)/2 + p_1(p_1 - r_1) + p_2(p_2-r_2)$, which is significantly greater than the Kronecker product and Kronecker sum structures. Our algorithm focuses on estimating the loading $U$ and $V$, i.e., the subspaces of the covariance. Notably, the error bound for \texttt{ASC}, which will be established in Section \ref{section_theo_analysis}, does not assume any exact distribution, while the error bound for \texttt{AP} requires the sub-Gaussianity assumption.

%%%%%%%%%%%%%%%%%%%%%%%%
\section{Theoretical Analysis} \label{section_theo_analysis}
%%%%%%%%%%%%%%%%%%%%%%%%

In this section, we provide the theoretical guarantees for the proposed algorithm. Specifically, we establish the estimation error bounds for \texttt{ASC} and \texttt{AP} in Sections \ref{section_initialization} and \ref{section_iterative_projection} respectively. The combination of these bounds allows us to derive the desired estimation error bound for the proposed \texttt{MOP-UP} estimator in Section \ref{section_global}. 

%%%%%%%%%%%%
\subsection{Error Bound for Initialization via \texttt{ASC}}\label{section_initialization}
%%%%%%%%%%%%

Recall that Corollary \ref{corollary_matrix_initialization} demonstrates that \texttt{ASC} achieves exact recovery of $U$ in the absence of noise. The subsequent theorem addresses the scenario where noise is present.
\begin{Theorem}[Error bound of \texttt{ASC} in the noisy case]\label{theorem_average_projection_noisy_bound}
Suppose $U\in \mathbb{O}_{p_1, r_1}, V \in \mathbb{O}_{p_2, r_2}$ are fixed semi-orthogonal matrices, $A$ and $B$ are random matrices with densities in $\RR^{r_1 \times p_2}$ and $\RR^{ p_1\times r_2}$ respectively, $Z$ is a random noise matrix with i.i.d. entries in $\RR^{p_1 \times p_2}$ independent of $A$ and $B$, the population matrix satisfies $X = UA + BV\T + Z$,  $\{X_i\}_{i=1}^n$ are i.i.d. copies of $X$, $p_2 \geq r_1 + r_2$, and $nr_2 \leq (n-1)(p_1- r_1) $. For any $0 \leq c \leq 1/2$, define $C^* := \frac{c^2}{8} + \PP(4\|Z\| > c \sigma_{r}(UA+ BV\T))$. 
If we further have 
\[
    n \geq c_1 \log p_1 \max\left\{C^{*-2} , {\left(1 - \lambda_{1}\left(\EE P_{U_\perp U_\perp\T B}\right) \right)^{-2}} \right\}
\] for some constant $c_1$, 
then with probability greater than $1- \exp\left\{-n\left(c_1 \log p_1 \max\left\{C^{*-2} , {\left(1 - \lambda_{1}\left(\EE P_{U_\perp U_\perp\T B}\right) \right)^{-2}} \right\}\right)^{-1} \right\}$, it follows that
\[
    \|\sin \Theta (\hat{U}, U)\| \leq c_2 \frac{C^*}{1 - \lambda_{1}\left(\EE P_{U_\perp U_\perp\T B}\right)}, \text{ for some constant $c_2>0$.}
\]
\end{Theorem}

The determination of the value $\lambda_{1}\left(\EE P_{U_\perp U_\perp\T B}\right)$ is of utmost importance in establishing Theorem \ref{theorem_average_projection_noisy_bound}. To illustrate the calculation of this value, consider the following example involving i.i.d. standard Gaussian variables.
\begin{Example}\label{example_iid_gaussian}
    Suppose the entries of $B$ are i.i.d. standard Gaussian distributed. 
    Then, we have $\EE P_{P_{U_\perp}B} = \min \{1, r_2 / (p_1 - r_1)\} \cdot P_{U_\perp}$ and hence $\lambda_{1}\left(\EE P_{U_\perp U_\perp\T B}\right) = \min \{1, r_2 / (p_1 - r_1)\}$. 
\end{Example}

%%%%%%%%%%%%
\subsection{Local Convergence of Iterations of \texttt{AP}} \label{section_iterative_projection}
%%%%%%%%%%%%

Next, we focus on the theoretical analysis for \texttt{AP}. To this end, we introduce the following assumptions. 
\begin{Assumption}[Conditions on Scores $A$ and $B$]\label{assumption_conditionnumber}
    Denote
    $$\lambda =\min \left\{\lambda_{\min }\left(\mathbb{E} A P_{V_\perp} A \T \right), \lambda_{\min }\left(\mathbb{E} B\T P_{U_\perp} B \right)\right\}.$$
    Assume in decomposition (\ref{eq:decomposition-equivalence-matrix}), $A$ and $B$ are independent and there is a constant $C$ such that
    \[\PP \left\{\max \{\left\| A \right\|^2, \left\| B \right\|^2 \}/\lambda \geq C \right\} \leq \nu, \quad \text{for some small $\nu < 1$. }\]
\end{Assumption}

In this context, $\lambda_{\min}(\mathbb{E} AP_{V_{\perp}}A^\top)$ represents the strength of the signal in $A$, excluding the interference from $B$ in the subspace $V$; a similar interpretation applies to $\lambda_{\min}(\mathbb{E} B^\top P_{U_{\perp}}B)$. Together, $\lambda$ essentially characterizes the overall signal strength, and the ratio $\mu^2/\lambda$ can be seen as a condition number that reflects the balance among the singular values of $A$ and $B$. Therefore, Assumption 1 essentially ensures that the condition number of the score matrices $A$ and $B$ is bounded.

Define the sub-Gaussian norm of a random variable $X$ as $\|X\|_{\psi_2}=\inf \left\{c>0: \mathbb{E}\left[\exp \left(X^2/c^2\right)\right] \leq 2\right\}$ \citep{vershynin2018high}.
\begin{Assumption}[Conditions on noise $Z$]\label{assumption_z}
$Z$ has i.i.d. sub-Gaussian entries with mean 0 and sub-Gaussian norm $\tau$. 
\end{Assumption}

Then we have the following result. 

\begin{Theorem}\label{theorem_convergence}
Let $\{X_i\}_{i=1}^n$ be a collection of matrices that satisfy the decomposition (\ref{eq:decomposition-equivalence-matrix}). Suppose the output of Algorithm \ref{algorithm_iterative_projection} is $\hat{U}^{(t)}, \hat{V}^{(t)}$ and define the errors as
$$\operatorname{Error}^{(t)} = \max\left\{\|\sin \Theta (U, \hat{U}^{(t)})\|, \|\sin \Theta (V, \hat{V}^{(t)})\|\right\}.$$
Assume that Assumptions \ref{assumption_conditionnumber} and \ref{assumption_z} hold. For any given $c_1>0$, there exist constants $c_2$, $c_3$, $c_4<1$, $c_5$ (all independent of any variable in the following inequalities) such that if initialization error $\operatorname{Error}^{(0)} \leq c_3$ and $n$ satisfies:
\[
    n \geq c_2 r_{\max} p_{\max} \max\left\{ \frac{p_{\max}^2 \tau^4}{p_{\min}^2 \mu^4}, \frac{p_{\max}^3 \tau^2}{p_{\min}^3 \mu^2}, \frac{p_{\max}^{3/2} \tau}{p_{\min}^{3/2} \mu}, 1\right\},
\]
then with a probability greater than $1-e^{-c_1r_{\min}p_{\max}} - \nu$, $\operatorname{Error}^{(t)}$ converges linearly with rate $c_4$: 
\[
\operatorname{Error}^{(t)} - \operatorname{Error} 
\leq c_4 \left(\operatorname{Error}^{(t-1)} - \operatorname{Error}\right),
\]
and the final error is bounded by 
\be\notag
\operatorname{Error} \leq c_5 \sqrt{\frac{\log p_{\max}}{n}} \max \left\{ \frac{p_{\max} \tau}{p_{\min} \mu}, \frac{p_{\max} \tau^2}{p_{\min} \mu^2}\right\}, 
\ee
where $r_{\max} = \max\{r_1, r_2\}, r_{\min} = \min\{r_1, r_2\}, p_{\min} = \min\{p_1, p_2\}$ and $p_{\max} = \max\{p_1, p_2\}$. 
\end{Theorem}

\begin{Remark}
    When $p_{1} \asymp p_{2} \asymp p$, the dimension $p$ has no effect on the final bound if we ignore the log term. To understand this, note that the number of parameters of $U$ is $O(p_1r_1)$, and that the number of effective samples to estimate $U$ is the total number of columns of all $X_i$'s, i.e., $np_2$. So when $r_1, r_2$ are fixed, $p_1, p_2$ both grow such that $p \asymp p_1 \asymp p_2$, both the effective dimension and sample size grow at the same rate and do not affect the final bound if we ignore the log term. 
\end{Remark}

\begin{Remark}
In the proof of Theorem \ref{theorem_convergence}, we adopt a two-step strategy to address the challenges involved. First, we establish a deterministic version of Theorem 1, assuming specific deterministic conditions for $A_i, B_i$, and $Z_i$. Subsequently, we demonstrate that these conditions are satisfied with high probability. The detailed proof is provided in the Supplementary Materials.

The proof of a deterministic version of Theorem \ref{theorem_convergence} relies on induction. In each induction step, we aim to give an estimation error upper bound for $U^{(t+1)}$ using the estimation error bound of $\hat{V}^{(t)}$ established from the previous induction step. A natural idea to achieve this is applying a matrix perturbation inequality. However, a direct application of the existing inequality, such as the Davis-Kahan Theorem \citep{davis1970rotation}, does not yield the desired results. We first focus on the noiseless case that $Z=0$ and $X = UA + BV\T$. In applying Davis-Kahan's Theorem, we consider $XX\T$ as the perturbed matrix derived from $UAA\T U\T$, which yields 
\be\label{eq_davis_kahan_proof_sketch}
    \|\sin \Theta (U, \hat{U}^{(t+1)})\| \leq \frac{\|X \hat V_\perp^{(t)} \hat V_\perp^{(t)\top} X\T - X V_\perp V_\perp\T X\T\|}{\min |\lambda_{r_2}(UA P_{V_\perp} A\T U) - \lambda_{r_2+1}(XX\T)|, |\lambda_{r_2}(UA P_{V_\perp} A\T U) - \lambda_{r_2-1}(XX\T)|}. 
\ee
Unfortunately, the right-hand side of \eqref{eq_davis_kahan_proof_sketch} may be significantly greater than $\|\sin \Theta (V, \hat{V}^{(t)})\|$. To see this, note that the numerator in \eqref{eq_davis_kahan_proof_sketch} can be roughly decomposed into: $\|UA (P_{ \hat V_\perp^{(t)}} - P_{V_\perp}) A\T U\|$, $\|UA P_{ \hat V_\perp^{(t)}} V B\T\|$, $\|BV\T P_{ \hat V_\perp^{(t)}} A\T U\|$, and $\|BV\T P_{ \hat V_\perp^{(t)}} V B\T\|$; the denominator involves the term $\lambda_{r_2}(UA P_{V_\perp^{(t)}} A\T U)$. Here, the first term from the numerator, $\|UA (P_{ \hat V_\perp^{(t)}} - P_{V_\perp}) A\T U\|$, can be at the same order of $\|UA P_{V_\perp} A\T U\|\|\sin \Theta (V, \hat{V}^{(t)})\|$ and the term $\|UA P_{V_\perp} A\T U\|$ is already greater than the denominator. Thus, it becomes difficult to prove that the right-hand side of \eqref{eq_davis_kahan_proof_sketch} is lower than $\|\sin \Theta (V, \hat{V}^{(t)})\|$. 
To overcome this issue, we develop a blockwise perturbation bound in the forthcoming Corollary \ref{Corollary_perturbation_bound}. After that, we apply matrix concentration inequalities to bound the terms in the numerator and denominator of the perturbation bound \eqref{inequality_eigenperturbation_bound}, including variants of matrix Bernstein (Lemma \ref{lemma_bernstein}) and matrix Chernoff (Lemma \ref{lemma_Chernoff}). 

When the noise $Z$ is non-zero, we instead prove that $\|\sin \Theta (U, \hat{U}^{(t+1)})\| \leq c_4 \|\sin \Theta (V, \hat{V}^{(t)})\| + K_1$ for some $K_1 = O(\|Z\|)$. $K_1$ can be further bounded by applying matrix concentration inequalities. As a result, we prove $\operatorname{Error}^{(t)} < c_4 \operatorname{Error}^{(t + 1)} + K_1$ for some constant $c_4 <1$, which can be equivalently written as $\operatorname{Error}^{(t)} - \operatorname{Error} < c_4 (\operatorname{Error}^{(t + 1)} - \operatorname{Error})$ where $\operatorname{Error} = K_1 / (1-c_4).$ Applying the reduction argument, we finish the proof of this theorem.

\end{Remark}

%%%%%%%%%%%%
\subsection{Overall Theory for \texttt{MOP-UP}} \label{section_global}
%%%%%%%%%%%%
The global convergence of Algorithms \ref{algorithm_average_projection} and \ref{algorithm_iterative_projection} can be summarised as follows.
\begin{Theorem}\label{theorem_global}
Suppose $U\in \mathbb{O}_{p_1, r_1}, V \in \mathbb{O}_{p_2, r_2}$ are some semi-orthogonal matrices, 
$A$ and $B$ are some random matrices with densities in $\RR^{r_1 \times p_2}$ and $\RR^{ p_1\times r_2}$ respectively, 
$Z$ is a random noise matrix with i.i.d. entries in $\RR^{p_1 \times p_2}$ independent of $A$ and $B$,
the population matrix satisfies $X = UA + BV\T + Z$, 
$\{X_i\}_{i=1}^n$ are i.i.d. copies of $X$, 
and $nr_2 \leq (n-1)(p_1- r_1) $. 
Assume the following hold in addition to Assumptions \ref{assumption_conditionnumber} and \ref{assumption_z}: 
\begin{enumerate}
    \item $\lambda_{1}\left(\EE P_{U_\perp U_\perp\T B}\right) < 1$;
    \item $\exists c \in [0,1/2]$ such that $C^* := \frac{c^2}{8} + \PP(4\|Z\| > c \sigma_{r}(UA+ BV\T))$ small enough;
\end{enumerate}
Then, for given constant $c_1$, there exist constants $c_2$ and $c_3$ (do not depend on any variable that appears in the following equations) such that if  
\[
    n \geq c_3 r_{\max} p_{\max} \max\left\{\frac{p_{\max}^2 \tau^4}{p_{\min}^2 \mu^4}, \frac{p_{\max}^3 \tau^2}{p_{\min}^3 \mu^2}, \frac{p_{\max}^{3/2} \tau}{p_{\min}^{3/2} \mu}, C^{*-2}, 1\right\}, 
\]
then with probability at least  $1 - e^{-c_1r_{\min} p_{\max}} - e^{-c_3} - \nu$, the estimation error at $t$th iteration of Algorithm \ref{algorithm_iterative_projection} initiated by Algorithm \ref{algorithm_average_projection} converges linearly to the final error which is bounded by
\be\label{equation_error_bound}
    \operatorname{Error} \leq c_2 \sqrt{\frac{\log p_{\max}}{n}} \max \left\{ \frac{p_{\max} \tau}{p_{\min} \mu}, \frac{p_{\max} \tau^2}{p_{\min} \mu^2}\right\}. 
\ee 
And hence, for some some $A_i$, $B_i$, and $Z_i$, the \texttt{MOP-UP}  estimation error of the signal can be bounded by
    \begin{align*}
            &\left\|P_{\hat U} X_i P_{\hat V} + P_{\hat U_\perp} X_i P_{\hat V} +  P_{\hat U} X_i P_{\hat V_\perp} - (UA_i+B_i V\T)  \right\| \\
            \leq &  \left\| Z_i - P_{U_\perp} Z_i P_{V_\perp}\right\| + \left( \left\|  X_i P_{V_\perp} \right\| + \left\|P_{U_\perp} X_i\right\|\right) \operatorname{Error} + \|X\| \operatorname{Error}^2. 
    \end{align*}
\end{Theorem}

%%%%%%%%%%%%%%%%%%%%%%%%
\subsection{A Key Technical Tool: Blockwise Eigenspace Perturbation Bound}\label{section_tech_lemma}
%%%%%%%%%%%%%%%%%%%%%%%%

The subsequent technical tool is crucial in establishing the validity of Theorem \ref{theorem_convergence} and possesses independent interests.
\begin{Theorem}[Blockwise Eigenspace Perturbation Bound]\label{Theorem_perturbation}
    Suppose $A \in \mathbb{R}^{p\times p}$ is a symmetric matrix, $\widetilde{V}=\left[V, V_{\perp}\right] \in \mathbb{O}_{p}$ are eigenvectors of $A$, where $V \in \mathbb{O}_{p, r}, V_{\perp} \in \mathbb{O}_{p, p-r}$ correspond to the first $r$ and last $\left(p-r\right)$ eigenvectors of $A$, respectively. $\widetilde{W}=\left[W, W_{\perp}\right] \in \mathbb{O}_{p}$ is any orthogonal matrix with $W \in \mathbb{O}_{p, r}, W_{\perp} \in \mathbb{O}_{p, p-r}$. Given that $\lambda_{r}(W\T A W)>\lambda_{r+1}(A)$, we have
	$$\|\sin \Theta(V, W)\|_{F} \leq \frac{\left\|W\T A W_{\perp}\right\|_F}{\lambda_{r}(W\T A W)-\lambda_{r+1}(A)} \wedge \sqrt{r}$$ 
and
    $$\|\sin \Theta(V, W)\| \leq \frac{\left\|W\T A W_{\perp}\right\|}{\lambda_{r}(W\T A W)-\lambda_{r+1}(A)} \wedge 1.$$
\end{Theorem}
\begin{Corollary}[Perturbation Bound]\label{Corollary_perturbation_bound}
    Denote the eigenvalue decompositions of $X$ and $X+Z$ as:
    \[
    X=\left[\begin{array}{ll}
        U & U_{\perp}
    \end{array}\right] \cdot\left[\begin{array}{cc}
        \Sigma_{1} & 0 \\
        0 & \Sigma_{2}
    \end{array}\right] \cdot\left[\begin{array}{c}
        U^{\top} \\
        U_{\perp}^{\top}
    \end{array}\right], 
    \]\[
    \hat{X}=X+Z=\left[\begin{array}{ll}
        \hat{U} & \hat{U}_{\perp}
    \end{array}\right] \cdot\left[\begin{array}{cc}
        \hat{\Sigma}_{1} & 0 \\
        0 & \hat{\Sigma}_{2}
    \end{array}\right] \cdot\left[\begin{array}{c}
        \hat{U}^{\top} \\
        \hat{U}_{\perp}^{\top}
    \end{array}\right].
    \]
    Then if $\lambda_{r} (P_U \hat{X} P_U)>\lambda_{r+1}(\hat{X})$, then\[
    \|\sin \Theta(U, \hat{U})\| \leq \frac{\left\|{P}_{U} Z {P}_{U_{\perp}}\right\|}{\lambda_{r} (P_U \hat{X} P_U)-\lambda_{r+1}(\hat{X})}\wedge 1. 
    \]
If further $\lambda_{r} (P_U \hat{X} P_U) > \|{P}_{U_{\perp}}\hat{X}{P}_{U_{\perp}}\|+\|{P}_{U} Z {P}_{U_{\perp}}\|$, 
    \be
    \label{inequality_eigenperturbation_bound}
    \|\sin \Theta(U, \hat{U})\| \leq \frac{\left\|{P}_{U} Z {P}_{U_{\perp}}\right\|}{\lambda_{r} (P_U \hat{X} P_U)-\|{P}_{U_{\perp}}\hat{X}{P}_{U_{\perp}}\|-\|{P}_{U} Z {P}_{U_{\perp}}\|}\wedge 1.
    \ee
\end{Corollary}
Compared to the classic Davis-Kahan Theorem \citep{davis1970rotation}
\[\left\|\sin \Theta\left(U, \hat U\right)\right\| \leq \frac{\|Z\|}{\min \{|\lambda_{r-1}(\hat{X})-\lambda_r(X)|, |\lambda_{r+1}(\hat{X})-\lambda_r(X)|\}},\]
our bound offers greater precision, particularly in the numerator of \eqref{inequality_eigenperturbation_bound}, which is $\|P_U ZP_{U_\perp}\|$. In our proof of Theorem \ref{theorem_convergence}, neither Davis-Kahan's nor Wedin's Theorem is sufficiently precise to establish the desired result. The reason is that, for example, in equation (\ref{eq:decomposition-equivalence-matrix}), a portion of $BV\T$ is noise when we attempt to recover $U$. Therefore, it becomes necessary to decompose $BV\T$ into blocks, namely $P_{U}BV\T$ and $P_{U_{\perp}}BV\T$, in order to separate the signal from the noise. As a result, a blockwise perturbation bound as described in \eqref{Corollary_perturbation_bound} can provide more appropriate bounds.

%%%%%%%%%%%%%%%%%%%%%%%%
\section{Real Data Analysis: MNIST}\label{section_real_data}
%%%%%%%%%%%%%%%%%%%%%%%%

%To demonstrate the practical applicability of our model, we applied it to two real-world datasets: MNIST and Brain fMRI. We compared our method, referred to as \texttt{MOP-UP}, with \texttt{MPCA}. The results highlight the advantages of our method.

%%%%%%%%%%%%
%\subsection{MNIST}\label{sec:MNIST}
%%%%%%%%%%%%

In this section, we apply the \texttt{MOP-UP} method to the MNIST (Modified National Institute of Standards and Technology) database. We select the first 6,000 images out of a total of 60,000 handwritten digit images as our training set. Additionally, we select all 10,000 testing images as our testing set. Each image is represented as a 28 by 28 bounded matrix $X \in [0,1]^{28\times 28}$, where each entry corresponds to the grayscale of a pixel in the image (ranging from 0 for white to 1 for black).

We apply \texttt{MOP-UP} to the images in the training set $\{X_i \in [0,1]^{28\times 28}\}_{i = 1}^{6000}$ for dimensional reduction. By utilizing Algorithms \ref{algorithm_average_projection} and \ref{algorithm_iterative_projection}, we obtain the loading estimates $\hat{U} \in \RR^{28\times r_1}$ and $\hat{V} \in \RR^{28\times r_2}$ in the decomposition $X_i = \bar X + B_iV^\top + U A_i + Z_i$ with certain rank values $(r_1, r_2)$, where $ \bar X = \sum_{i = 1}^{6000} X_i / 6000$ is the mean matrix of the training set. After that, we map each $X_i$ to $\{\hat{U}\T (X_i - \bar X) \hat{V}, \hat{U}\T (X_i - \bar X) \hat{V}_\perp, \hat{U}_\perp\T (X_i - \bar X) \hat{V}\}$, where the dimension of the right-hand side is $28(r_1+r_2) - r_1r_2$. Similarly, we map the test images $\{\tilde X_i \in [0,1]^{28\times 28}\}_{i = 1}^{10000}$ to $\tilde{X}_i \mapsto \{\hat{U}\T (\tilde{X}_i - \bar{\tilde{X}}) \hat{V}, \hat{U}\T (\tilde{X}_i - \bar{\tilde{X}}) \hat{V}_\perp, \hat{U}_\perp\T (\tilde{X}_i - \bar{\tilde{X}}) \hat{V}\}$.

To illustrate the effectiveness of our model, we utilize the training set after dimension reduction, denoted as $\{\hat{U}\T (X_i - \bar X) \hat{V}, \hat{U}\T (X_i - \bar X) \hat{V}\perp, \hat{U}\perp\T (X_i - \bar X) \hat{V}\}_{i = 1}^{6000}$, along with their corresponding labels $\{Y_i \in {0, \cdots, 9}\}_{i = 1}^{6000}$ to train different classifiers, including \texttt{SVM} (Support Vector Machine), \texttt{KNN} (K-Nearest Neighbor), and \texttt{XGB} (extreme gradient boosting \citep{chen2015xgboost}). Subsequently, we randomly divide the test set after dimension reduction into 10 folds. For each fold, we evaluate the test accuracy of the classifier, defined as the number of correctly classified samples divided by the total number of samples. We repeat this process for all 10 folds and calculate the mean and variance of the accuracy across the folds. It is important to note that we did not tune the hyperparameters of all the classifiers, except for selecting the best kernel among linear, polynomial, radial, and sigmoid for \texttt{SVM}. Based on our evaluation, the polynomial kernel yielded the best performance for the dimension-reduced data processed by \texttt{MOP-UP}. 
\begin{figure}[ht]
		\centering
		\includegraphics[width=0.8\columnwidth]{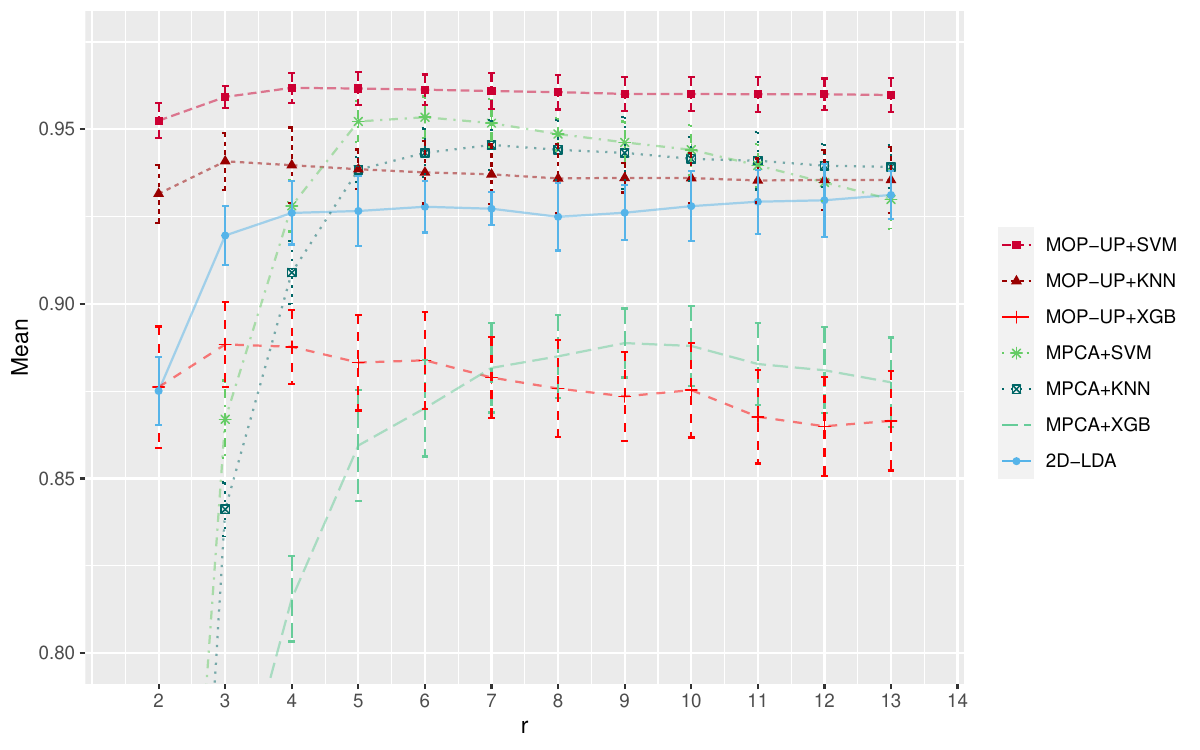}
        \caption{Comparison of accuracy: Mean accuracy across 10 folds versus rank $r = r_1 = r_2$ used as a hyperparameter in \texttt{MPCA}, \texttt{2D-LDA}, and our proposed \texttt{MOP-UP}. The length of the error bar represents the standard deviation.} 
	\label{mnist_comparison}
\end{figure}

We have also followed the same procedure, but this time we replaced \texttt{MOP-UP} with \texttt{MPCA}. For \texttt{MPCA}, the best kernel across all folds was found to be radial. We set $r:= r_1 = r_2$ in both our model and \texttt{MPCA}, and varied the value of $r$ from 2 to 14. 
Furthermore, we considered \texttt{2D-LDA} (2-Dimensional Linear Discriminant Analysis \citep{li20052d}), which is a supervised-learning variation of \texttt{MPCA} and two-dimensional generalization of Linear Discriminant Analysis. 
The results of our comparison are presented in Figure \ref{mnist_comparison}. Note that both \texttt{MPCA} and \texttt{MOP-UP} usually converge within 5 iterations. 
\begin{figure}[ht]
	\centering
		\includegraphics[width=0.8\columnwidth]{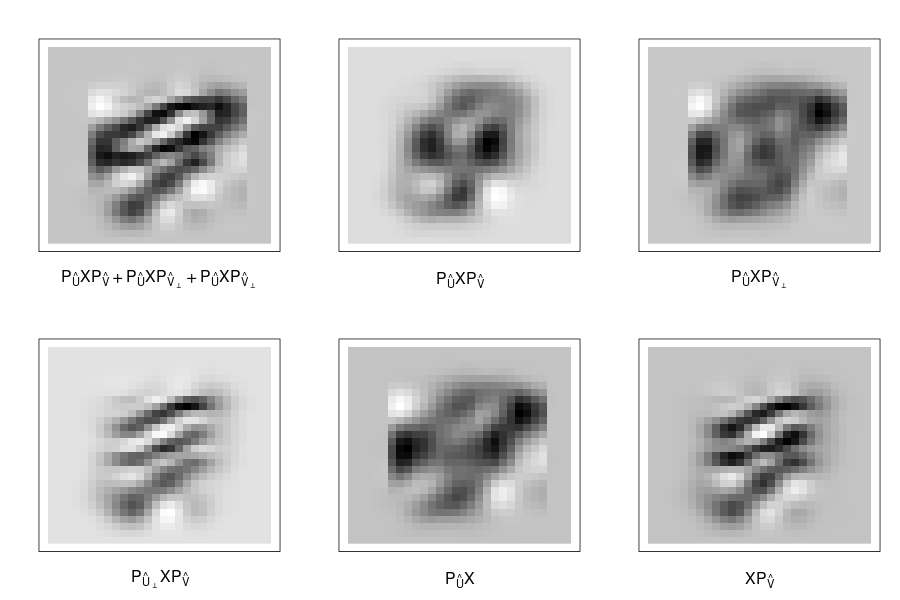}
		\caption{Visualization of dimension reduced digit ``9" images by \texttt{MOP-UP} with $r = 3$}
	\label{mnist_example}
\end{figure}

\begin{figure}[!ht]
		\centering
		\includegraphics[width=0.8\columnwidth]{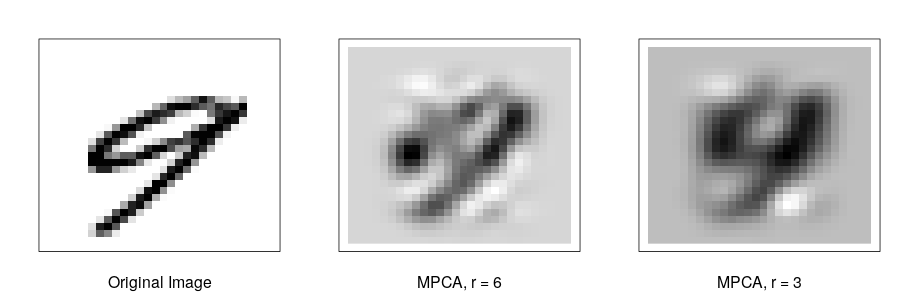}
		\caption{Visualization of dimension reduced digit ``9" images by \texttt{MPCA}. The dimension-reduced digit ``9" images by \texttt{MPCA} with $r = 3$ or $6$ ($r=6$ matches the top-left image of Figure \ref{mnist_example}) is unidentifiable.} 
		\label{mnist_example_mpca}
\end{figure}

In Figure \ref{mnist_example}, we visualize the dimension-reduced digit ``9" images by \texttt{MOP-UP} with $r = 3$. We observe that $P_{\hat{U}}X$ captures the column information of the digit ``9" image, while $XP_{\hat{V}}$ captures the row information. It is also worth noting that the top-left image in Figure \ref{mnist_example} corresponds to a rank $6$ matrix that captures the main features of the digit ``9". To provide a comparison, we also plot the same digit ``9" image after applying \texttt{MPCA} with $r=6,3$ in Figure \ref{mnist_example_mpca}. Notably, the dimension-reduced digit ``9" images by \texttt{MPCA} with $r = 3$ or $6$ ($r=6$ matches the top-left image of Figure \ref{mnist_example}) is unidentifiable.

%%%%%%%%%%
\section{\texttt{MOP-UP} for Higher-order Tensors}\label{sec:extend_to_tensor}
%%%%%%%%%%

In this section, we briefly discuss how the framework of \texttt{MOP-UP} can be extended to higher-order tensor data. Suppose we observe a collection of order-$d$ tensors $\X_1,\ldots, \X_n\in \mathbb{R}^{p_1\times \cdots \times p_d}$. Matrix data corresponds to $d=2$ and we shall now consider the case when $d\geq 3$. We aim to identify mode-wise subspaces $U_k\in \mathbb{O}_{p_k, r_k}$ such that each tensor observation can be decomposed approximately as:
\begin{equation*}
    \X_i = \M + \sum_{k=1}^d \A_{ki} \times_k U_k + \Z_i, \quad i=1,\ldots, n.
\end{equation*}
To provide a rigorous statistical interpretation for the \texttt{MOP-UP} framework, we discussed briefly the higher-order spiked covariance model in Section \ref{sec:high-order-PCA}. Denote $\I_{\mathbf{p}_d}$ as the order-$(2d)$ tensor in $\mathbb{R}^{\mathbf{p} \times \mathbf{p}}$ with entries $(\I_{\mathbf{p}_d})_{\mathbf{q}, \mathbf{q}} = 1$, where $\mathbf{q} = (q_1, q_2, \cdots, q_d)$, $q_k \in \{1, \cdots, p_k\}$, and 0 elsewhere. Then, the order-$d$ spiked covariance model can be defined as
\begin{Definition}[Order-$d$ Spiked Covariance Model]\rm \label{definition_spiked_covariance_tensor}
    Suppose $\X \in \mathbb{R}^{\bp}$ is an order-$d$ random tensor with $\EE \X = 0$. We say $\X$ has a \emph{rank-$\br$ high-order spiked covariance}, if there exists $\sigma^2>0$, $U_k\in \mathbb{O}_{p_k, r_k}$, such that
    \[
    \Cov(\X) = \SSigma_0 + \sigma^2 {\I_{\bp_d}},\ \SSigma_0 \in \mathbb{R}^{\bp \times \bp}, 
    \]
    \begin{equation}
        \SSigma_0 \times_{k = 1}^d U_{k\perp} = 0. \notag
    \end{equation}
\end{Definition}

Many of the methods and theories presented in this paper for the matrix spiked covariance model can be extended to the higher-order case. One way to approach this is by considering the order-$d$ spiked covariance model as equivalent to a decomposition form.
\begin{Theorem}[Equivalent forms for order-$d$ spiked covariance model]\label{theorem_equivalence_tensor}
    $\X \in \mathbb{R}^{\bp}$ has a rank-$\br$ high-order spiked covariance (Definition \ref{definition_spiked_covariance_tensor}) if and only if $\X$ can be decomposed as
    \begin{equation}\label{eq:decomposition-equivalence}
        \X = \sum_{k=1}^d \A_k \times_k U_k + \Z,
    \end{equation}
    where $U_k\in \mathbb{O}_{p_k, r_k}$ are fixed semi-orthogonal matrices, $\A_k \in \mathbb{R}^{p_1\times \cdots \times p_{k-1}\times r_k\times p_{k+1}\times \cdots \times p_d}$ are random tensors with mean 0, and $\Z \in \mathbb{R}^{\bp}$ is a noise tensor, where all entries of $\Z$ has mean 0, covariance $\sigma^2 \I_{\bp_d}$, and is uncorrelated with random tensors $\A_1,\ldots, \A_d$. 
\end{Theorem}
Furthermore, the concept of identifiability can be extended to the tensor case, allowing for the generalization of Theorem \ref{theorem_identifiability}. This generalization guarantees the identifiability of the mode-wise principal subspaces $\Span(U_k)$, where $k = 1,\cdots, d$. The specific details and proof of this result can be found in Supplementary Materials, stated as Theorem \ref{theorem_identifiability_tensor}.

However, in the case of order-$d$ tensors ($d\geq 3$), the \texttt{ASC} algorithm (Algorithm \ref{algorithm_average_projection}) does not work as effectively as it does in the matrix case. In the matrix case, when recovering $U$, \texttt{ASC} requires two steps of singular value decomposition (SVD). The first SVD involves taking the first $r_1 + r_2$ singular vectors of $X_i$, where $r_1 + r_2$ is chosen to match the rank of $X_i$. The second SVD is performed on the average of some projectors. To ensure that the projectors are nontrivial (i.e., not identity operators), we require $r_1 + r_2 < p_1$ (which is implicitly enforced by the condition $nr_2 \leq (n-1)(p_1- r_1)$ in Corollary \ref{corollary_matrix_initialization}). In the case of order-$d$ tensors ($d\geq 3$), ensuring the almost sure exact recovery of $U_1$ would require $r_1 + \sum_{k = 2}^d (r_k \prod_{h\notin {1, k}} p_h) < p_1$, which is impractical to satisfy. A possible method for initialization is the classic high-order singular value decomposition (\texttt{HOSVD}), represented as
\begin{equation*}
    \begin{split}
        \hat{U}_k^{(0)} = \SVD_{r_k}\left(\begin{bmatrix}
            \mathcal{M}_k(\X_1) ~ \cdots ~ \mathcal{M}_k(\X_n)
        \end{bmatrix}\right)
    \end{split}.
\end{equation*}
In this context, a possible approach is to matricize or unfold all tensor data along their $k$-th mode, combining them into a single matrix, and then applying singular value decomposition (SVD). However, the effectiveness of such a method \texttt{HOSVD} is not yet clearly understood. To overcome this limitation and tackle the challenges posed by higher-order spiked covariance models, it would be beneficial for future research to explore initialization methods. Such investigations could potentially lead to the development of more suitable approaches for addressing these challenges.

\begin{algorithm}
\caption{Alternating Projection \texttt{AP} for Order-$d$ Data}
    \label{algorithm_iterative_projection_tensor}
        \algrenewcommand\algorithmicensure{\textbf{Output:}}
        \algrenewcommand\algorithmicrequire{\textbf{Input:}}
    \begin{algorithmic} 		\Require Data tensors $\{\X_i\}_{i = 1}^n \in \mathbb{R}^{ p_1\times \cdots \times p_d}$, target rank $(r_1, r_2, \cdots r_d)$, initialization $\{\hat U_j^{(0)}\}_{j = 1}^d$, maximal number of iteration $t_0$.
        \Ensure{Estimation $\{\hat U_j^{(t)}\}_{j = 1}^d$} 
            \State Centralization: $\X_i \gets \X_i - \bar{\X}$
        %\State $\epsilon \gets \infty, t \gets 1$
        \For{$t$ in $1:t_0$}
        \For{$j$ in $1:d$}
        \State  $\hat{U}_j^{(t)} \gets \Eigen_{r_j}\left( \sum_{i = 1}^{n}\MM_j \left(  \X_i \times_{k \neq j} \left( \hat{U}_{k\perp}^{(t-1)} \right)\T  \right) \MM_j \left(  \X_i \times_{k \neq j} \left( \hat{U}_{k\perp}^{(t-1)} \right)\T  \right) \T \right)$
        \EndFor
            \State Break the for loop if converged or maximum number of iteration $t_0$ reached
            \EndFor\\
        \Return $\{\hat U_j^{(t)}\}_{j = 1}^d$
    \end{algorithmic}
\end{algorithm}

Lastly, it is worth mentioning that Algorithm \ref{algorithm_iterative_projection}, referred to as \texttt{AP}, remains applicable and can be further generalized to the tensor case as Algorithm \ref{algorithm_iterative_projection_tensor}. The resulting algorithm, when applied to tensors, provides an iterative projection-based approach for estimating the principal subspaces $U_k$. The corresponding final error bound in this tensor setting would be
$$\operatorname{Error} \lesssim \sqrt{\frac{\log p_{\max}}{n}} \max \left\{\theta {\frac{u}{\mu}}, {\frac{u^2}{\mu^2}}\right\}, $$
where $\theta = \max\left\{ 1, \sqrt{\frac{p_h}{\prod_{k\neq h} p_k}}; h = 1, \cdots, d\right\}$, $u = \left\|\frac{\MM_h(\Z)}{\sqrt{\prod_{k\neq h}p_k}}\right\|_{\psi_2}$ and $\mu$ is a high-probability upper bound of $\frac{\max_{k} \left\| \MM_h \left(\A_{k}\right) \right\|}{\sqrt{\prod_{k \neq h} p_k}}$. 
This result is formally stated as Theorem \ref{theorem_convergence_tensor} in Supplementary Materials. In summary, the local convergence of Algorithm \ref{algorithm_iterative_projection_tensor} is guaranteed with high probability given a proper initialization to be studied in the future. 

\section*{Conflict of Interests} None declared.

\section*{Data Availability}

The authors thank Christina Meade and Ryan Bell for providing the functional MRI data from cocaine users and for helpful discussions. More details on data processing can be found at \cite{zhang2023cocaine}. This dataset is available upon request to Anru R. Zhang and Christina Meade. 

The MNIST dataset is publicly available at \url{https://yann.lecun.com/exdb/mnist/}.

\section*{Funding}

M. Yuan was supported in part by NSF Grants DMS-2015285 and DMS-2052955. A. R. Zhang was supported in part by NSF Grant CAREER-2203741 and NIH Grants R01HL169347 and R01HL168940.

\bibliographystyle{apalike}
\bibliography{reference}
\listoffigures

%%%%%%%%%%%%%%
\newpage
%\begin{appendices}
\appendix

\begin{center}
\textbf{\Large{}{}{}{}{}{} Supplementary materials for ``Mode-wise Principal Subspace Pursuit and Matrix Spiked Covariance Model"}{\Large{}{}}\\
 {\huge{}{} }{\huge\par}
\par\end{center}

\begin{center}
\large{}{}{}Runshi Tang,~ Ming Yuan,~ and ~ Anru R. Zhang{\large\par}
\par\end{center}

\begin{abstract}
     We collect the simulation studies, additional real data analysis on functional MRI of cocaine users, and all technical proofs in these supplementary materials.
\end{abstract}

%%%%%%%%%%%%%%%%%%%%%%%%
\section{Simulation Study}\label{section_simulation}
%%%%%%%%%%%%%%%%%%%%%%%%

In this section, we assess the performance of the proposed \texttt{MOP-UP} through simulated data in different settings.

In the following experimental setup, we investigate the estimation error under varying values of $n$, $p_1$, and $\tau$. For each combination of $n$, $p_1$, $R$, and the distribution of $Z$, we conduct 10 simulations. In each simulation, we fix $p_2 = 30$, $r_1 = 5$, $r_2 = 7$, and generate independent samples for all entries of $A_i$ and $B_i$ from a uniform distribution over the interval $(-1, 1)$ for $i = 1, \ldots, n$. We also independently generate a pair of orthogonal matrices $U$ and $V$. The noise matrices $Z_i$ are sampled independently in three different settings: bounded, normal, and heavy-tail distributions. Specifically, $Z_i$ follows a uniform distribution over the interval (-$R$, $R$), a Gaussian distribution with mean 0 and variance $R^2$, or $R$ times a random sample from a central $t$-distribution with 3 degrees of freedom. Next, we apply Algorithm \ref{algorithm_average_projection} and \ref{algorithm_iterative_projection} with 10 iterations and compute the mean and standard deviation of the estimation error over the 10 simulations. It was found that the algorithm usually converges within 5 iterations. Notably, in Equation \eqref{equation_error_bound}, $\tau \propto R$ and $\mu \propto 1$. Since we only take $R \leq 1$ and $p_1 \geq p_2$, we can simplify our theoretical upper bound \eqref{equation_error_bound} as follows:
\be\label{equation_bound}
\operatorname{Error} = \max\left\{\|\sin \Theta (U, \hat{U})\|, \|\sin \Theta (V, \hat{V})\|\right\} \lesssim R p_{1} \sqrt{\frac{\log p_{1}}{n}}. 
\ee

\begin{figure}[!ht]
    \centering 
    \includegraphics[width=0.8\columnwidth]{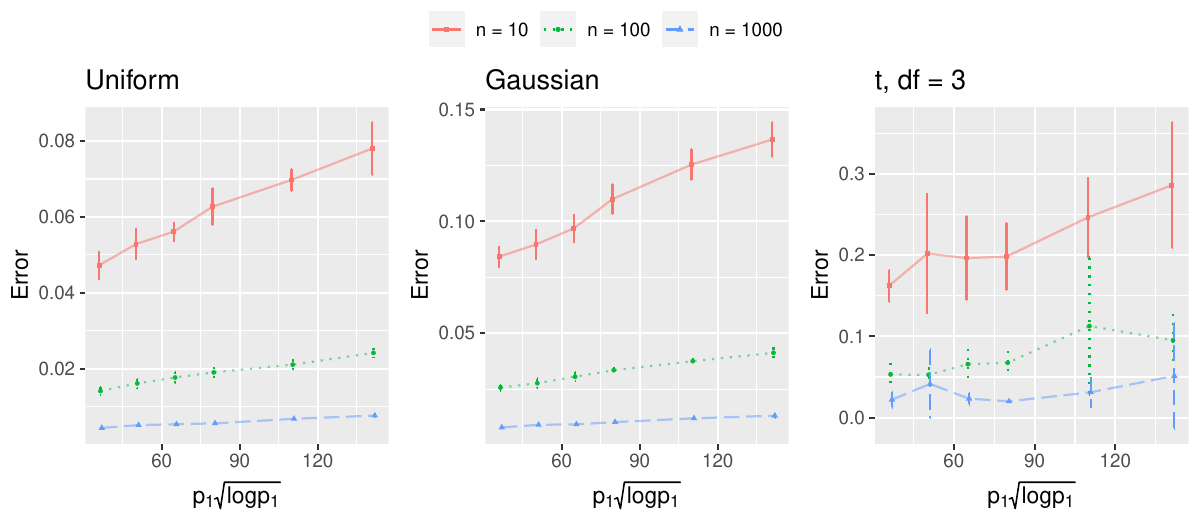}
    \caption{Estimation error of \texttt{MOP-UP} with the varying value of $p_1$
    }\label{fig_chang_on_p1}
\end{figure}

\begin{figure}[!ht]
    \centering 
    \includegraphics[width=0.8\columnwidth]{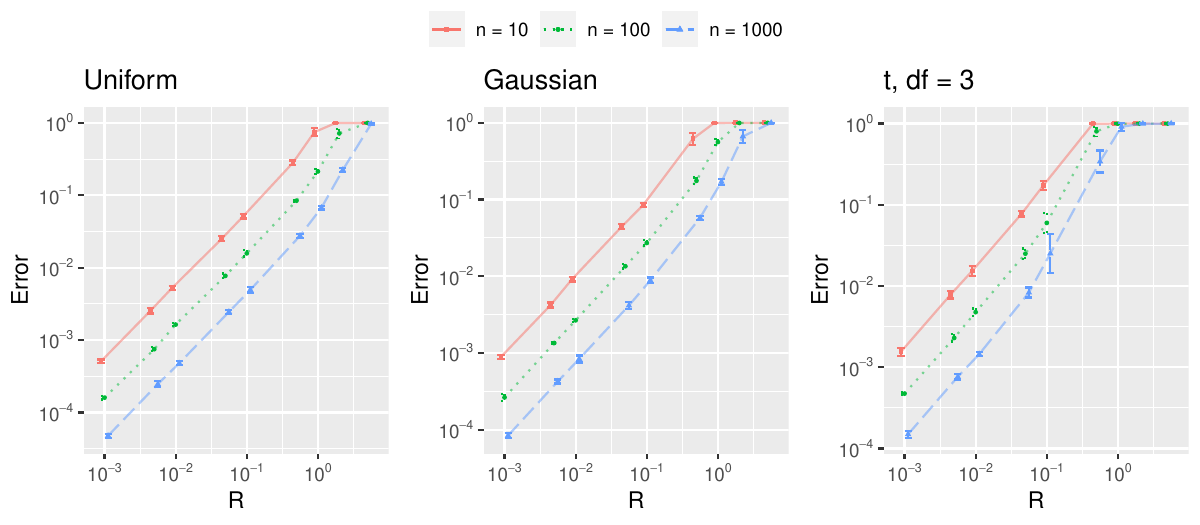}
    \caption{Estimation error of \texttt{MOP-UP} with the varying value of $R$}\label{fig_chang_on_R}
\end{figure}

\begin{figure}[!ht]
    \centering 
    \includegraphics[width=0.8\columnwidth]{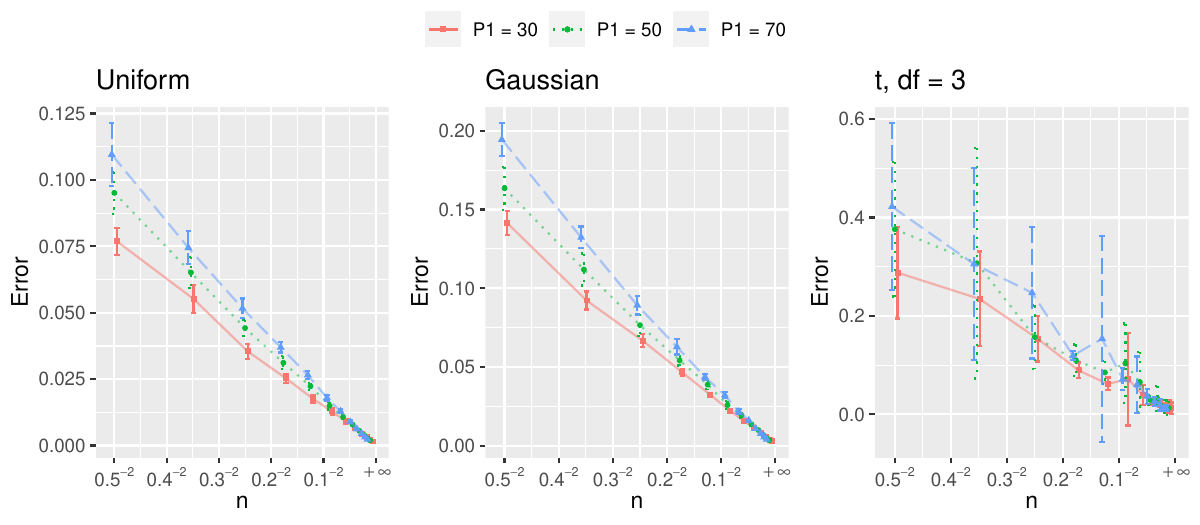}
    \caption{Estimation error of \texttt{MOP-UP} with the varying value of $n$}\label{fig_chang_on_n}
\end{figure}

We plot the error mean versus parameters of interest and the length of the interval at each point is twice the standard deviation of Errors in Figures \ref{fig_chang_on_p1}, \ref{fig_chang_on_R} and \ref{fig_chang_on_n}. We also scale the axis according to the corresponding order on the right-hand side of the bound (\ref{equation_bound}). In Figure \ref{fig_chang_on_p1}, the x-axis is scaled by $p_1 \sqrt{\log p_1}$, $R$ is set to be 0.1, and $p_1$ varies across the values of 30, 40, 50, 60, 80, and 100. In Figure \ref{fig_chang_on_R}, both axes are scaled by logarithm, $p_1$ is set to be 40, and $R$ varies across the values of 0.001, 0.005, 0.01, 0.05, 0.1, 0.5, 1, 2 and 5. In Figure \ref{fig_chang_on_n},  x-axis is scaled by $|-{n}^{-1/2}|$, $R$ is also set to be 0.1, and $n$ varies across the values of $\{2^i\}_{i=2}^{12}$. We can see the trending in plots are mostly linear, especially for large $n$ and small $p_1$. Since we have scaled the axis according to the right-hand side of \eqref{equation_bound}, it indicates the simulation results are consistent with our error bound. \\

We evaluated the performance of the BIC criterion for rank selection, denoted as $(\hat r_1, \hat r_2)$, as discussed in Section \ref{section_rank_selection}. We generated $Z_i$ with i.i.d. $N(0, R^2)$ entries, setting $n = 5, r_1 = 3,$ and $r_2 = 4$ with different noise levels $R$ and sizes $p_1 = p_2 = p$. We conducted 100 simulations for each noise level and distribution. In each simulation, we computed the BIC for all pairs $(\tilde r_1, \tilde r_2)$ with $\tilde r_i = 2, \cdots, 9$, denoted $(\hat r_1, \hat r_2)$ as the pair that achieved the minimum value, and calculated the absolute loss $|r_i - \hat r_i|$. We then computed the mean of the absolute losses over 100 simulations and presented the results in Table \ref{tab_r_hat}. It is noteworthy that the BIC %is calculated based on normally distributed $Z_i$, and when $Z_i$ follows a normal distribution, the BIC 
can accurately identify the true $(r_1, r_2)$ when the noise level is moderate.

\begin{table}
    \centering
    \scalebox{0.95}{
    {\footnotesize
   \begin{tabular}{l|lllll|lllll|lllll|lllll}
        \hline
        $R$ & \multicolumn{5}{c}{0.05} & \multicolumn{5}{c}{0.1} & \multicolumn{5}{c}{0.15} & \multicolumn{5}{c}{0.2} \\ 
        \hline
        $p$ & 30 & 50 & 70 & 90 & 110 & 30 & 50 & 70 & 90 & 110 & 30 & 50 & 70 & 90 & 110 & 30 & 50 & 70 & 90 & 110 \\ 
        \hline
        $|r_1 - \hat r_1|$ & 0 & 0 & 0 & 0 & 0 & 0 & 0 & 0 & 0 & 0 & 0.01 & 0 & 0 & 0 & 0 & 0.19 & 0.23 & 0.44 & 0.61 & 0.67 \\ 
        $|r_2 - \hat r_2|$ & 0 & 0 & 0 & 0 & 0 & 0 & 0 & 0 & 0 & 0 & 0.02 & 0 & 0 & 0 & 0 & 0.23 & 0.57 & 0.74 & 1.13 & 1.31 \\ 
        \hline
    \end{tabular}}}
    \caption{Mean of $|r_i - \hat r_i|$ over all simulations for different noise levels and $p$. }
    \label{tab_r_hat}
\end{table}

%%%%%%%%%%
\section{Real Data Analysis: Functional MRI of Cocaine Users}\label{sec:mri}
%%%%%%%%%%

In this section, we present the performance of \texttt{MOP-UP} on a Magnetic Resonance Imaging (MRI) dataset, which was derived from a clinical study conducted by Duke University \citep{hall2021human, CocaineConnectome}. The study enrolled adults aged 18-60 with or without a history of cocaine use. Cocaine use was defined as regular cocaine use for more than 1 year, with a minimum of 2 days of use in the past 30 days. Non-cocaine use was defined as follows: no lifetime cocaine use (abuse or dependence), no history of regular cocaine use, no cocaine use in the past year, and a cocaine-negative urine drug screen. The study comprised a total of $n = 293$ subjects, with 94 of them identified as cocaine users. For each subject, an MRI scan was performed, and after preprocessing, a functional MRI (fMRI) matrix was obtained for each subject. Each fMRI matrix $X_i$, where $i = 1, \ldots, 293$, is a symmetric matrix of size 246-by-246. Each row (or column) represents a Region of Interest (ROI), which corresponds to a group of neural nodes in a specific area of the human brain. The entry $(X_i)_{j,k}$ of the matrix represents the connection strength between two ROIs, namely ROI$_j$ and ROI$_k$. For further details regarding data acquisition, MRI processing, and background information, please refer to \cite{hall2021human, CocaineConnectome}.	

\noindent{\bf Classification.}~ The objective of this subsection is to predict cocaine use based on dimension-reduced data obtained using our matrix spiked covariance model and the \texttt{MOP-UP} method. Given that this is a binary classification problem, we evaluate the performance using the Area Under the Receiver Operating Characteristic curve (AUROC) and the Area Under the Precision-Recall Curve (AUPRC). These metrics are selected because they consider not only accuracy but also factors such as true positive rate and false positive rate, providing a comprehensive evaluation of the model's performance.

We employ a 10-fold cross-validation procedure to evaluate the performance of our \texttt{MOP-UP} and \texttt{MPCA} methods. The process is as follows: First, we divide all samples into 10 folds, selecting one fold as the test set while pooling the remaining folds into a training set. We then specify the target rank, denoted as $r$, where in both \texttt{MOP-UP} and \texttt{MPCA}, we set $r_1 = r_2 = r$. The training set is fed into either \texttt{MOP-UP} or \texttt{MPCA}, resulting in the output matrices ${\hat{U}\T (X_i - \bar{X}) \hat{U}, \hat{U}\T (X_i - \bar{X}) \hat{U}_\perp}$ (due to symmetry) for \texttt{MOP-UP} or ${S_i}$ for \texttt{MPCA}, where $X_i$ represents the fMRI matrix of subject $i$, and $\bar{X}$ denotes the sample mean.

Subsequently, we train a support vector machine (SVM) classifier using the output from either \texttt{MOP-UP} or \texttt{MPCA}, and evaluate the classifier's performance on the test set, recording the AUROC and AUPRC metrics. This procedure is repeated 10 times, with each fold serving as the test set, and we report the mean values of the predictive measures across the 10 tests. We utilize the SVM classifier with four different kernels: linear, polynomial, radial, and sigmoid. The hyperparameters are set to their default values in the R package \texttt{e1071} without further tuning. For both \texttt{MPCA} and \texttt{MOP-UP}, we find that the radial and sigmoid kernels perform better, and thus we present the results for these two kernels in Tables \ref{table_HCP_HOSD} and \ref{table_HCP_MPCA}.
	 
The combination of \texttt{MPCA} and \texttt{SVM} achieves the best AUROC and AUPRC values of 0.756 and 0.692, respectively, with $r = 23$ and the radial kernel. On the other hand, the combination of \texttt{MOP-UP} and \texttt{SVM} yields the best AUROC and AUPRC values of 0.762 and 0.710, respectively, with $r = 3$ and the sigmoid kernel.

In a related study by \cite{gowin2019using}, fMRI data along with demographic and clinic variables were used to train various linear models and a random forest classifier without employing dimension reduction techniques. The AUROC values reported in their study ranged from 0.53 to 0.65 for linear models and 0.62 for the random forest classifier. Both \texttt{MOP-UP} and \texttt{MPCA} outperform the results reported in \cite{gowin2019using}, with \texttt{MOP-UP} demonstrating slightly superior performance compared to \texttt{MPCA}. It is important to note that our approach does not utilize demographic or clinic information, and we did not perform parameter tuning for the \texttt{SVM} classifier.	

	\begin{table}[!ht]
		\centering
		\begin{tabular}{cccc}\toprule
			\textbf{r} & \textbf{Kernel} & \textbf{AUROC} & \textbf{AUPRC} \\ \hline
			2 & radial & 0.758 & 0.669 \\ 
			2 & sigmoid & 0.757 & 0.704 \\ 
			3 & radial & 0.761 & 0.696 \\ 
			3 & sigmoid & {\bf 0.762} & {\bf 0.710} \\ 
			4 & radial & 0.746 & 0.689 \\ 
			4 & sigmoid & 0.739 & 0.670 \\ 
			5 & radial & 0.739 & 0.667 \\ 
			5 & sigmoid & 0.733 & 0.652 \\ \hline
		\end{tabular}
        \caption{ Classification result by \texttt{MOP-UP}}
		\label{table_HCP_HOSD}
	\end{table}
	
	\begin{table}[!ht]
		\centering
		\begin{tabular}{cccc|cccc}\toprule
			\textbf{r} & \textbf{Kernel} & \textbf{AUROC} & \textbf{AUPRC}  & \textbf{r} & \textbf{Kernel} & \textbf{AUROC} & \textbf{AUPRC}  \\ \hline
			4 & radial & 0.631 & 0.515 & 22 & radial & 0.750 & 0.687 \\ 
			4 & sigmoid & 0.629 & 0.542 & 22 & sigmoid & 0.742 & 0.644  \\ 
			12 & radial & 0.675 & 0.610 & 23 & radial & {\bf 0.756} & {\bf 0.692} \\ 
			12 & sigmoid & 0.627 & 0.542 & 23 & sigmoid & 0.739 & 0.637 \\ 
			14 & radial & 0.694 & 0.603 & 24 & radial & 0.742 & 0.663 \\ 
			14 & sigmoid & 0.650 & 0.537 & 24 & sigmoid & 0.750 & 0.644 \\ 
			16 & radial & 0.710 & 0.604 & 25 & radial & {0.734} & {0.661} \\ 
			16 & sigmoid & 0.680 & 0.573 & 25 & sigmoid & 0.746 & 0.646 \\ 
			17 & radial & 0.726 & 0.623 & 27 & radial & 0.738 & 0.664 \\ 
			17 & sigmoid & 0.697 & 0.580 & 27 & sigmoid & 0.736 & 0.650 \\ 
			18 & radial & 0.712 & 0.613 & 31 & radial & 0.731 & 0.665 \\ 
			18 & sigmoid & 0.684 & 0.566 & 31 & sigmoid & 0.714 & 0.638 \\ 
			20 & radial & 0.736 & 0.657 & 35 & radial & 0.720 & 0.642 \\ 
			20 & sigmoid & 0.729 & 0.626 & 35 & sigmoid & 0.701 & 0.615 \\
			 \hline
		\end{tabular}\caption{Classification result by \texttt{MPCA}}
	\label{table_HCP_MPCA}
	\end{table}

\noindent{\bf Clustering.}~ We further perform unsupervised learning by clustering ROIs (Regions of Interest) based on the output of \texttt{MOP-UP}. As mentioned earlier, each row or column of the fMRI matrix corresponds to an ROI, and the matrix's entries represent the connections between these ROIs. Therefore, our goal is to cluster the rows (or columns) of the fMRI matrix.

\begin{figure*}[!ht]
		\centering
		\includegraphics[width = 0.7\textwidth]{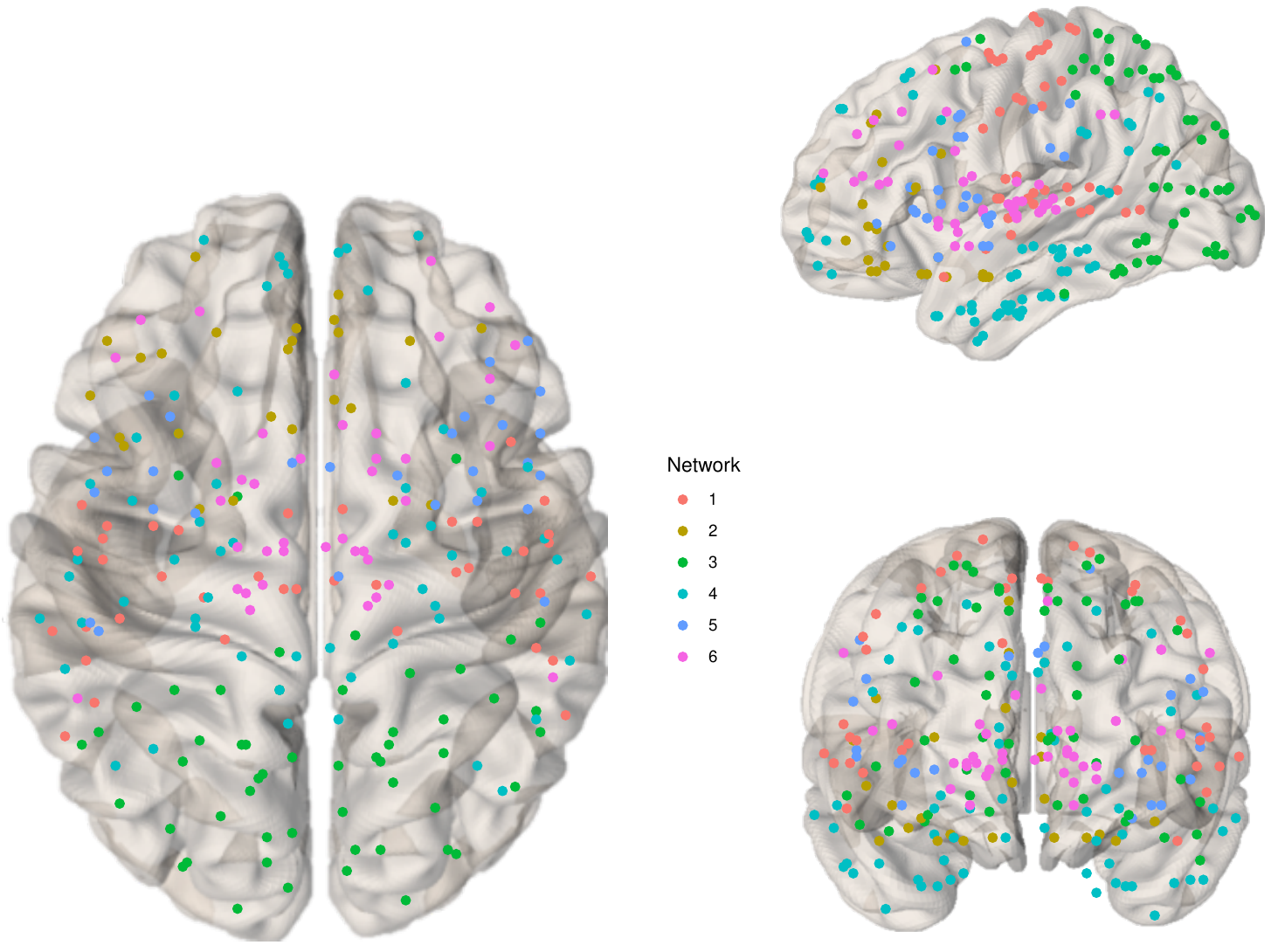}
		\caption{Clustering result of brain regions by \texttt{MOP-UP} and k-means ($k=6$)}
\label{fig_HCP_Cluster}
\end{figure*}

Given that the \texttt{MOP-UP} model achieved high performance in classification with $r=3$, we expect it to preserve a significant amount of information from the original data. Consequently, we utilize the output $\hat{U} \in \RR^{246\times 3}$, which consists of 246 three-dimensional vectors. Each vector represents an ROI, and its entries represent the loadings of the ROI. We feed this output into the K-means clustering algorithm, which assigns a label vector $l$ to the ROIs. To visualize the clustering result, we map the ROIs to their physical locations in the human brain, assign different colors to each cluster, and plot the result for $K = 6$ in Figure \ref{fig_HCP_Cluster}. It is important to note that no prior information about the physical locations of the ROIs was used in the clustering process. However, Figure \ref{fig_HCP_Cluster} demonstrates that the ROIs belonging to the same clusters according to our method tend to be physically closer to each other and the ROIs clustered by our \texttt{MOP-UP} are closely related to the brain networks in the literature \citep{riedl2016metabolic}.
	
%%%%%%%%%%%%%%%%%%%%%%%%
\section{Proof of Theorem \ref{theorem_equivalence}}
%%%%%%%%%%%%%%%%%%%%%%%%

We first provide the following lemma: 
\begin{Lemma}\label{Lemma_existence}
    Given a positive definite (semi-positive definite and non-singular) matrix $$
    \Sigma = \left[               
    \begin{array}{cc}  
        \Sigma_x & \Sigma_{xz}\\  
        \Sigma_{xz}\T & \Sigma_{z}\\  
    \end{array}
    \right]              
    $$ and a random vector $x$ with $\mathbb{E}x = 0$ and $\var(x) = \Sigma_x$, there exists a random vector $z$ s.t. $\mathbb{E}z = 0$ and the variance matrix of the joint distribution of $(x, z)$ is $\Sigma$. 
\end{Lemma}
\begin{proof}
    Notice 
    $$
    \left[               
    \begin{array}{cc}  
        I & -\Sigma_x^{-1}\Sigma_{xz}\\  
        0 & I\\  
    \end{array}
    \right]\T
    \left[               
    \begin{array}{cc}  
        \Sigma_x & \Sigma_{xz}\\  
        \Sigma_{xz}\T & \Sigma_{z}\\  
    \end{array}
    \right] 
    \left[               
    \begin{array}{cc}  
        I & -\Sigma_x^{-1}\Sigma_{xz}\\  
        0 & I\\  
    \end{array}
    \right] 
    = \left[               
    \begin{array}{cc}  
        \Sigma_x & 0\\  
        0 & \Sigma_z - \Sigma_{xz}\T\Sigma_x^{-1} \Sigma_{xz}\\  
    \end{array}
    \right]    .        
    $$So, $\Sigma_x$ and $\Sigma_z - \Sigma_{xz}\T\Sigma_x^{-1} \Sigma_{xz}$ are positive definite. 
    
    There exist $z_0 \sim N(0, I)$ such that $z_0 \perp x$. Let $z = Az_0+Bx$. Then $\mathbb{E}z = 0$ and $\operatorname{cov}(x,z) = \Sigma_x B\T = \Sigma_{xz}$ if $B = (\Sigma_x^{-1} \Sigma_{xy})\T$. 
    
    Moreover, we have $\mathbb{E}zz\T = AA\T + B\Sigma_x B\T = \Sigma_z \Rightarrow AA\T = \Sigma_z - B\Sigma_x B\T = \Sigma_z - \Sigma_{xz}\T\Sigma_x^{-1} \Sigma_{xz}$. The existence of such $A$ is guaranteed, since $\Sigma_z - \Sigma_{xz}\T\Sigma_x^{-1} \Sigma_{xz}$ is positive definite.
\end{proof}

Now let's go back to the proof of Theorem \ref{theorem_equivalence}. 

\begin{proof}
Without loss of generality, assume $\mathbb{E} X = 0$ and $\sigma = 1$. For tensors $\A$ and $\B$, $\A \otimes \B$ refers to their tensor product in this section. 

Assume the decomposition equation\[
X = UA + BV\T +Z
\] holds. Hence, the covariance tensor 
$$\begin{aligned}
    \boldsymbol\Sigma 
    =& \mathbb{E} (X \otimes X)\\
    =& \mathbb{E} \left(UA + BV\T+Z\right) \otimes \left(UA + BV\T+Z\right)\\
    =& \mathbb{E} \left(UA + BV\T\right) \otimes \left(UA + BV\T\right) + \I_{(p_1\times p_2)_2}.
\end{aligned}$$
Denote $\boldsymbol{\Sigma}_{0} = \mathbb{E} \left(UA + BV\T\right) \otimes \left(UA + BV\T\right)$. Thus, 
$$
\begin{aligned}
    &\boldsymbol{\Sigma}_{0} \times_1 U_{\perp}\T \times_2 V_{\perp}\T\\
    = &\mathbb{E} \left((UA) \otimes (BV\T) + (UA) \otimes (UA) + (BV\T) \otimes (UA)+ (BV\T) \otimes (BV\T)\right)\times_1 U_{\perp}\T \times_2 V_{\perp}\T \\
    =& 0, 
\end{aligned}$$which proves the sufficiency. 

To prove the necessity, assume $X$ has spiked covariance. Define $$
A = U\T (X-Z), \qquad 
B = P_{U_{\perp}}(X-Z) V,
$$
where $Z$ is some random matrix with $\mathbb{E}Z = 0,\mathbb{E}(Z\otimes Z) = \mathbb{E}(X\otimes Z) = \I_{(p_1\times p_2)_2}$. So $\mathbb{E}(A \otimes Z) = 0$ and $\mathbb{E}(B \otimes Z) = 0$. Here, to see the existence of such $Z$, we can vectorize $X$ as a random vector. Then, the existence of $z$ is guaranteed by lemma \ref{Lemma_existence}.

Now denote ${Y} = {X}- {Z}$. Notice $\RR^{p_1} = \Span(U) + \Span(U_{\perp})$ implies that for any $u \in \mathbb{R}^{p_1}$ we have decomposition $u = u_1 + u_2$ with $u_1 \in \Span(U), u_2 \in \Span(U_{\perp})$ and similar results hold for the case replacing $U$ by $V$. Thus,
\begin{align*}
    &u\T \left(Y - UA - BV\T\right)v\\
    =&u_1\T \left(Y - UA - BV\T\right)v + u_2\T \left(Y - UA - BV\T\right)v\\
    =&u_1\T \left(Y - P_{U} Y - P_{U_\perp} Y P_{V} \right)v + u_2\T \left(Y - P_{U} Y - P_{U_\perp} Y P_{V} \right)v\\
    =&u_2\T \left(Y - P_{U} Y - P_{U_\perp} Y P_{V} \right)(v_1 + v_2)\\
    =& u_2\T \left(Y - P_{U} Y - P_{U_\perp} Y P_{V} \right)v_2\\
    =& u_2\T Y v_2
\end{align*}
Notice $\boldsymbol{\Sigma}_{0} \times_1 U_{\perp} \times_2 V_{\perp}=0$ implies that for any $u_2 \in \Span(U_{\perp}), v_2 \in \Span(V_{\perp}), u_3 \in \mathbb{R}^{p_1}$ and $v_3 \in \mathbb{R}^{p_2},$ we have $\boldsymbol{\Sigma}_{0} (u_2, v_2, u_3, v_3) = 0$. Thus, for any vectors $u,v, u', v'$ with suitable dimensions, 
\begin{align*}
    &\left(\mathbb{E} \left(Y - UA - BV\T\right)\otimes \left(Y - UA - BV\T\right)\right)(u, v, u', v')\\
    =& \mathbb{E} u\T \left(Y - UA - BV\T\right) v  {u'}\T \left(Y - UA - BV\T\right)v'\\
    =& \mathbb{E} u_2\T Y v_2 {u_2'}\T Yv'_2 \\
    =& \boldsymbol{\Sigma}_{0} (u_2, v_2, u_2', v_2') \\
    =& 0,
\end{align*} 
i.e., the covariance tensor is 0, and hence, the decomposition equation holds a.s. Thus, the theorem is proved. 
\end{proof}

%%%%%%%%%%%%%%%%%%%%%%%%
\section{Proof of Theorem \ref{theorem_identifiability}}
%%%%%%%%%%%%%%%%%%%%%%%%

\begin{proof}
Assume that for some $U',V'$ we have $P_{U'_\perp}  YP_{V'_{\perp}} = 0.$ Notice that $\|\sin \Theta (U, U')\| = 0$ is equivalent to $P_{U'_{\perp}} U = 0$. Now assume $P_{U'_{\perp}} U \neq 0$. We have 
\begin{align*}
    0=&P_{U'_\perp}  YP_{V'_{\perp}} \\
    =& P_{U'_\perp} (UA + BV\T) P_{V'_{\perp}} \\
    =& P_{U'_\perp} UA P_{V'_{\perp}} + P_{U'_\perp} BV\T P_{V'_{\perp}}. 
\end{align*}
Intuitively, to make the last line 0, we need its two terms to cancel out with each other. However, by the condition in the theorem, the probability for $P_{U'_\perp} UA P_{V'_{\perp}}$ to cancel out with $P_{U'_\perp} BV\T P_{V'_{\perp}}$ for any given $B$ is strictly less than 1. 

To make the statement rigorous, notice that it follows
$\Span\left( UA P_{V'_{\perp}} + BV\T P_{V'_{\perp}}\right) \subseteq \ker(P_{U'_\perp})$, which implies for any $v$ such that $P_{V'_{\perp}} v\neq 0$, 
we have 
\[
     UA P_{V'_{\perp}} v + BV\T P_{V'_{\perp}} v \in \ker(P_{U'_\perp}),
\]and hence
\[
    UA P_{V'_{\perp}} v \in \mathcal{A}, 
\]where $\mathcal{A}$ represents the affine space $\{u - BV\T P_{V'_{\perp}} v: \forall u \in \ker(P_{U'_\perp})\}$.

For given $B$, if $BV\T P_{V'_{\perp}} v= 0$, we have $\mathcal{A} = \ker(P_{U'_{\perp}}) = \Span(U') \neq \Span(U)$, and thus $\Span(U) \not \subseteq \mathcal{A}$. 

If $BV\T P_{V'_{\perp}} v \neq 0$, then $\mathcal{A}$ is a shifted $r_1$ dimensional space. 
If the shift direction is in the subspace, i.e., $BV\T P_{V'_{\perp}} v \in \ker(P_{U'_{\perp}})$, then $\mathcal{A} = \ker(P_{U'_{\perp}})$, and hence $\Span(U) \not \subseteq \mathcal{A}$. 
If the shift direction is not in the subspace, i.e., $BV\T P_{V'_{\perp}} v \notin \ker(P_{U'_{\perp}})$, then for any $u \in \Span(U)$ with small enough $\|u\|$, we have $u\notin \mathcal{A}$. Thus, $\Span(U) \not \subseteq \mathcal{A}$.

By the discussion above, we always have $\Span(U) \not \subseteq \mathcal{A}$. Hence by the condition in the theorem, we have $\PP(UA P_{V'_{\perp}} v \in \mathcal{A} | B) < 1$, which concludes that 
\[
    \PP\left(P_{U'_\perp}  YP_{V'_{\perp}} = 0\right) \leq 
    \EE \left( \PP (UA P_{V'_{\perp}} v \in \mathcal{A} | B)\right) < 1. 
\]
Thus, by Theorem \ref{theorem_equivalence}, if the covariance tensor $\SSigma$ of $Y$ satisfies $\SSigma \times_1 U'_\perp \times_2  V'_{\perp} = 0$, then there exist some $A', B'$ such that $Y = U'A' + B'{V'}\T$, and hence there must be $\PP\left(P_{U'_\perp}  YP_{V'_{\perp}} = 0 \right) = 1$. Contradict. Thus, $\SSigma \times_1 U'_\perp \times_2  V'_{\perp} \neq 0$.
\end{proof}

%%%%%%%%%%%%%%%%%%%%%%%%
\section{Proof of Theorem \ref{theorem_span_sum}}
%%%%%%%%%%%%%%%%%%%%%%%%

We first introduce the following technical lemmas. 
\begin{Lemma}\label{lemma_span_sum}
    Let $A$ and $B$ be two $m$-by-$n$ matrices. If $\ker(B) + \ker(A) = \RR^n$, then we have $\Span(A)\subseteq \Span(A+B)$. If $A$ and $B$ further satisfy $\Span (A) \cap \Span (B) = \{0\}$, then $\Span(A) \subseteq \Span(A+B)$ if and only if $\ker(B) + \ker(A) = \RR^n$. 
\end{Lemma}

\begin{proof}
For any $u \in \RR^n$, we have 
$$
\begin{aligned}
&u \in \ker(B) + \ker(A)\\
\Leftrightarrow& \exists a_0 \in \ker(A), b_0 \in \ker(B): u = a_0 + b_0\\
\Leftrightarrow& -a_0 \in \ker(A) \text{ and } u-a_0 \in \ker(B)\\
\Rightarrow & Au = (A+B)(u-a_0)\\
\Rightarrow & Au \in \Span(A+B).
\end{aligned}$$
Thus, as $u$ can be arbitrary, it follows that $\Span(A) \subseteq \Span(A+B)$. 

If we further have $\Span A \cap \Span B = \{0\}$, then for any $u \in \RR^n$, it follows that 
$$
\begin{aligned}
&\Span(A) \subseteq \Span(A+B) \\
\Rightarrow &\exists v: (A+B)v = Au\\
\Leftrightarrow& A(u-v) = Bv\\
\Rightarrow & u-v\in \ker(A), v \in \ker(B)\\
\Leftrightarrow &u \in \ker(B) + \ker(A).
\end{aligned}$$    
\end{proof}

\begin{Lemma}\label{lemma_span_sum_random}
    If $A$ is a $p$-by-$r_A$ random matrix with density in $\RR^{pr_A}$ and $B$ is a $p$-by-$r_B$ deterministic matrix, $r_A, r_B \leq p$, then we have 
    $$\PP\left(\dim(\Span(A)\cap \Span(B)) = \max\{0, r_A + r_B - p\}\right) = 1.$$
\end{Lemma}        

\begin{proof}
Denote $A = [a_1, \cdots, a_{r_A}]$ and $B = [b_1, \cdots, b_{r_B}]$. Let's prove this by induction. Consider the case that $r_A = 1$. The case when $p = 1$ is trivial. If $p \geq 2$, we have $$\PP(\dim(\Span(A)\cap \Span(B)) = 1) = \PP(a_1 \in \Span(B) \subsetneq \RR^p)  = 0.$$ 
Thus, 
$$\PP(\dim(\Span(A)\cap \Span(B)) = 0 = \max\{ 0, r_A + r_B - p\}) = 1. $$

Before going to the induction step, notice the following fact: as the joint density of $a_1, \cdots, a_{r_A}$ exists, we have the conditional density of $a_{r_A}|A_{-r_A}:= a_{r_A}|a_1, \cdots, a_{r_A - 1}$, i.e., the conditional distribution $a_{r_A}|A_{-r_A}$ is almost surely absolutely continuous with respect to Lebesgue measure. 
    
Now we assume the lemma holds for $r_A = m_A-1$ and $r_B = m_B$. Consider the case $r_A = m_A \geq 2$ and $r_B = m_B$. 

If $p \geq r_A + r_B$, then denote $A_{-r_A} = [a_1, \cdots, a_{r_A-1}]$ and by induction, we have 
$$\PP(\dim(\Span(A_{-r_A})\cap \Span(B)) = 0) = 1.$$ 
And by $p > r_B + r_A-1,$ we have 
$$\Span(B) + \Span(A_{-r_A})\subsetneq \RR^p.$$ 
Thus, as the Lebesgue measure of any nontrivial subspace is 0 and $a_{r_A}|A_{-r_A}$ is absolutely continuous with respect to the Lebesgue measure almost surely, we know 
$$\PP(a_{r_A} \in \Span(B) + \Span(A_{-r_A})|A_{-r_A}) = 0,$$ 
and 
$$\PP(a_{r_A} \in \Span(A_{-r_A})|A_{-r_A}) = 0.$$ 
Further notice if for some $u$ such that $0 \neq u \in \Span(A) + \Span(B)$ but $u \notin \Span(A_{-r_A}) + \Span(B)$, then $u = a+b$ for some nonzero $a\in \Span(A)$, $a\notin \Span(A_{-r_A})$ and $b\in \Span(B)$. Then, we have $a = a_0 + c \cdot a_{r_A}$ with $a_0\in\Span(A_{-r_A})$ and some nonzero c $\in \RR$. Thus, $c \cdot a_{r_A} = u - a_0 - b \in \Span(B) + \Span(A_{-r_A})$, which yields that 
$$
\begin{aligned}
    &\PP(\dim(\Span(A) \cap \Span(B)) \geq 1 + \dim(\Span(A_{-r_A}) \cap \Span(B))) \\
    =&\PP(\Span(A) \cap \Span(B)  \supsetneq \Span(A_{-r_A}) \cap \Span(B)) \\
    \leq& \EE (\PP(a_{r_A} \in \Span(B) + \Span(A_{-r_A})|A_{-r_A})) \\
    =& 0.
\end{aligned}
$$
Hence, $\PP(\dim(\Span(A) \cap \Span(B)) = 0) = 1$. 

If $p \leq r_A + r_B -1$, then we have $\PP(\dim(\Span(A_{-r_A})\cap \Span(B)) = 0) = 1$. But this time we have $\PP(\dim(\Span(A_{-r_A})+ \Span(B)) = \RR^p) = 1$ and thus $\PP(a_{r_A} \in \Span(A_{-r_A}) + \Span(B)|A_{-r_A}) = 1$. We can similarly prove that 
$$
\begin{aligned}
    &\PP(\Span(A) \cap \Span(B) = \Span(A_{-r_A}) \cap \Span(B)) \\
    \leq& \EE(\PP(a_{r_A} \notin \Span(B) + \Span(A_{-r_A})|A_{-r_A})) \\
    =& 0,
\end{aligned}
$$
which indicates that $\PP(\dim(\Span(A) \cap \Span(B)) = \dim(\Span(A_{-r_A}) \cap \Span(B)) + 1 = r_A + r_B - p) = 1$. These complete the induction step and the lemma holds. 
\end{proof}

Now we are ready to prove Theorem \ref{theorem_span_sum}. 

\begin{proof}
    Let's first prove $\Span(U) \subseteq \bigcap_{i = 1}^n \Span(X_i)$. 
    
    By comparing the dimension of both sides, we have the following facts: 1. $\ker(V\T) = \ker(BV\T)$; 2. $\Span(U) = \Span(UA)$; and 3. $\ker(A) = \ker(UA)$. 

    Notice Lemma \ref{lemma_span_sum_random} applies to $A\T$ and $V$, which yields $\Span(A\T) \cap \Span(V) = \{0\}$ almost surely. Thus, 
    \begin{align*}
        &\Span(A\T) \cap \Span(B\T) = \{0\}\\
        \Rightarrow &\RR^{p_2} = \{0\}^\perp = (\Span(A\T) \cap \Span(V))^\perp = \Span(A\T)^\perp + \Span(V)^\perp \\
         & \quad = \ker(A) + \ker(V\T) = \ker(UA) + \ker(BV\T)\\
        \overset{\text{(Lemma \ref{lemma_span_sum})}}{\Rightarrow} &\Span(U) = \Span(UA) \subseteq \Span(UA+BV\T).
    \end{align*}
    It proves $\Span(U) \subseteq \bigcap_{i = 1}^n \Span(X_i)$ almost surely. 
    
    Before prove the other direction, let's firstly prove $\bigcap_{i = 1}^n \Span(U_\perp\T B_i) = \{0\}$. Notice the density of $U_\perp\T B$ exists, and $\{B_i\}$ are i.i.d. copies of $B$. So, by Lemma \ref{lemma_span_sum_random}, we have $\dim(\Span(U_\perp\T B_1) \cap \Span(U_\perp\T B_2)) = \max\{0, 2r_2 - (p_1- r_1)\}$ almost surely. If $2r_2 - (p_1- r_1) \leq 0$, then it is done. If $2r_2 - (p_1- r_1) > 0$, then we consider $\dim((\Span(U_\perp\T B_1) \cap \Span(U_\perp\T B_2)) \cap \Span(U_\perp\T B_3))$. Notice given $B_1$ and $B_2$, $(\Span(U_\perp\T B_1) \cap \Span(U_\perp\T B_2)) = \Span(D)$ for some $(p_1 - r_1)$-by-$(2r_2 - (p_1- r_1))$ matrix $D$. Then apply Lemma \ref{lemma_span_sum_random} to $U_\perp\T B_3$ and $D$, which yields $\dim(\Span(U_\perp\T B_1) \cap \Span(U_\perp\T B_2) \cap \Span(U_\perp\T B_3)) = \max\{0, 3r_2 - 2(p_1- r_1)\}$. We repeat this procedure until $kr_2 - (k-1)(p_1- r_1) \leq 0$ for some $k$. $nr_2 - (n-1)(p_1- r_1) \leq 0$ guarantees that it will stop before (or at) $k = n$, which will yield $\bigcap_{i = 1}^n \Span(U_\perp\T B_i) = {0}$ almost surely. 
    
    Now let's go back to prove $\Span(U) \supseteq \bigcap_{i = 1}^n \Span(X_i)$. Notice for some vectors $v_i \in \RR^{p_2}$, if $X_i v_i = X_j v_j$ for all $i,j \in \{1,\cdots n\}, $ then it follows 
    \begin{align*}
        &U_\perp\T X_1 v_1 = \cdots = U_\perp\T X_n v_n\\
        \Rightarrow &U_\perp\T B_1 V\T v_1 = \cdots = U_\perp\T B_nV\T  v_n. 
    \end{align*}
    Recall that we have proved $\bigcap_{i = 1}^n \Span(U_\perp\T B_i) = \{0\}$ almost surely, which implies $U_\perp\T B_i V\T v_i = 0$ and hence $X_i v_i = (P_U + P_{U_\perp}) X_i v_i = P_U X_i v_i \in \Span(U)$. Thus, $\Span(U) \supseteq \bigcap_{i = 1}^n \Span(X_i)$ almost surely, which finishes the proof.
\end{proof}

%%%%%%%%%%%%%%%%%%%%%%%%
\section{Proof of Theorem \ref{theorem_average_projection_noisy_bound}}
%%%%%%%%%%%%%%%%%%%%%%%%

\begin{proof}
    Denote $r = r_1 +r_2$, $Q_i = \SVD_{r}\left( X_i \right)$, $Y_i = UA_i + B_iV\T$, $\tilde Q_{i} = \SVD_{r}\left( Y_i \right)$, $\tilde{U} = \Eigen_{r_1}\left( \frac{1}{n}\sum_{i = 1}^n \tilde Q_{i} {\tilde Q_{i}}\T\right)$ and $\tilde{P}_i = \tilde Q_{i} {\tilde Q_{i}}\T$. Then by Theorem \ref{theorem_span_sum}, we have 
$\tilde{P}_i = P_U + \hat{P}_i$, where $\hat{P}_i$ is some projection matrix such that $\Span(\hat{P}_i) \subseteq \Span(U_\perp)$ almost surely. Thus, $\lambda_{r_1}(\frac{1}{n}\sum_{i = 1}^n\tilde{P}_i ) = 1$ almost surely. 
    
    Then we have 
\be
    \|\sin \Theta (\hat{U}, U)\| \leq \|\sin \Theta (\hat{U}, \tilde{U})\| + \|\sin \Theta (\tilde{U}, U)\| = \|\sin \Theta (\hat{U}, \tilde{U})\|, \notag
\ee
where $\|\sin \Theta (\tilde{U}, U)\| = 0$ is yielded by Corollary \ref{corollary_matrix_initialization}. To bound $\|\sin \Theta (\hat{U},\tilde{U})\|$, we can use the matrix perturbation theory. In the setup of $\hat X = X + Z$, we can let $ X = \frac{1}{n}\sum_{i = 1}^n \tilde{P}_i$ and $\hat X = \frac{1}{n}\sum_{i = 1}^n  Q_{i} {Q_{i}}\T$. Davis-Khan Theorem (e.g. Corollary 2.8 in \cite{chen2021spectral}) yields 
\be \label{eq_initialization_DK}
    \|\sin \Theta (\hat{U}, \tilde{U})\| \leq \frac{\sqrt{2}\left\| \frac{1}{n}\sum_{i = 1}^n (\tilde{P}_i -   Q_{i} {Q_{i}}\T) \right\|}{\lambda_{r_1}(\frac{1}{n}\sum_{i = 1}^n\tilde{P}_i) - \lambda_{r_1+1}(\frac{1}{n}\sum_{i = 1}^n\tilde{P}_i)}. 
\ee
We are going to upper bound the numerator and lower bound the denominator of \eqref{eq_initialization_DK}. 

\subsubsection{Numerator}

we can use Matrix Chernoff bound (e.g., \cite{tropp2012user}):  
	
\begin{Lemma} [Matrix Chernoff] \label{lemma_Chernoff}
    Consider a finite sequence $\left\{\boldsymbol{X}_k\right\}$ of independent, random, self-adjoint matrices that satisfy
    $$
    \boldsymbol{X}_k \succcurlyeq \mathbf{0} \quad \text { and } \quad \lambda_{\max }\left(\boldsymbol{X}_k\right) \leq R \quad \text { almost surely. }
    $$
    Compute the minimum and maximum eigenvalues of the sum of expectations,
    $$
    \mu_{\max }:=\lambda_{\max }\left(\sum_k \mathbb{E} \boldsymbol{X}_k\right) \text {. }
    $$
    Then
    $$
    \begin{aligned}
        & \mathbb{P}\left\{\lambda_{\max }\left(\sum_k \boldsymbol{X}_k\right) \geq(1+\delta) \mu_{\max }\right\} \leq d \cdot\left[\frac{\mathrm{e}^\delta}{(1+\delta)^{1+\delta}}\right]^{\mu_{\max } / R} \quad \text { for } \delta \geq 0 .
    \end{aligned}
    $$
\end{Lemma}

Notice here our matrices $\tilde{P}_i -   Q_i {Q_i}\T$ are not necessarily p.s.d. but they are bounded by $\|\tilde Q_i \tilde{Q}_{i}\T -   Q_i {Q_i}\T\| \leq 1$ (the operator norm of the difference of projectors are bounded by 1). So, we can let $W_i = \tilde Q_i \tilde{Q}_{i}\T -   Q_i {Q_i}\T + I$, then $W_i$ are p.s.d. and we can apply Matrix Chernoff bound to $W_i$, which yields \be\label{eq_initailization_chernoff}
\begin{aligned}
    & \mathbb{P}\left\{\lambda_{1}\left(\frac{1}{n}\sum_k W_k\right) \geq \frac{(1+\delta)\mu_{\max }}{n}\right\} \leq p_1 \cdot\left[\frac{\mathrm{e}^\delta}{(1+\delta)^{1+\delta}}\right]^{\mu_{\max } / 2} \quad \text { for } \forall \delta \geq 0 .
\end{aligned}
\ee

To bound $\mu_{\max} = n\lambda_1(\EE W_1) = n\lambda_1(\EE \tilde{P}_1 - \EE  Q_{1} {Q_{1}}\T) + n$, we use self adjoint dilation and Theorem 1 in \cite{xia2021normal}, where we let
$$
\hat{A}=\left(\begin{array}{cc}
    0 &  X_1  \\
     X_1 \T & 0
\end{array}\right), \quad A=\left(\begin{array}{cc}
    0 & Y_1  \\
    Y_1 ^{\top} & 0
\end{array}\right),$$
and	
$$X=\left(\begin{array}{cc}
    0 & Z_1 \\
    Z_1\T & 0
\end{array}\right).
$$
Denote $\hat{\Theta} \hat{\Theta}^{\top} := \diag(Q_{i} {Q_{i}}\T, H_{i} {H_{i}}\T)$, $\Theta \Theta^{\top} := \diag(\tilde{P}_i, \tilde H_{i} {\tilde H_{i}}\T)$, $H_i, \tilde H_{i}$ are the first $r_1 + r_2$ right singular vectors of $UA + BV\T + Z$ and $UA + BV\T$ respectively, the event $\mathcal{A} := \{ 4\|Z\| \leq c \sigma_{r}(Y) \}$ for some $c <1$ and $I_{\{\mathcal{A}\}}$ is the indicator function of event $\mathcal{A}$. 
Then it yields that 
\begin{align} 
    \mu_{\max} - n
    = &n\left\|\EE Q_{i} {Q_{i}}\T -  \EE\tilde{P}_i\right\| \notag\\
    \leq &n\left\|\EE \left( \hat{\Theta} \hat{\Theta}^{\top}-\Theta \Theta^{\top} \right)\right\| \notag\\
    = & n\left\|\EE \left( \hat{\Theta} \hat{\Theta}^{\top}-\Theta \Theta^{\top} \right) \diag(I_{\{\mathcal{A}\}}) 
        + \EE \left( \hat{\Theta} \hat{\Theta}^{\top}-\Theta \Theta^{\top} \right) \diag(I_{\{\mathcal{A}^c\}})\right\| \notag\\
    \leq & n\left\|\EE \left(\hat{\Theta} \hat{\Theta}^{\top}- \Theta \Theta^{\top}\right) \diag(I_{\{\mathcal{A}\}}) \right\|
        + n\EE \left\| \left(\hat{\Theta} \hat{\Theta}^{\top}- \Theta \Theta^{\top}\right) \diag(I_{\{\mathcal{A}^c\}}) \right\| \notag\\
    \leq & n\left\|\EE \left(\hat{\Theta} \hat{\Theta}^{\top}- \Theta \Theta^{\top}\right) \diag(I_{\{\mathcal{A}\}}) \right\|
        + n\PP(\mathcal{A}^c) \notag\\
    = & n\left\|\EE \left(\hat{\Theta} \hat{\Theta}^{\top}- \Theta \Theta^{\top} - \mathcal{S}_{A,1} \right) \diag(I_{\{\mathcal{A}\}}) \right\| + n\PP(\mathcal{A}^c) \notag\\
    \leq & n\EE \left\| \diag(I_{\{\mathcal{A}\}}) \left(\hat{\Theta} \hat{\Theta}^{\top}- \Theta \Theta^{\top} - \mathcal{S}_{A,1} \right)\right\| + n\PP(\mathcal{A}^c) \notag\\
    = & n\EE \left\| \diag(I_{\{\mathcal{A}\}})\sum_{k = 2}^\infty \mathcal{S}_{A,k}\right\| + n\PP(\mathcal{A}^c)\notag\\
    \leq & n\EE I_{\{\mathcal{A}\}} \sum_{k = 2}^\infty \left\| \mathcal{S}_{A,k}\right\| + n\PP(\mathcal{A}^c)\notag\\
    \leq & n\EE I_{\{\mathcal{A}\}} \sum_{k = 2}^\infty \left(\frac{4\|Z\|}{\lambda_{r}(Y)}\right)^k + n\PP(\mathcal{A}^c)\notag\\
    \leq& n\EE \frac{I_{\{\mathcal{A}\}}}{1-\frac{4\|Z\|^2}{\lambda_{r}(Y)^2}} \frac{\|Z\|^2}{\lambda_{r}(Y)^2} + n\PP(\mathcal{A}^c), \label{eq_initailization_bias}
\end{align}
where $\mathcal{S}_{A,k}(X)$ are defined as in \cite{xia2021normal}, 
the inequality in the fifth line holds because $\left\| \hat{\Theta} \hat{\Theta}^{\top}- \Theta \Theta^{\top}\right\| \leq 1$, 
the equality in the sixth line holds because $\EE \mathcal{S}_{A,1}\diag(I_{\{\mathcal{A}\}}) = \EE(\EE(\mathcal{S}_{A,1}\diag(I_{\{\mathcal{A}\}})|A,B)) = 0$, 
the equality in the eighth line holds by Theorem 1 in \cite{xia2021normal},
and the inequality in the tenth line holds by the fact $\left\|\mathcal{S}_{A, k}\right\| \leq \left(\frac{4\|Z\|}{\lambda_{r}(Y)}\right)^k$ given in the discussion after Theorem 1 in \cite{xia2021normal} when the event $\mathcal{A}$ happens. 

Additionally, we have $ \tr (\EE \tilde{P}_1 - \EE  Q_{1} {Q_{1}}\T) = \EE \tr(\tilde{P}_1 - \EE  Q_{1} {Q_{1}}\T) = 0$. Thus, there must be $\lambda_1(\EE \tilde{P}_1 - \EE  Q_{1} {Q_{1}}\T) \geq 0, $ which yields that $\mu_{\max} \geq n$. 

Finally, combine \eqref{eq_initailization_chernoff} and \eqref{eq_initailization_bias}, we have 
\begin{align}
    &\mathbb{P}\left\{\lambda_{1 }\left(\frac{1}{n}\sum_i (\tilde{P}_i -   Q_{i} {Q_{i}}\T)\right)\geq{(1+\delta)C^* + \delta}\right\} \notag\\
    =&\mathbb{P}\left\{\lambda_{1 }\left(\frac{1}{n}\sum_i (\tilde{P}_i -   Q_{i} {Q_{i}}\T + I)\right)\geq{(1+\delta)C^* + \delta +1}\right\} \notag\\
    \leq &\mathbb{P}\left\{\lambda_{1 }\left(\frac{1}{n}\sum_i W_i\right)\geq\frac{(1+\delta)\mu_{\max }}{n}\right\} \notag\\ 
     \leq& p_1 \cdot\left[\frac{\mathrm{e}^\delta}{(1+\delta)^{1+\delta}}\right]^{\mu_{\max } / 2} \notag\\
     \leq& p_1 \cdot\left[\frac{\mathrm{e}^\delta}{(1+\delta)^{1+\delta}}\right]^{n / 2} \notag, 
\end{align}
where $C^* := \EE \frac{I_{\mathcal{A}}}{1-\frac{4\|Z\|}{\sigma_{r}(Y)}} \frac{\|Z\|^2}{\sigma_{r}(Y)^2} + \PP\{ \mathcal{A} ^c\}$. Similarly, 
\[
    \mathbb{P}\left\{\lambda_{1 }\left(\frac{1}{n}\sum_k (Q_{i} {Q_{i}}\T - \tilde{P}_i)\right)\geq{(1+\delta)C^* + \delta}\right\} \leq 
     p_1 \cdot\left[\frac{\mathrm{e}^\delta}{(1+\delta)^{1+\delta}}\right]^{n/ 2} , 
\]and hence, 
\be \label{eq_initailization_bound1}
    \mathbb{P}\left\{\left\|\frac{1}{n}\sum_k (\tilde{P}_i -   Q_{i} {Q_{i}}\T)\right\| \geq{(1+\delta)C^* + \delta}\right\} \leq 
    2 p_1 \cdot\left[\frac{\mathrm{e}^\delta}{(1+\delta)^{1+\delta}}\right]^{n / 2}. 
\ee
Further notice that $\frac{\mathrm{e}^\delta}{(1+\delta)^{1+\delta}} = \exp\left\{ \delta - (1+\delta)\log(1+\delta) \right\} = \exp\{-\delta^2/ 2 + o(\delta^2)\}$. Thus, it can be summarized as follows: 

For any $\delta$ satisfying $\min\{1, C^*\} \geq \delta > 0$, there is constant $c_1$ such that if $n>c_1 C^{*-2} \log p_1 $, with high probability, the following holds: 
\be \label{eq_initialization_numerator}
    \left\|\frac{1}{n}\sum_{i = 1}^n (\tilde{P}_i -   Q_{i} {Q_{i}}\T)\right\| \lesssim C^*
\ee

\subsubsection{Denominator}

By Lemma \ref{lemma_Weyl_eigenvalues}, it follows
\[
    \lambda_{r_{1}+1}\left(\frac{1}{n}\sum_{i = 1}^n\tilde{P}_i\right) \leq \lambda_{r_{1}+1}\left(P_U\right) + \lambda_1\left(\frac{1}{n}\sum_{i = 1}^n\hat{P}_i\right) = \lambda_1\left(\frac{1}{n}\sum_{i = 1}^n\hat{P}_i\right). 
\]
We need to prove $\lambda_1\left(\frac{1}{n}\sum_{i = 1}^n\hat{P}_i\right) \leq 1-c$ with high probability. 

Notice $\EE \left( \frac{1}{n} \hat{P}_i - \EE \frac{1}{n} \hat{P}_i\right) = 0$, $\left\|\frac{1}{n} \hat{P}_i - \EE \frac{1}{n} \hat{P}_i \right\| \leq \frac{2}{n}$ and $\left\| \sum_{i = 1}^n \EE \left( \frac{1}{n} \hat{P}_i - \EE \frac{1}{n} \hat{P}_i\right) \left( \frac{1}{n} \hat{P}_i - \EE \frac{1}{n} \hat{P}_i\right)\T \right\| \leq \frac{4}{n}$. Thus, by Matrix Bernstein, we have 
\[
    \PP\left( \lambda_1\left( \sum_{i = 1}^n \left(\frac{1}{n} \hat{P}_i - \EE \frac{1}{n} \hat{P}_i\right) \right) \geq t\right) 
    \leq p_1 \exp\left( \frac{-3nt^2}{32}\right). 
\]We have proved the concentration. 
Hence, for constant $c' = (1 - \|\EE \hat{P}_1\|)/2$, we have 
\begin{align}
    &\mathbb{P}\left\{\lambda_{r_1}\left(\frac{1}{n}\sum_{i = 1}^n\tilde{P}_i\right) - \lambda_{r_1+1}\left(\frac{1}{n}\sum_{i = 1}^n\tilde{P}_i\right) \leq (1 - \|\EE \hat{P}_1\|)/2\right\} \notag\\
    =&\mathbb{P}\left\{1 - \lambda_{r_1+1}\left(\frac{1}{n}\sum_{i = 1}^n\tilde{P}_i\right) \leq c' \right\} \notag\\
    = &\mathbb{P}\left\{\lambda_1\left(\frac{1}{n}\sum_{i = 1}^n\hat{P}_i\right) \geq 1 - c' \right\} \notag\\
    \leq &\mathbb{P}\left\{\lambda_1\left(\frac{1}{n}\sum_{i = 1}^n\hat{P}_i - \EE \hat{P}_1 \right) \geq 1 - c' - \lambda_1\left(\EE \hat{P}_1 \right)\right\} \notag\\
     \leq& p_1 \exp\left( \frac{-3n\left( 1 - \lambda_1\left(\EE \hat{P}_1 \right) \right)^2}{128}\right) \label{eq_upper_bound_denominator}. 
\end{align}

Finally, combining \eqref{eq_initialization_DK}, \eqref{eq_initialization_numerator} and \eqref{eq_upper_bound_denominator}, we obtained that if 
\[
    n \gtrsim \log p_1 \max\left\{ C^{*-2} , \frac{1}{\left(1 - \lambda_1\left(\EE \hat{P}_1 \right) \right)^2} \right\},
\]
then with high probability, it follows
\[
    \|\sin \Theta (\hat{U}_j^{(0)}, U_j)\| \lesssim \frac{C^*}{1 - \lambda_1\left(\EE \hat{P}_1 \right)}. 
\]
For $0\leq x<1/4$, the function $\frac{x^2}{1-4x}$ is convex (its second order derivative is $\frac{2}{(1-4x)^3}$.) Hence, we have 
    \[
	\EE \frac{I_{\{\mathcal{A}\}}}{1-\frac{4\|Z\|}{\sigma_{r}(UA + BV\T)}} \frac{\|Z\|^2}{\sigma_{r}(UA + BV\T)^2}
        = \EE \frac{I_{\{\mathcal{A}\}}^2}{1-\frac{4I_{\{\mathcal{A}\}} \|Z\|}{\sigma_{r}(UA + BV\T)}} \frac{\|Z\|^2}{\sigma_{r}(UA + BV\T)^2}
	\leq \frac{ {\rho}^2}{1-4\rho}, 
    \]where $\rho = \EE \frac{I_{\{\mathcal{A}\}} \|Z\|}{\sigma_{r}(UA+BV\T))} \leq c/4 < 1/4$. Thus, $C^* \leq \frac{c^2}{16(1-c)} + \PP(4\|Z\| > c \sigma_{r}(UA+ BV\T))$. If further $c\leq 1/2$, then $C^* \leq \frac{c^2}{8} + \PP(4\|Z\| > c \sigma_{r}(UA+ BV\T))$. 

Recall that we have $\Span (\hat{P}_1) \subseteq \Span(U_\perp)$ almost surely. Thus $\|\EE \hat{P}_1\| = \| \EE P_{U_\perp}\hat{P}_1\| = \|\EE P_{U_\perp}\tilde{P}_1\|$. Also recall that $\tilde{P}_1$ is the projector to $\Span(UA_1 + B_1 V\T) = \Span(U) + \Span(B_1) = \Span(U) + \Span(P_{U_\perp}B_1)$. So, for any vector $v\in \RR^{p_1},$ we have $v = b + u + z\in \RR^{p_1},$ where $b\in \Span(P_{U_\perp} B_1)$, $u \in \Span(U)$ and $z\in (\Span(UA_1 + B_1 V\T))^\perp$, and hence we have $P_{U_\perp} \tilde{P}_1 v = b$. Thus, we have $P_{U_\perp}\tilde{P}_1 =  P_{U_\perp U_\perp\T B_1}$ and $\|\EE P_{U_\perp U_\perp\T B_1}\| = \|\EE \hat{P}_1\|$. Hence, the therorem holds. 
\end{proof}

%%%%%%%%%%%%%%%%%%%%%%%%
\section{Proof of Example \ref{example_iid_gaussian}}
%%%%%%%%%%%%%%%%%%%%%%%%

\begin{proof}
    Notice $P_{U_\perp U_\perp\T B} = P_{U_\perp}B (B\T P_{U_\perp}P_{U_\perp}B)^\dagger B\T P_{U_\perp} = U_\perp U_\perp\T B (B\T U_\perp U_\perp\T B)^\dagger B\T U_\perp U_\perp\T = U_\perp P_{U_\perp\T B} U_\perp\T$, where $A^\dagger$ is the generalized inverse of matrix $A$. Further, notice the entries of $U_\perp\T B$ are i.i.d. standard Gaussian. Denote $W = U_\perp\T B$. Then, for any orthogonal matrix $O \in \mathcal{O}_{p_1 - r_1}$, we have $OW \overset{d}{=} W$. Thus, $P_W \overset{d}{=} P_{O W}$. Hence, we have $\EE P_W = \EE P_{O W} = O\EE  P_W O\T$. As $O$ can be chosen arbitrarily in $\mathcal{O}_{p_1 - r_1}$, we have $\EE  P_W = aI$ for some $a$. By calculating trace $a(p_1 - r_1) = \tr(\EE  P_W) = \EE \tr(P_W) = \min \{r_2, p_1 - r_1\}$, we have $a = \min \{1, r_2 / (p_1 - r_1)\}$. Thus, $\EE P_{U_\perp U_\perp\T B} =  U_\perp\EE P_{U_\perp\T B} U_\perp\T = \min \{1, r_2 / (p_1 - r_1)\} \cdot P_{U_\perp}$. 
\end{proof}

%%%%%%%%%%%%%%%%%%%%%%%%
\section{Proof of Theorem \ref{theorem_convergence}}
%%%%%%%%%%%%%%%%%%%%%%%%

Let's first introduce notations of sets $\mathcal{A}^U$ and $\mathcal{A}^V$ of conditions on $A_i, B_i$, $Z_i$, and initialization. 
\begin{Notation}\label{notation_W5678} \label{notation_kappa}
    Denote
    \begin{align*}
        T_U = & \left\{ W \in \RR^{p_2 \times p_2}: \|W\| \leq 1 , \tr(W) = 0, W = W\T, \rank(W) \leq 2 r_2 \right\}, \\
        L_U = & \frac{1}{np_2}\sigma_{r_1}\left(\sum_{i = 1}^{n}A_i P_{V_\perp}A_i \T   \right),\\
        M_{1,U} = &\frac{1}{np_2} \sup_{E\in T_h} \left\|\sum_{i = 1}^{n} A_i E \left( B_iV\T \right)\T \right\| , \\
        M_{3,U} = &\frac{1}{np_2}\sup_{E\in T_h}\left\|\sum_{i = 1}^{n} Z_i E \left( UA_i +B_i V\T \right)\T   \right\|  , \\
        M_{4,U} = &\frac{1}{np_2}\sup_{E\in T_h} \left\|\sum_{i = 1}^{n} Z_i E  Z_i \T \right\|, \\
        \Delta_U = & L_U - (M_{1,U} + M_{3,U} + M_{4,U}),\\
        \xi_U =& 10 +M_{1,U}+ 3M_{3,U} +3M_{4,U},\\
        \Gamma_U = & \Delta_U - \varepsilon_0 \xi_U,\\
        \kappa_U =& \frac{2}{np_2}\left\|\sum_{i = 1}^{n} Z_i P_{V_\perp} \left( UA_i \right)\T \right\| +\frac{1}{np_2}\left\|\sum_{i = 1}^{n} Z_i P_{V_\perp} Z_i\T - n\sigma^2(p_2 - r_2) I \right\|,\\
        \xi_V, &\Gamma_V, \kappa_V, L_V \text{ are defined similarly by replacing $U, p_2, r_2$ by $V, p_1, r_1$ respectively, }\\
        \mathcal{A}_1 = & \{\Gamma_U > 0, \Gamma_V > 0\},\\
        \mathcal{A}_2 = & \left\{\frac{4\kappa_U}{L_U - \varepsilon_0\xi_U -  3 \kappa_U} \leq \frac{\Gamma_U}{L_U} \varepsilon_0, \frac{4\kappa_V}{L_V - \varepsilon_0\xi_V -  3 \kappa_V} \leq \frac{\Gamma_V}{L_V} \varepsilon_0\right\},\\
        \mathcal{A}_3^U = & \left\{\max_{i}\{\left\|A_i \right\|, \left\|B_i \right\|\} \ \leq \sqrt{p_2}\right\},\\
        \mathcal{A}_3^V = & \left\{\max_{i}\{\left\|A_i \right\|, \left\|B_i \right\|\} \ \leq \sqrt{p_1}\right\},\\
        \mathcal{A}^U = & \mathcal{A}_1 \cap \mathcal{A}_2 \cap\mathcal{A}_3^U,\\
        \mathcal{A}^V = & \mathcal{A}_1 \cap \mathcal{A}_2 \cap\mathcal{A}_3^V
    \end{align*}
\end{Notation}

The strategy of this proof is to first establish a deterministic upper bound for estimation error given that $A_i, B_i$, $Z_i$ are nonrandom satisfying conditions $\{A_i, B_i, Z_i; i = 1,\cdots n\} \subseteq \mathcal{A}$ (in Section \ref{section_deterministic_bound_proof}), and then prove these conditions hold with high probability (in Section \ref{section_statistical_bound}). 

%%%%%%%%%%%
\subsection{Deterministic Bound}\label{section_deterministic_bound_proof}
%%%%%%%%%%%
We first introduce the following technical lemmas that will be used in this section: 
\begin{Lemma}[Weyl's eigenvalue inequality]\label{lemma_Weyl_eigenvalues}For Hermitian matrices $A$ and $B$, we have
    \be \label{inequality_eigenvalue}
    \lambda_{i+j-1}(A+B) \leq \lambda_i(A)+\lambda_j(B).
    \ee 
    As a result, \be \label{inequality_eigenvalue_sum}
    \lambda_{r}(A) - \|B\| \leq \lambda_{r}(A+B) \leq \lambda_{r}(A) + \|B\|. 
    \ee
\end{Lemma}	
\begin{proof}
    See page 40 of \cite{tao2012topics} and its references for (\ref{inequality_eigenvalue}). Then, letting $i = r$ and $j = 1$, we have: \[
    \lambda_{r}(A+B) \leq \lambda_{r}(A) + \lambda_{1}(B) \leq \lambda_{r}(A) + \|B\|. 
    \]Similarly, we have: \[
    \lambda_{r}(A) - \|B\| \leq \lambda_{r}(A) +\lambda_{r}(B) = \lambda_{r}(A) - \lambda_{1}(-B) \leq \lambda_{r}(A+B), 
    \]which finishes the proof of (\ref{inequality_eigenvalue_sum}). 
\end{proof}

\begin{Lemma}[Ky Fan singular value inequality, \cite{fan1951maximum}]\label{lemma_KyFan_Singular}
    For any $n\times n$ matrices $A, B$,
    $$
    \sigma_{r+t+1}(A+B) \leq \sigma_{r+1}(A)+\sigma_{t+1}(B),
    $$
    where $t \geq 0, r \geq 0, r+t+1 \leq n$. Specially, let $r = r_1 -1$ and $t = 0$: \[
    \sigma_{r_1}(A+B) \leq \sigma_{r_1}(A)+\sigma_{1}(B).
    \]
\end{Lemma}

\begin{Lemma}[Exercise VII.I.11 in \cite{bhatia1997matrix}]\label{sintheta_distance}
    $$\|\sin \Theta (V, \hat{V}^{(t)})\| = \|V\T \hat{V}^{(t)}_{\perp}\| = \| \hat{V}^{(t) \top}V_\perp \|= \| P_{V_{\perp}} - P_{\hat{V}^{(t)}_{\perp}} \|.$$
\end{Lemma}

Define 
\[
    S_i = UA_i + B_i V\T = X_i - Z_i,
\]
\[
    H = \frac{1}{np_2}\sum_{i = 1}^{n} S_i V_\perp V_\perp\T S_i = \frac{1}{np_2}\sum_{i = 1}^{n} UA_i P_{V_\perp} A_i\T U\T,
\]
\[
    \widehat H^{(t)} = \frac{1}{np_2}\sum_{i = 1}^{n} X_i \hat{V}_\perp^{(t)} \hat{V}_\perp^{(t)\top} X_i.
\]
Then we have the following lemma: 

\begin{Lemma}\label{parameters_2}
    In the context of Corollary \ref{Corollary_perturbation_bound}, let $\widehat{X} = \widehat{H}^{(t+1)} - \frac{p_2 - r_2}{p_2}\sigma^2 I, {X} = {H},$ and denote 
    $$\varepsilon_t = \max\left\{\|\sin \Theta (U, \widehat{U}^{(t)})\|, \|\sin \Theta (V, \widehat{V}^{(t)})\|\right\}, \alpha = \lambda_{r_1} (P_{U} \widehat{X} P_{U}), \beta = \|{P}_{U_{\perp}}\widehat{X}{P}_{U_{\perp}}\|,z_{21} = \|{P}_{U} (X - \widehat{X}) {P}_{U_{\perp}}\|.$$ When $A_i, B_i$ are nonrandom satisfying condition $\{A_i, B_i; i = 1,\cdots n\} \subseteq \mathcal{A}_3^U$, we have: 
    $$
    \begin{aligned}
        \alpha \geq& L_U - \varepsilon_t(4 + 2M_{3,U} + M_{4,U}) - \kappa_U,\\
        \alpha \leq&L_U + \varepsilon_t(4 + 2M_{3,U} + M_{4,U}) + \kappa_U,\\
        \beta \leq& \kappa_U + \varepsilon_t (5 + M_{4,U}),\\
        z_{21} \leq& \kappa_U + \varepsilon_t ( M_{1,U} + M_{3,U} + M_{4,U} + \varepsilon_t). 
    \end{aligned}
    $$
\end{Lemma}

\begin{proof}
First notice that $\rank(P_{\hat{V}_\perp^{(t)}} - P_{V_\perp}) = \rank(P_{\hat{V}^{(t)}} - P_{V}) \leq 2r_2. $
    Then by Lemma \ref{lemma_Weyl_eigenvalues}, 
    \begin{align*}
          &n p_2 \alpha\\
        =&n p_2 \lambda_{r_1}\left(U\T \widehat{H}^{(t+1)} U -  \sigma^2\frac{p_2 - r_2}{p_2}\sigma^2 \cdot U_1\T I U_1\right) \\
        \overset{(\ref{inequality_eigenvalue_sum})}{\leq} & \lambda_{r_1}\left(\sum_{i = 1}^{n}\left(  UA_i \hat{V}_\perp^{(t)}\right) \left(  UA_i \hat{V}_\perp^{(t)}\right)\T\right) 
        + 2\left\|\sum_{i = 1}^{n} \left(  UA_i \hat{V}_\perp^{(t)}\right) \left(  B_iV\T \hat{V}_\perp^{(t)}\right)\T U\right\|\\
        & + \left\|U\T\sum_{i = 1}^{n} \left(  B_iV\T \hat{V}_\perp^{(t)}\right) \left(  B_iV\T \hat{V}_\perp^{(t)}\right)\T U\right\| 
        + 2\left\|U\T \sum_{i = 1}^{n} \left(UA_i + B_i V\T \right) \hat{V}_\perp^{(t)} \left( Z_i \hat{V}_\perp^{(t)} \right) \T U\right\| \\
        & + \left\|U\T\sum_{i = 1}^{n} \left(Z_i \hat{V}_\perp^{(t)}\right) \left(Z_i \hat{V}_\perp^{(t)}\right) \T U - n\sigma^2 (p_2 - r_2)U\T I U\right\|\\
        \leq & \sigma_{r_1}\left(\sum_{i = 1}^{n} A_i P_{V_\perp} A_i\T   \right) + \varepsilon_t\left\| \sum_{i = 1}^{n}A_i \frac{P_{\hat{V}_\perp^{(t)}} - P_{V_\perp}}{\varepsilon_t}  A_i\T \right\| + 2\left\|\sum_{i = 1}^{n} A_i P_{V_\perp} \left( B_iV\T \right)\T \right\|\\
        &+ 2\varepsilon_t\left\|\sum_{i = 1}^{n} A_i \frac{P_{\hat{V}_\perp^{(t)}} - P_{V_\perp}}{\varepsilon_t}  \left( B_iV\T \right)\T \right\| 
        + np_2\varepsilon_t^2 
        + 2\left\|\sum_{i = 1}^{n} Z_i P_{V_\perp} \left( UA_i + B_iV\T\right)\T   \right\|\\
        &+ 2\varepsilon_t \left\|\sum_{i = 1}^{n} Z_i  \frac{P_{\hat{V}_\perp^{(t)}} - P_{V_\perp}}{\varepsilon_t} \left( UA_i + B_iV\T\right)\T   \right\| 
        +\left\|\sum_{i = 1}^{n} Z_i P_{V_\perp}  Z_i  \T - n\sigma^2(p_2 - r_2) I \right\|\\
        &+ \varepsilon_t \left\|\sum_{i = 1}^{n} Z_i \frac{P_{\hat{V}_\perp^{(t)}} - P_{V_\perp}}{\varepsilon_t}  Z_i  \T \right\| \\
        \leq& np_2\left( L_U + \varepsilon_t(4 + 2M_{3,U} + M_{4,U}) + \kappa_U \right), 
    \end{align*}
    where the last inequality holds by $\{A_i, B_i, Z_i; i = 1,\cdots n\} \subseteq \mathcal{A}\cap\mathcal{B}$. 
    
    By the same procedure, we can similarly derive a lower bound of $\alpha$: (details are presented in the proof of Lemma \ref{parameters_2_tensor} in more general setting)
    \[
        \alpha \geq  L_U - \varepsilon_t(4 + 2M_{3,U} + M_{4,U}) - \kappa_U ; 
    \]
    a upper bound of $\beta$: 
    \[
        \beta \leq \kappa_U + \varepsilon_t (5 + M_{4,U}); 
    \]and a upper bound of $z_{21}$:
    \[
        z_{21} \leq \kappa_U + \varepsilon_t ( M_{1,U} + M_{3,U} + M_{4,U} + \varepsilon_t). 
    \]
\end{proof}

We are ready to prove the following:

\begin{Theorem}\label{theorem_convergence_deterministic}
    In Algorithm \ref{algorithm_iterative_projection}, let $X_i = UA_i + B_i V\T + Z_i \in \mathbb{R}^{p_1\times p_2}, i = 1, \cdots n$. 
    Denote the estimation error $\varepsilon_t = \max\left\{\|\sin \Theta (U, \widehat{U}^{(t)})\|, \|\sin \Theta (V, \widehat{V}^{(t)})\|\right\}$. 
    When $\{A_i, B_i, Z_i; i = 1,\cdots n\}$ are nonrandom satisfying condition $\{A_i, B_i, Z_i; i = 1,\cdots n\} \subseteq \mathcal{A}^U$, there is a constant $\chi_U = \frac{L_U - \Gamma_U}{L_U}<1$,  such that for $t=0,1,\ldots, m$, 
    \begin{equation}
        \|\sin \Theta (U, \widehat{U}^{(t+1)})\| \leq \chi_U \|\sin \Theta (V, \widehat{V}^{(t)})\| + K_1^U\leq \varepsilon_0 , \notag
    \end{equation}
    where $$K_1^U = \frac{4\kappa_U}{L_U - \xi_U -  3 \kappa_U}. $$
    Similar results hold by switching $U$ and $V$. Overall, if $\{A_i, B_i, Z_i; i = 1,\cdots n\} \subseteq \mathcal{A}^U \cap \mathcal{A}^V$, then 
    there is a constant $\chi = \max\{ \frac{L_U - \Gamma_U}{L_U}, \frac{L_V - \Gamma_V}{L_V}\}<1$, which does not depend on $t$, such that for $t=0,1,\ldots, m$, 
    \begin{equation}
        \varepsilon_{t+1} \leq \chi \varepsilon_t + K_1\leq \varepsilon_0 , \notag
    \end{equation}
    where $$K_1 = \max\left\{\frac{4\kappa_U}{L_U - \xi_U -  3 \kappa_U}, \frac{4\kappa_V}{L_V - \xi_V -  3 \kappa_V}\right\}. $$
    Consequently, 
    \begin{align}
        \varepsilon_t \leq \chi^t \varepsilon_0 + K_2,\notag 
    \end{align}where $K_2 = K_1 \frac{1-\chi^t}{1-\chi}$. 
\end{Theorem}

\begin{proof}
    We are going to prove this by induction. Assume the statement holds for $m<t$. Consider $m = t$ and we are calculating $\widehat U^{(t+1)}$. 
    
    By Corollary \ref{Corollary_perturbation_bound}:
    $$
    \|\sin \Theta(U, \widehat{U}^{(t+1)})\| \leq \frac{z_{21}}{\alpha-\beta-z_{21}}. 
    $$
    Denote$$
    \begin{aligned}
        \alpha_1 =& L_U - \varepsilon_t(4 + 2M_{3,U} + M_{4,U}),\\
        \beta_1 =& \varepsilon_t(5 + M_{4,U}),\\
        z_{1} =& \varepsilon_t(M_{1,U} + \varepsilon_t + M_{3,U} + M_{4,U}). 
    \end{aligned}
    $$
    By Lemma \ref{parameters_2}, 
    \begin{align}
        \frac{z_{21}}{\alpha-\beta-z_{21}}
        & \leq \frac{z_1 + \kappa_1}{\alpha_1 - \beta_1 - z_1 - 3\kappa_1} \notag\\
        & = \frac{z_1}{\alpha_1 - \beta_1 - z_1}+ \frac{\kappa_1}{\alpha_1 - \beta_1 - z_1-3\kappa_1} + \frac{z_1}{\alpha_1 - \beta_1 - z_1}\frac{3\kappa_1}{\alpha_1 - \beta_1 - z_1-3\kappa_1}. \label{inequality_zab}
    \end{align}
    To bound the first term on right hand side of (\ref{inequality_zab}), notice that the function $f(y) = (x-y)/x$ is monotone decreasing for $y<x$ with any given $x$ and that by $\{A_i, B_i, Z_i; i = 1,\cdots n\} \subseteq \mathcal{A}$, we have
    $$
    \begin{aligned}
        \alpha_1 - \beta_1 - z_1 - \frac{z_1}{\varepsilon_t }
        =&  L_U - (M_{1,U} + M_{3,U} + M_{4,U})\\
        &- \varepsilon_t (10 +M_{1,U}+ 3M_{3,U} +3M_{4,U})\\
        \geq& \Gamma_U > 0. 
    \end{aligned}
    $$So, 
    \be
    \label{inequality_term1_1}
    \frac{z_1/\varepsilon_t}{\alpha_1 - \beta_1 - z_1} \leq \frac{\alpha_1 - \beta_1 - z_1 - \Gamma_U}{\alpha_1 - \beta_1 - z_1}.
    \ee
    Further notice that the function $g(x) = (x-y)/x$ is monotone increasing on $x>y$ for fixed $y$ and that we have $\alpha_1 - \beta_1 - z_1 \leq L_U$. So, 
    \be
    \label{inequality_term1_2}
    \frac{\alpha_1 - \beta_1 - z_1 - \Gamma_U}{\alpha_1 - \beta_1 - z_1} \leq \frac{L_U - \Gamma_U}{L_U}. 
    \ee
    Thus, combining (\ref{inequality_term1_1}) and (\ref{inequality_term1_2}), we have
    \be
    \label{inequality_term1}
    \frac{z_1}{\alpha_1 - \beta_1 - z_1} \leq \frac{L_U - \Gamma_U}{L_U}\varepsilon_t. 
    \ee
    To bound the remaining two terms of (\ref{inequality_zab}), we have \[
    \frac{\kappa_U}{\alpha_1 - \beta_1 - z_1-3\kappa_U} + \frac{z_1}{\alpha_1 - \beta_1 - z_1}\frac{3\kappa_U}{\alpha_1 - \beta_1 - z_1-3\kappa_U} \leq \frac{4\kappa_U}{\alpha_1 - \beta_1 - z_1-3\kappa_U}, 
    \]
    and
    \[
    \begin{aligned}
        \alpha_1 - \beta_1 - z_1 
        \geq  L_U - \varepsilon_0 \xi_U,
    \end{aligned}
    \]
    which yield
    \be
    \label{inequality_term2}
        \frac{\kappa_U}{\alpha_1 - \beta_1 - z_1-3\kappa_U} + \frac{z_1}{\alpha_1 - \beta_1 - z_1}\frac{3\kappa_U}{\alpha_1 - \beta_1 - z_1-3\kappa_U} \leq \frac{4\kappa_U}{L_U - \varepsilon_0\xi_U -  3 \kappa_U}. 
    \ee
    Combining (\ref{inequality_zab}), (\ref{inequality_term1}) and (\ref{inequality_term2}), we finally have \[
    \|\sin \Theta(U, \widehat{U}^{(t+1)})\| \leq \frac{z_{21}}{\alpha-\beta-z_{21}} \leq \frac{L_U - \Gamma_U}{L_U} \varepsilon_t + \frac{4\kappa_U}{L_U - \varepsilon_0\xi_U -  3 \kappa_U} \leq \chi \varepsilon_t + K_1. 
    \]
    By $\{A_i, B_i, Z_i; i = 1,\cdots n\} \subseteq \mathcal{A}$, we further have\[\begin{aligned}
        \frac{L_U - \Gamma_U}{L_U} \varepsilon_t + \frac{4\kappa_U}{L_U - \varepsilon_0\xi_U -  3 \kappa_U}
        &\leq \frac{L_U - \Gamma_U}{L_U} \varepsilon_0 + \frac{\Gamma_U}{L_U} \varepsilon_0 = \varepsilon_0. 
    \end{aligned}
    \]Thus, we have proved\[
    \|\sin \Theta(U, \widehat{U}^{(t+1)})\| \leq \chi \varepsilon_t + K_1 \leq \varepsilon_0. 
    \]Similarly, we can prove $\|\sin \Theta(V, \widehat{V}^{(t+1)})\| \leq \chi \varepsilon_t + K_1 \leq \varepsilon_0$. So the statement holds by induction.
\end{proof}

%%%%%%%%%%%
\subsection{Statistical Bound}\label{section_statistical_bound}
%%%%%%%%%%%
In this section, we are going to argue that when $\{A_i, B_i, Z_i; i = 1,\cdots n\}$ are random matrices satisfying Assumption \ref{assumption_conditionnumber} and \ref{assumption_z} with proper initialization, the probability of $\mathcal{A}$ is high and the estimation error converges to 0 in probability.  

For convenience, let's rewrite Assumption \ref{assumption_conditionnumber} as following: 
\begin{Assumption}\label{assumption_conditionnumber_star}
    Assume in decomposition (\ref{eq:decomposition-equivalence-matrix}), $A$ and $B$ are independent and there are constants $\mu_U$ and $\mu_V$ such that
    \[\PP \left(\left\{\mu_U \leq \frac{\max \{\left\| A \right\|, \left\| B \right\|\}}{\sqrt{p_2}} \right\}\cup\left\{\mu_V \leq \frac{\max \{\left\| A \right\|, \left\| B \right\|\}}{\sqrt{p_1}} \right\}\right) \leq \nu, \quad \text{for some small $\nu < 1$. }\]
    Denote $\lambda_1$ and $\lambda_2$ as 
    $$\lambda_U =\lambda_{\min }\left(\frac{1}{p_2}\mathbb{E} A P_{V_\perp} A \T \right), \quad
    \lambda_V =\lambda_{\min }\left(\frac{1}{p_1}\mathbb{E} B\T P_{U_\perp} B \right).$$
    We have $\max\{\mu_U^2/\lambda_U, \mu_V^2/\lambda_V\}\leq C$ for some constant $C>0$.
\end{Assumption}

We additionally assume $\mu_1 = 1$ and $\nu = 0$ for now, i.e., $\PP(\mathcal{A}_3^U) = 1$. Denote $\sigma^2$ and $\zeta^4$ as the variance and fourth moments of each entry $Z_i$. 
Define $u = \max \left\{ \left\| Z/\sqrt{p_1}\right\|_{\psi_2}, \left\|Z/\sqrt{p_2}\right\|_{\psi_2} \right\}$. 
The following lemma bounds the sub-Gaussian norm of $\|Z_i\|$. 

\begin{Lemma}\label{lemma_subGaussian_norm}
    Assume $Z$ is a $m$ by $n$ random matrix with i.i.d. sub-Gaussian entries with sub-Gaussian norm $K$. Then $\|Z\|$ is sub-Gaussian and $\|\|Z\|\|_{\psi_2} \lesssim K(\sqrt{m} + \sqrt{n})$. 
\end{Lemma}
\begin{proof}
    Notice the following two facts: 
    \begin{enumerate}
        \item There is an absolute constant $C_1$ such that for any $t>0$, \be
        \label{eq_equivalent_def_sugGaussian}
        \PP\left(\|Z\| > C_1K(\sqrt{m} + \sqrt{n} + t)\right) \leq 2e^{-t^2}
        \ee
        \item That a random variable $X$ is sub-Gaussian is equivalent to the following: \[
        \mathbb{P}\{|X| \geq t\} \leq 2 \exp \left(-t^2 / K_1^2\right) \quad \text { for all } t \geq 0
        \]Furthermore, there are absolute constants $C_2$, $C_3$ such that $C_2K_1 \leq K \leq C_3K_1$. 
    \end{enumerate}
    In (\ref{eq_equivalent_def_sugGaussian}), let $y = C_1K(\sqrt{m} + \sqrt{n} + t)$. Then we have \[
    \PP\left(\|Z\| > y\right) \leq 2e^{-\frac{y^2}{C_1^2K^2} + 2(\sqrt{m} + \sqrt{n})\frac{y}{C_1K}}\leq 2e^{-\frac{y^2}{2C_1^2K^2}}
    \]for $y\geq 4C_1K(\sqrt{m} + \sqrt{n})$. When $y \leq 4C_1K(\sqrt{m} + \sqrt{n})$, we have \[
    \PP\left(\|Z\| > y\right) \leq 1 \leq 2e^{-\frac{y^2}{(4C_1K(\sqrt{m} + \sqrt{n}))^2}}.
    \]Hence, by the second fact, $\|\|Z\|\|_{\psi_2} \lesssim K(\sqrt{m} + \sqrt{n})$. 
\end{proof}

let's firstly bound the term $L_U$ in Notation \ref{notation_W5678} by the well-known Matrix Chernoff inequality (Lemma \ref{lemma_Chernoff}). In our setting,  it yields: 
\begin{align*}
    \PP\{L_U \leq (1-\delta) \lambda_U\}
    \leq&
    r_1 \cdot \left[\frac{\mathrm{e}^{-\delta}}{(1-\delta)^{1-\delta}}\right]^{n\lambda_U}. 
\end{align*}
Taking $\delta = 1/2$, since $e^{-0.5}/0.5^{0.5} \leq 0.86$, we have 
\begin{Corollary}\label{corollary_L}
    \[
    \PP\{L_U \leq \lambda_U/2\}
    \leq
    r_1 \cdot  \exp\left\{\log(0.86)n\lambda_U\right\}.
    \]
\end{Corollary}
To bound the terms $M$'s in Notation \ref{notation_W5678}, we are going to use the strategy called ``union bound''. To that end, let's first estimate the covering number. 

\begin{Lemma}[Lemma 7 in \cite{zhang2018tensor}]\label{lemma_covering_number}
    Let $\mathcal{X}_{p_{1}, p_{2}, r}=\left\{X \in \mathbb{R}^{p_{1} \times p_{2}}: \operatorname{rank}(X) \leq r,\|X\| \leq 1\right\}$ be the class of low-rank matrices under spectral norm. Then there exists an $\varepsilon$-net $\overline{\mathcal{X}}_{r}$ for $\mathcal{X}_{p_{1}, p_{2}, r}$ with cardinality at most $((4+\varepsilon) / \varepsilon)^{\left(p_{1}+p_{2}\right) r}$. Specifically, there exists $X^{(1)}, \ldots, X^{(N)}$ with $N \leq((4+\varepsilon) / \varepsilon)^{\left(p_{1}+p_{2}\right) r}$, such that for all $X \in \mathcal{X}_{p_{1}, p_{2}, r}$, there exists $i \in\{1, \ldots, N\}$ satisfying $\left\|X^{(i)}-X\right\| \leq \varepsilon$. 
\end{Lemma}
In our setting, recall 
$$T_U = \left\{ W \in \RR^{p_2 \times p_2}: \|W\|\leq 1 , \tr(W) = 0, W = W\T, \rank(W) \leq 2r_2 \right\}.$$
We have
\begin{Corollary}\label{corollary_epsilon_net}
    There exists subset $\bar{T}_U$ of ${T}_U$ such that for some absolute constant $C_0$, $$|\bar{T}_U|\leq \exp(4 r_2 p_2 \log (C_0/\varepsilon)),$$
    \[
    M_{1,U} \leq \frac{1}{np_2} \sup_{E\in \bar T_1} \left\|\sum_{i = 1}^{n} A_i E \left(B_i V\T\right)\T \right\| + \varepsilon, 
    \] and similar bounds hold for other $M_{i,U}$ for $i= 3,4$ in Notation \ref{notation_W5678}. 
\end{Corollary}
Then, we need the following concentration inequality: 
\begin{Lemma}\label{lemma_bernstein}[Matrix Bernstein, subexponential non-symmetric version]
    Let $Z, Z_1, \ldots, Z_n$ be i.i.d. random matrices with dimensions $m_1 \times m_2$ that satisfy $\mathbb{E}(Z)=0$. Suppose that $U_Z^{(\alpha)}<\infty$ for some $\alpha \geq 1$. Then there exists a constant $C>0$ such that, for all $t>0$, with probability at least $1-e^{-t}$
    $$
    \left\|\frac{Z_1+\cdots+Z_n}{n}\right\| \leq C \max \left\{\sigma_Z \sqrt{\frac{t+\log (m)}{n}}, U_Z^{(\alpha)}\left(\log \frac{U_Z^{(\alpha)}}{\sigma_Z}\right)^{1 / \alpha} \frac{t+\log (m)}{n}\right\},
    $$
    where $m=m_1+m_2$, 
    $$U_Z^{(\alpha)}\in \inf \left\{u>0: \mathbb{E} \exp \left(\|Z\|^\alpha / u^\alpha\right) \leq 2\right\}, \quad \alpha \geq 1, $$ 
    $$\sigma_Z \geq \max \left\{\left\|\frac{1}{n} \sum_{i=1}^n \mathbb{E}\left(Z_i Z_i^{\top}\right)\right\|^{1 / 2},\left\|\frac{1}{n} \sum_{i=1}^n \mathbb{E}\left(Z_i^{\top} Z_i\right)\right\|^{1 / 2}\right\}$$ and \[
    \sigma_Z \leq U_Z^{(\alpha)}
    \]
\end{Lemma}
\begin{proof}
    This is a slight generalization of Proposition 2 in \cite{koltchinskii2011nuclear} and its reference \cite{koltchinskii2011neumann}. First, we can directly generalize the choice of $U_Z^{(\alpha)}$. Then it can be shown that we really only need an upper bound for $\sigma_Z$ such that equation (3.7) in \cite{koltchinskii2011neumann} to be well defined. 
\end{proof}
In our case, using union bound, Corollary \ref{corollary_epsilon_net} and Lemma \ref{lemma_bernstein} yields the following corollaries for different $M$'s in Notation \ref{notation_W5678}. 
\begin{Corollary}\label{corollary_M1}
    For given $d$, $c_1$ and $c_2$, there exists constant $c_3$ such that if $$n \geq  c_3\left(\frac{1}{\lambda_U}\right)^2(r_2) (p_1 + p_2),$$ then with probability at least $1-e^{-(c_2)r_2 (p_1 + p_2)}$, 
    we have
    \[
    M_{1,U} \leq c_1\lambda_U. 
    \]
\end{Corollary}

\begin{proof}
    Let $W_i = \frac{1}{p_2}A_i E \left(B_i V\T\right)\T$, then $\|W_i\| \leq 1$ and hence $\exp((\log 2) \|W_i\|) \leq 2$. Also, $\|W_i W_i^{\top}\| = \|W_i\T W_i\| \leq 1$. So in Lemma \ref{lemma_bernstein}, for any given $\varepsilon>0$, constant $C$ and $E \in T_1$ , let $t = C(r_2) (p_1 + p_2)$, $\sigma_Z = 1$, $\alpha = 1$ and $U_Z^{(\alpha)} = 1/\log2$, which yields that there exists constant $C_2$ such that with probability at least $1-e^{-Cr_2 (p_1 + p_2)}$, the following holds: 
    \[\begin{aligned}
        \frac{1}{n p_2} \left\|\sum_{i = 1}^{n} A_i E \left(B_i V\T\right)\T \right\| 
        \leq C_2 \max \left\{\sqrt{\frac{r_2 (p_1 + p_2)}{n}}, \frac{r_2 (p_1 + p_2)}{n}\right\}. 
    \end{aligned}
    \]Then by union bound from Corollary \ref{corollary_epsilon_net}, it yields that with probability at least
    $$1-e^{-(C-\log(C_0/\varepsilon))r_2 (p_1 + p_2)},$$ we have 
    \[
        M_{1,U} \leq C_2 \max \left\{\sqrt{\frac{r_2 (p_1 + p_2)}{n}}, \frac{r_2 (p_1 + p_2)}{n}\right\} + \varepsilon. 
    \]
    So for given constants $C_4$ and $C_5$, let $\varepsilon$ be a constant multiplier of $\lambda_U$ and hence, if 
    $$n \geq  C_3\left(\frac{1}{\lambda_U}\right)^2(r_2) (p_1 + p_2), $$ 
    for some large enough $C_3$, then with probability at least  $1-e^{-(C_4)(r_2) (p_1 + p_2)}$, we have
    \[
        M_{1,U} \leq C_5\lambda_U. 
    \]
\end{proof}
To deal with $M_{3,U}$ $M_{4,U}$ and $\kappa_U$, first notice the following fact: 
\begin{Lemma}\label{lemma_Z}
    Assume $Z$ is a $p$ by $q$ random matrix whose entries $Z_{ij}$ are i.i.d. with mean 0, variance $\sigma^2$ and forth moment $\zeta^4$. $D$ is a fixed $q$ by $q$ symmetric matrix. Then we have \[
    \EE Z DZ\T = \sigma^2 tr(D) I, 
    \]\[
    \EE Z DZ\T Z\T DZ \preccurlyeq p_2 \zeta^4 tr(D^2) I. 
    \]
\end{Lemma}
\begin{proof}
    Denote 
    \[Z\T = \begin{bmatrix}
        z_{1}        
        \cdots
        z_{p}      
    \end{bmatrix},\] where $z_{j}$ is $j$th column vector of $Z\T$. Hence, 
    \[DZ\T = \begin{bmatrix}
        D z_{1}
        \cdots 
        D z_{p} 
    \end{bmatrix}.\]Furthermore, 
    \begin{align*}
        \EE Z DZ\T
        =&  \begin{bmatrix} \left(\EE z_{h}\T Dz_{k}\right)_{h,k} \end{bmatrix}_{p \times p}\\
        =& \diag \left(\EE(D\T z_{h})\T(z_{h})\right)_{h = 1, \cdots p},
    \end{align*}
    and \[
    \EE(D\T z_{h})\T(z_{h}) = \tr(D) \sigma^2.
    \]
    Thus, \[
    \EE Z DZ\T = \sigma^2 tr(D) I. 
    \]
    Further notice $(Z DZ\T Z\T DZ)_{i,h} = \sum_{j = 1}^p z_i \T D z_j z_j\T D z_h$, $\EE  \sum_{j = 1}^p z_i \T D z_j z_j\T D z_h = 0$ if $i\neq h$,  
    \[
    \EE((D\T z_{h})\T(z_{h}))((D\T z_{h})\T(z_{h}))\T = \EE(\sum_{i,j} z_{hi}z_{hj}D_{ij})^2 = \sum_{i}D_{ii}^2\EE z_{hi}^4 + \sum_{i\neq j} D_{ij}^2\EE z_{hi}^2 \EE z_{hj}^2 \leq  \tr(D\T D)\zeta^4, 
    \]and
    \[
    \EE((D\T z_{h})\T(z_{j}))((D\T z_{j})\T(z_{h}))\T = \sigma^2 \EE(z_h\T D^2 z_h)  \leq  \tr(D\T D)\zeta^4. 
    \]Thus, the second statement in the lemma holds. 
\end{proof}

\begin{Corollary}\label{corollary_M3}
    For given $c_1$ and $c_2$, there exists constant $c_3$ such that if 
    $$n \geq c_3 (r_2) (p_1 + p_2) \max\left\{\frac{\theta^2u^2}{\lambda_U^2}, \frac{\theta u}{\lambda_U}\right\},$$ 
    where $\theta = \max\left\{ \sqrt{\frac{p_1}{p_2}}, 1 \right\}$, then with probability at least  $1-e^{-c_2 r_2 (p_1 + p_2)}$, we have
    \[
    M_{3,U} \leq c_1\lambda_U. 
    \]
\end{Corollary}
\begin{proof}
    Let $W_i = \frac{1}{p_2} Z_i E \left( UA_i + B_i V\T \right)\T$, then $\|W_i\| \leq \frac{\|Z_i\|}{\sqrt{p_2}}$ and 
    \[
        \sqrt{\log 2}\cdot \left\| \frac{\|Z_i\|}{\sqrt{p_2}} \right\|_{\psi_1} \leq \left\| \frac{\|Z_i\|}{\sqrt{p_2}} \right\|_{\psi_2}. 
    \]
    Hence, 
    $\mathbb{E} \exp \left(\frac{\sqrt{\log 2}\|W_i\|}{u}\right) \leq \mathbb{E} \exp \left(\frac{\sqrt{\log 2}\|Z_i\|}{u\sqrt{p_2}}\right) \leq 2$. Also, for some absolute constant $C_6>1$, we have
    $$
    \begin{aligned}
        \left\|\EE W_i W_i^{\top}\right\|
        =& \frac{1}{p_2} \left\|\EE \left(\EE \left(\left.  Z_i E \left( UA_i + B_i V\T \right)\T \left( UA_i + B_i V\T \right)  E  Z_i\T \right| A_i, B_i \right)\right)\right\|\\
        =& \frac{1}{p_2} \left\|\EE \left( \sigma^2 \tr\left( E \left( UA_i + B_i V\T \right)\T \left( UA_i + B_i V\T \right) E \right) I \right) \right\|\\
        \leq& \frac{1}{p_2}\EE \left( \sigma^2 \left\| UA_i + B_i V\T \right\|^2 \right) \\
        \leq& 4\sigma^2\\
        \leq& 4C_6^2u^2, 
    \end{aligned}$$and 
    $$
    \begin{aligned}
         \left\|\EE W_i^{\top} W_i \right\|
        =& \frac{1}{p_2} \left\|\EE \left(\EE \left(\left. \left( UA_i + B_i V\T\right) E Z_i\T Z_i E \left(UA_i + B_i V\T \right)\T  \right| \A_{i,k} \right)\right)\right\|\\
        =& \frac{1}{p_2} \left\|\EE \left( \sigma^2 p_1  \left( UA_i + B_i V\T\right) E^2 \left( UA_i + B_i V\T\right)\T \right) \right\|\\
        \leq& \frac{1}{p_2}\EE \left( \sigma^2 p_1 \left\| UA_i + B_i V\T \right\|^2 \right) \\
        \leq& \frac{4 p_1}{p_2}\sigma^2\\
        \leq& C_6^2\frac{4 p_1}{p_2} u^2. 
    \end{aligned}$$
    So in Lemma \ref{lemma_bernstein}, for any given $\varepsilon>0$, constant $C$ and $E \in T_U$ , let $t = C(r_2) (p_1 + p_2)$, $\sigma_Z = 2C_6 u \theta$, $\alpha = 1$ and $U_Z^{(\alpha)} = 2C_6u\theta/\sqrt{\log 2}$, where $\theta = \max\left\{ \sqrt{\frac{p_1}{p_2}}, 1 \right\}$. It yields that, there exists constant $C_2$ such that, with probability at least  $1-e^{-Cr_2 (p_1 + p_2)}$, the following holds: 
    \[\begin{aligned}
        &\frac{1}{np_2} \left\|\sum_{i = 1}^{n} Z_i E \left(UA_i + B_iV\T \right)\T   \right\| \\
        \leq& C_2 2u\theta \max \left\{\sqrt{\frac{r_2 (p_1 + p_2)}{n}}, \frac{r_2 (p_1 + p_2)}{n}\right\},
    \end{aligned}
    \]where $\theta = \max\left\{ \sqrt{\frac{p_1}{p_2}}, 1 \right\}$. Then by union bound from Corollary \ref{corollary_epsilon_net}, it yields that with probability at least  $1-e^{-(C-\log(C_0/\varepsilon))r_2 (p_1 + p_2)}$: 
    \[
    M_{3,U} \leq C_2 2u\theta \max \left\{ \sqrt{\frac{r_2 (p_1 + p_2)}{n}}, \frac{r_2 (p_1 + p_2)}{n}\right\} + \varepsilon.
    \]
    So, for given constants $C_4$ and $C_5$, let $\varepsilon$ be a constant multiplier of $\lambda_U$ and hence, if 
    $$n \geq C_3 r_2 (p_1 + p_2) \max\left\{\frac{\theta^2u^2}{\lambda_U^2}, \frac{\theta u}{\lambda_U}\right\}$$
    for some large enough $C_3$, then with probability at least  $1-e^{-C_4 r_2 (p_1 + p_2)}$, we have\[
    M_{3,U} \leq C_5\lambda_U. 
    \]
\end{proof}

\begin{Corollary}\label{corollary_M4}
    For given $c_1$ and $c_2$, there exists constant $c_3$ such that if $$n \geq c_3 r_2 (p_1 + p_2) \max\left\{\frac{u^4}{\lambda_U^2}, \frac{u^2}{\lambda_U}\right\},$$  then with probability at least  $1-e^{-c_2 r_2 (p_1 + p_2)}$, we have
    \[
    M_{4,U} \leq c_1\lambda_U. 
    \]
\end{Corollary}

\begin{proof}
    Let $W_i = \frac{1}{p_2} Z_i E Z_i\T$, then $\|W_i\| \leq \frac{\|Z_i\|^2}{p_2}$ and hence $\mathbb{E} \exp \left(\|W_i\|/u^2\right) \leq \mathbb{E} \exp \left(\frac{\|Z_i\|^2}{u^2p_2}\right) \leq 2$. Also, for some absolute constant $C_6 > 1$, we have
    
    $$
    \begin{aligned}
        \left\|\EE W_i W_i^{\top}\right\|
        =& \frac{1}{p_2^2} \left\|\EE  \left( Z_i E Z_i\T Z_i E Z_i \T \right)\right\|\\
        \overset{\text{Lemma \ref{lemma_Z}}}{\leq}& \frac{1}{p_2}\zeta^4 tr(E^2) \leq \zeta^4 \|E\|^2\\
        \leq&  \zeta^4 \leq C_6^2 u^4.
    \end{aligned}$$
    So in Lemma \ref{lemma_bernstein}, for any given $\varepsilon>0$, constant $C$ and $E \in T_U$ , let $t = Cr_2 (p_1 + p_2)$, $\sigma_Z = C_6 u^2$, $\alpha = 1$ and $U_Z^{(\alpha)} = 2C_6u^2$, which yields that there exists constant $C_2$ such that with probability at least  $1-e^{-Cr_2 (p_1 + p_2)}$, the following holds: 
    \[\begin{aligned}
        &\frac{1}{np_2} \left\|\sum_{i = 1}^{n} Z_i E Z_i\T   \right\|
        \leq& C_2 u^2 \max\left\{\sqrt{\frac{r_2 (p_1 + p_2)}{n}}, \frac{r_2 (p_1 + p_2)}{n}\right\}.
    \end{aligned}
    \]
    Then by union bound from Corollary \ref{corollary_epsilon_net}, it yields that with probability at least  $$1-e^{-(C-\log(C_0/\varepsilon))r_2 (p_1 + p_2)},$$ we have 
    \[
    M_{4,U} \leq C_2 u^2 \max\left\{\sqrt{\frac{r_2 (p_1 + p_2)}{n}}, \frac{r_2 (p_1 + p_2)}{n}\right\} + \varepsilon. 
    \]
    So for given $d$ and constants $C_4$ and $C_5$, let $\varepsilon$ be a constant multiplier of $\lambda_U$ and hence, if 
    $$n \geq C_3 r_2 (p_1 + p_2) \max\left\{\frac{u^4}{\lambda_U^2}, \frac{u^2}{\lambda_U}\right\}$$ 
    for some large enough $C_3$, then with probability at least  $1-e^{-(C_4)r_2 (p_1 + p_2)}$, we have
    \[
    M_{4,1} \leq C_5\lambda_U. 
    \]
\end{proof}

\begin{Corollary}\label{corollary_kappa1}
    For given $c_1$ and $c_2$, there exists constant $c_3$ such that if $$n \geq c_3 \log p_1 \max\left\{\frac{\theta^2 u^2}{\lambda_U^6}, \frac{\theta u}{\lambda_U^3}\right\},$$where $\theta = \max\left\{ \sqrt{\frac{p_1}{p_2}}, 1 \right\}$, then with probability at least  $1-e^{-c_2\log p_1}$, we have
    \[
    \frac{1}{n p_2} \left\|\sum_{i = 1}^{n} Z_i P_{{V}_{\perp}}  \left(UA_i\right)\T   \right\| \leq c_1\lambda_U^3. 
    \]
\end{Corollary}
\begin{proof}
    Let $W_i = \frac{1}{\prod_{k \neq 1}p_k} Z_i P_{{V}_{\perp}} \left(UA_i\right)\T$, then $\|W_i\| \leq \frac{\|Z_i\|}{\sqrt{p_2}}$ and \[
    \sqrt{\log 2} \cdot \left\| \frac{\|Z_i\|}{\sqrt{p_2}} \right\|_{\psi_1} \leq \left\| \frac{\|Z_i\|}{\sqrt{p_2}} \right\|_{\psi_2}. 
    \]
    Hence, 
    $\mathbb{E} \exp \left(\frac{\|W_i\|\sqrt{\log 2}}{u}\right) \leq \mathbb{E} \exp \left(\frac{\|Z_i\|\sqrt{\log 2}}{u\sqrt{p_2}}\right) \leq 2$. Also, there exists an absolute constant $C_6>1$ such that
    $$
    \begin{aligned}
        & \left\|\EE W_i W_i^{\top}\right\|\\
        =& \frac{1}{p_2^2} \left\|\EE \left(\EE \left(\left.  Z_iP_{V_\perp} \left(UA_i\right)\T \left(UA_i\right) P_{V_\perp} Z_i \T \right| A_i, B_i \right)\right)\right\|\\
        =& \frac{1}{p_2^2} \left\|\EE \left( \sigma^2 \tr\left( P_{V_\perp} \left(UA_i\right)\T \left(UA_i\right) P_{V_\perp} \right) I \right) \right\|\\
        \leq& \frac{1}{p_2}\EE \left( \sigma^2 \left\| \left(UA_i\right) \right\|^2 \right) \leq \sigma^2\leq C_6^2 u^2,
    \end{aligned}$$and 
    $$
    \begin{aligned}
        & \left\|\EE W_i^{\top}W_i \right\|\\
        =& \frac{1}{p_2^2} \left\|\EE \left(\EE \left(\left. \left(UA_i\right) P_{V_\perp}\T Z_i\T Z_iP_{V_\perp} \left(UA_i\right)\T  \right| A_i, B_i \right)\right)\right\|\\
        =& \frac{1}{p_2^2} \left\|\EE \left( \sigma^2 p_1  \left(UA_i\right) P_{V_\perp}^2 \left(UA_i\right)\T \right) \right\|\\
        \leq& \frac{1}{p_2}\EE \left( \sigma^2 p_1 \left\| \left(UA_i\right) \right\|^2 \right) \\
        \leq& \frac{p_1}{p_2}\sigma^2\\
        \leq& C_6^2\frac{p_1}{p_2}u^2.
    \end{aligned}$$
    So in Lemma \ref{lemma_bernstein}, for any given $\varepsilon>0$, constant $C$ and $E \in T_U$, let $t = C\log p_1$, $\sigma_Z = C_6\theta u$, $\alpha = 1$ and $U_Z^{(\alpha)} = C_6 \theta u/\sqrt{\log 2}$, where $\theta = \max \left\{ \sqrt{\frac{p_1}{p_2}}, 1\right\}$. It yields that there exists constant $C_2$ such that with probability at least  $1-e^{-C\log p_1}$, the following holds: 
    \begin{align}
        &\frac{1}{np_2} \left\|\sum_{i = 1}^{n} Z_i P_{V_\perp} \left(UA_i\right)\T   \right\| \notag \\
        \leq& C_2 u\theta \max \left\{\sqrt{\frac{\log p_1}{n}},  \frac{\log p_1}{n}\right\} \label{inequality_kappa_1},
    \end{align}
    where $\theta = \max\left\{ \sqrt{\frac{p_1}{p_2}}, 1 \right\}$.	So, for given constants $C_4$ and $C_5$ if 
    $$n \geq C_3\log p_1 \max\left\{\frac{\theta^2 u^2}{\lambda_U^6}, \frac{\theta u}{\lambda_U^3}\right\}$$ 
    for some large enough $C_3$, then with probability at least  $1-e^{-C_4\log p_1}$, we have
    \[
    \frac{1}{n p_2} \left\|\sum_{i = 1}^{n} Z_i P_{V_\perp} \left(UA_i\right)\T   \right\| \leq C_5\lambda_U^3. 
    \]
\end{proof}
\begin{Corollary}\label{corollary_kappa2}
    For given $c_1$ and $c_2$, there exists constant $c_3$ such that if $$n \geq c_3 \log p_1 \max\left\{\frac{u^4}{\lambda_U^6}, \frac{u^2}{\lambda_U^3}\right\}, $$ then with probability at least  $1-e^{-c_2\log p_1}$, we have
    \[
    \frac{1}{n p_2} \left\|\sum_{i = 1}^{n} Z_i P_{V_\perp} Z_i\T- n\sigma^2\left({p_2 - r_2}\right) I  \right\| \leq c_1\lambda_U^3. 
    \]
\end{Corollary}
\begin{proof}
    Let $W_i = \frac{1}{p_2}\left(Z_i P_{V_\perp} Z_i\T- \sigma^2\left({p_2 - r_2}\right) I \right)$, then $\|W_i\| \leq \max\left\{\frac{\|Z_i\|^2}{p_2}, \sigma^2r\right\}$, where $r = \frac{p_2 - r_2}{p_2}$. 
    Hence, 
    $\mathbb{E} \exp \left(\frac{\|W_i\|}{\tau}\right) \leq \mathbb{E} \exp \left(\max\left\{\frac{\|Z_i\|^2}{\tau p_2}, \frac{\sigma^2r}{\tau}\right\}\right) \leq 2$, 
    where $\tau = \max\left\{u^2, \frac{\sigma^2 r}{\log 2}\right\} \leq u^2 / \log2$. Also, 
    $$
    \begin{aligned}
         \EE W_i W_i^{\top}
        =&\frac{1}{p_2^2}\left(\EE Z_i P_{V_\perp} Z_i \T Z_i P_{V_\perp} Z_i \T\right)\\
        &-\frac{2\sigma^2\left(p_2 - r_2\right)}{p_2^2}\left(\EE Z_i P_{V_\perp} Z_i \T \right) + \frac{\sigma^4\left(p_2 - r_2\right)^2}{p_2^2}I\\
        \overset{\text{Lemma \ref{lemma_Z}}}{\preccurlyeq} & \frac{\zeta^4}{p_2}\tr \left(P_{V_\perp}^2 \right)I
        -\frac{2\sigma^4\left(p_2 - r_2\right)}{p_2^2} \tr\left(P_{V_\perp}\right)I 
        + \frac{\sigma^4\left(p_2 - r_2\right)^2}{p_2^2}I\\
        \preccurlyeq & \frac{p_2 - r_2}{p_2}\zeta^4 I.
    \end{aligned}$$
    Thus, for some absolute constant $C_6> 1$,
    \[
    \left\|\EE W_i W_i^{\top}\right\| \leq \frac{p_2 - r_2}{p_2}\zeta^4 \leq \zeta^4 \leq C_6^2 u^4. 
    \]
    So in Lemma \ref{lemma_bernstein}, for any given $\varepsilon>0$, constant $C$ and $E \in T_U$ , let $t = C\log p_1$, $\sigma_Z =C_6 u^2$, $\alpha = 1$ and $U_Z^{(\alpha)} =C_6 u^2 / \log2$, which yields that there exists constant $C_2$ such that with probability at least  $1-e^{-C\log p_1}$, the following holds: 
    \begin{align}
        \frac{1}{n p_2} \left\|\sum_{i = 1}^{n} Z_i  P_{V_\perp} Z_i \T - n\sigma^2\left({p_2 - r_2}\right) I \right\|
        \leq C_2 \max \left\{u^2 \sqrt{\frac{\log p_1}{n}}, u^2\frac{\log p_1}{n}\right\} \label{inequality_kappa_2}.
    \end{align}
    So, for given $d$ and constants $C_4$ and $C_5$ if 
    $$n \geq C_3\log p_1 \max\left\{\frac{u^4}{\lambda_U^6}, \frac{u^2}{\lambda_U^3}\right\}$$ 
    for some large enough $C_3$, then with probability at least  $1-e^{-C_4\log p_1}$, we have
    \[
    \frac{1}{n p_2} \left\|\sum_{i = 1}^{n} Z_i P_{V_\perp} Z_i\T- n\sigma^2\left({p_2 - r_2}\right) I  \right\| \leq C_5\lambda_U^3. 
    \]
\end{proof}

Corollaries \ref{corollary_L}, \ref{corollary_M1}, \ref{corollary_M3} and \ref{corollary_M4} imply that with proper choice of $c_1$, we have
$$\Delta_U  = L_U - (M_{1,U} + M_{3,U} + M_{4,U})\gtrsim \lambda_U,$$
and 
$$\xi_U \leq 10 + 3(M_{1,U} + M_{3,U} + M_{4,U}) \lesssim 1$$
with high probability. If $d$ is fixed and we set $\varepsilon_0$ as a small constant multiplier of $\lambda_U$, we have 
$$\Gamma_U = \Delta_U - \epsilon_0 \xi_U \gtrsim \lambda_U. $$ 
So, by Corollaries \ref{corollary_kappa1} and \ref{corollary_kappa2}, we have 
$$\frac{4\kappa_U}{L_U - \varepsilon_0 \xi_U - 3\kappa_U} \lesssim \lambda_U^2 \lesssim \frac{\Gamma_U \varepsilon_0}{L_U}$$with high probability. 
Additionally, we can bound the error $K_2$ in Theorem \ref{theorem_convergence_deterministic} by (\ref{inequality_kappa_1}) and (\ref{inequality_kappa_2}). 

The argument above can be summarized as the following lemma.
\begin{Lemma}
    Given constants $c>0$ and assuming Assumption \ref{assumption_z} and \ref{assumption_conditionnumber_star} hold with $\mu_U = 1$ in Assumption \ref{assumption_conditionnumber_star}, then there exists constants $c_1$, $c_4$, $c_5<1$, $c_6$ (does not depend on any variable appeared in the following equations) such that if $\varepsilon_0 \leq c_4 \lambda_U$, \[
    n \geq c_1 \log p_{\max} \max\left\{\frac{u^4}{\lambda_U^6}, \frac{u^2}{\lambda_U^3}, \frac{\theta u^2}{\lambda_U^6}, \frac{\theta u}{\lambda_U^3}\right\},
    \]
    and
    \[
    n \geq c_1 r_{\max} (p_1 + p_2) \max\left\{\frac{1}{\lambda_U^2}, \frac{u^4}{\lambda_U^2}, \frac{u^2}{\lambda_U}, \frac{\theta^2 u^2}{\lambda_U^2}, \frac{\theta u}{\lambda_U}\right\}, 
    \]where $\theta = \max\left\{ \sqrt{\frac{p_2}{p_1}}, \sqrt{\frac{p_1}{p_2}} \right\}$, 
    then with probability at least  $1- e^{-c_2r_{\min} (p_1 + p_2)}$, $\{A_i, B_i, Z_i; i = 1,\cdots n\} \subseteq \mathcal{A}_U$. 
\end{Lemma}

Now let's consider general $\mu_U$ (the previous discussion is based on $\mu_U = 1$). We can let $Y_i = X_i / \mu_U(\{X_i\})$ and then $\mu_U(\{Y_i\} ) = 1$ and $\lambda_U(\{Y_i\}) = \lambda_U(\{X_i\}) / \mu_U(\{X_i\})^2 \geq 1/C$, where $\mu_U(\{X_i\})$ refers the $\mu_U$ of $X_i$ in Assumption \ref{assumption_conditionnumber_star}, $\mu_U(\{Y_i\})$ refers the $\mu_U$ of $Y_i$ and $\lambda_U(\{Y_i\})$ are defined similarly. 
Then we can apply the above lemma and Theorem \ref{theorem_convergence_deterministic} to $Y_i$. And for general $\nu$, we take out the probability that $\mu_U \leq \frac{\max \{\left\| A \right\|, \left\| B \right\|\}}{\sqrt{p_2}}$. Finally, we switch $U$ by $V$ and combine them: 

\begin{Lemma}
    Given constant $c_1>0$, matrices $\{X_i\}$ satisfying decomposition (\ref{eq:decomposition-equivalence-matrix}) and Assumption \ref{assumption_conditionnumber} and \ref{assumption_z}, then when applying Algorithm \ref{algorithm_iterative_projection}, there exists constants $c_2$, $c_3$, $c_4<1$, $c_5$ (do not depend on any variable appeared in the following equations) such that if $\varepsilon_0 \leq c_3$ and $n$ satisfies the following: 
    \[
    n \geq c_2 r_{\max} (p_1 + p_2) \max\left\{ u^4/\mu^4, \theta^2 u^2/\mu^2, \theta u/\mu, 1\right\},
    \]
    then with probability at least  $1-e^{-c_1r_{\min} p_{\max}} - \nu$, the estimation error defined as $$\operatorname{Error}^{(t)} = \max\left\{\|\sin \Theta (U, \hat{U}^{(t)})\|, \|\sin \Theta (V, \hat{V}^{(t)})\|\right\},$$ in Algorithm \ref{algorithm_iterative_projection} converges linearly with rate $\chi \leq c_4$: 
    \[
    \operatorname{Error}^{(t)} - \operatorname{Error} 
    \leq \chi \left(\operatorname{Error}^{(t-1)} - \operatorname{Error}\right) ,
    \]
    and the final error is bounded by 
    $$\operatorname{Error} \leq c_5 \sqrt{\frac{\log p_{\max}}{n}} \max \left\{\theta u/\mu, u^2/\mu^2\right\},$$ 
    where $\theta = \sqrt{\max\left\{1, p_1/p_2, p_2/p_1\right\}}, r_{\max} = \max\{r_1, r_2\}, r_{\min} = \min\{r_1, r_2\}$ and $p_{\max} = \max\{p_1, p_2\}$. 
\end{Lemma}

As all the entries of $Z$ are i.i.d. sub-Gaussian distributed as $z$ with sub-Gaussian norm $\tau$, by Lemma \ref{lemma_subGaussian_norm}, we have $u = \max \left\{ \left\|Z/\sqrt{p_1}\right\|_{\psi_2}, \left\|Z/\sqrt{p_2}\right\|_{\psi_2} \right\} \lesssim \tau \sqrt{p_{\max}/p_{\min}}$. The condition for sample size $n$ can be expressed as
\[
    n \geq c_2 r_{\max} p_{\max} \max\left\{ \frac{p_{\max}^2 \tau^4}{p_{\min}^2 \mu^4}, \frac{p_{\max}^3 \tau^2}{p_{\min}^3 \mu^2}, \frac{p_{\max}^{3/2} \tau}{p_{\min}^{3/2} \mu}, 1\right\},
\]
and the final error bound becomes 
\be\notag
\operatorname{Error} \leq c_5 \sqrt{\frac{\log p_{\max}}{n}} \max \left\{ \frac{p_{\max} \tau}{p_{\min} \mu}, \frac{p_{\max} \tau^2}{p_{\min} \mu^2}\right\}. 
\ee

%%%%%%%%%%%%%%%%
\section{Proof of Theorem \ref{Theorem_perturbation}}\label{section_propostion1_proof}
%%%%%%%%%%%%%%%%

Without loss of generality, assume $A$ is positive semi-definite, as we can add $aI$ to $A$ without changing its eigen-structure. 
let's first consider the scenario that $A$ has the block form
\be\label{form_A}
A= \left[\begin{array}{cc} 
    A_{11} & A_{12}\\
    A_{21} & A_{22}
\end{array}\right], \quad A_{11} = 
\left[\begin{array}{ccc} 
    \lambda_{1}(A_{11}) & & \\
    & \ddots &\\
    &  & \lambda_{r}(A_{11}) \\
\end{array}\right]
\ee
and $W$, $W_\perp$ satisfy\be\label{form_W}
W = \begin{bmatrix}
    I_{r\times r}\\
    0
\end{bmatrix}, \quad 
W_\perp = \begin{bmatrix}
    0\\
    I_{(p-r)\times (p-r)}
\end{bmatrix}. 
\ee
Hence, We also have $A_{11} = W\T A W,$ $A_{12} = W\T A W_{\perp} = A_{21}\T$ and $A_{22} = W_{\perp}\T AW_{\perp}$. Denote the $k$th eigenvector of $A$(corresponding to $\lambda_k(A)$) as $$v^{(k)} = \begin{bmatrix}
    \alpha^{(k)}\\
    \beta^{(k)}
\end{bmatrix}$$where $\alpha^{(k)}, \beta^{(k)}$ are the first $r$ elements and the rest $p-r$ elements of $v^{(k)}$. Noticing $\lambda_k(A) v^{(k)}= A v^{(k)}$, by comparing coefficients of this identity, we have for $1\leq i\leq r, r+1 \leq k\leq p$, \be
\lambda_i(A_{11}) \alpha^{(k)}_i + \left(A_{12} \beta^{(k)}\right)_i = \lambda_k(A) \alpha^{(k)}_i, \quad\text{i.e., }\quad
\alpha^{(k)}_i = \frac{-\left(A_{12} \beta^{(k)}\right)_i}{\lambda_i(A_{11}) - \lambda_k(A)}. \label{equation_alpha}
\ee
By the assumption that $\lambda_{r}(A_{11})= \lambda_{r}(W\T A W)>\lambda_{r+1}(A)$, we have
\[
\left(\alpha^{(k)}_i\right)^2 \leq \frac{\left(A_{12} \beta^{(k)}\right)_i^2}{\left(\lambda_r(A_{11}) - \lambda_{k+1}(A)\right)^2}. 
\] Then, we can bound the Frobenius norm of $\sin \Theta(V, W)$ via 
\begin{align}
    \left\|\begin{bmatrix}\alpha^{(r+1)}, \cdots \alpha^{(p)}\end{bmatrix}\right\|_F^2 
    \leq& \frac{\sum_{1\leq i\leq r, r+1 \leq k\leq p} \left(A_{12} \beta^{(k)}\right)_i^2}{\left(\lambda_r(A_{11}) - \lambda_{k+1}(A)\right)^2}\notag\\ 
    =& \frac{\left\|A_{12}\begin{bmatrix}\beta^{(r+1)}, \cdots \beta^{(p)}\end{bmatrix}\right\|_F^2}{\left(\lambda_r(A_{11}) - \lambda_{k+1}(A)\right)^2}\notag\\
    =& \frac{\sum_{1\leq i\leq r} \left\|A_{12,i} \begin{bmatrix}\beta^{(r+1)}, \cdots \beta^{(p)}\end{bmatrix} \right\|_{l_2}^2}{\left(\lambda_r(A_{11}) - \lambda_{k+1}(A)\right)^2}\notag\\
    \leq &\frac{\sum_{1\leq i\leq r} \left\|\begin{bmatrix}\beta^{(r+1)}, \cdots \beta^{(p)}\end{bmatrix} \right\|^2 \left\|A_{12,i}\T \right\|_{l_2}^2}{\left(\lambda_r(A_{11}) - \lambda_{k+1}(A)\right)^2}\notag \\
    \leq & \frac{\|A_{12}\|^2_F}{\left(\lambda_r(A_{11}) - \lambda_{k+1}(A)\right)^2} \notag,
\end{align}where $A_{12,i}$ is $i$th row of $A_{12}$. Thus, \begin{align*}
    \|\sin \Theta(V, W)\|_{F} 
    = &\|W\T V_\perp\|_{F}\\
    = &\left\|\begin{bmatrix}\alpha^{(r+1)}, \cdots \alpha^{(p)}\end{bmatrix}\right\|_F\\
    \leq & \frac{\|A_{12}\|_F}{\left(\lambda_r(A_{11}) - \lambda_{k+1}(A)\right)}\\ 
    =& \frac{\left\|W\T A W_{\perp}\right\|_F}{\lambda_{r}(W\T A W)-\lambda_{r+1}(A)},
\end{align*}where the first equality holds as a result of Exercise VII.I.11 in \cite{bhatia1997matrix}.

To prove the upper bound of the spectral norm of $\sin \Theta(V, W)$, let $s=\left[s_{r+1}, \cdots, s_{p}\right]\T \in \mathbb{R}^{p-r}$ be any vector with $\|s\|_{l_2}=1$. We have
$$
\begin{aligned}
    & \sum_{k=r+1}^{p} s_{k} \alpha_{i}^{(k)} \\
    \overset{(\ref{equation_alpha})}{=}& \sum_{k=r+1}^{p} \frac{-s_{k}\left(A_{12} \beta^{(k)}\right)_i}{\lambda_i(A_{11}) - \lambda_k(A)}\\
    =&\sum_{k=r+1}^{p} \frac{-s_{k}}{\lambda_i(A_{11})} \frac{1}{1-\lambda_{k}(A) / \lambda_i(A_{11})} \left(A_{12} \beta^{(k)}\right)_i \\
    = & \sum_{k=r+1}^{p} \sum_{l=0}^{\infty} \frac{-s_{k} \lambda_{k}^{ l}(A)}{\lambda_{i}^{l+1}(A_{11})} \left(A_{12} \beta^{(k)}\right)_i\\
    =&\sum_{l=0}^{\infty} \frac{-A_{12,i}}{\lambda_{i}^{l+1}(A_{11})}\left(\sum_{k=r+1}^{p} s_{k} \lambda_{k}^{ l}(A) \beta^{(k)}\right), 
\end{aligned}
$$where $A_{12,i}$ is the $i$th row of $A_{12}$. 

Hence, by the assumption that $\lambda_{r}(W\T A W)>\lambda_{r+1}(A)\geq\lambda_{p}(A)>0$, we have
\begin{align*}
    &\left\|\begin{bmatrix}\alpha^{(r+1)}, \cdots \alpha^{(p)}\end{bmatrix}s\right\|_{l_2}\\
    = & \left\|\sum_{k=r+1}^{p} \alpha^{(k)}s_{k} \right\|_{l_2} \\
    = & \left\|\sum_{k=r+1}^{p} \left[\begin{array}{c}
        \alpha^{(k)}_1 s_{k} \\
        \vdots \\
        \alpha^{(k)}_r s_{k}
    \end{array}\right] \right\|_{l_2} \\
    \leq & \sum_{l=0}^{\infty}\left\|\left[\begin{array}{c}
        {-A_{12,1}}/{\lambda_{1}^{l+1}(A_{11})} \\
        \vdots \\
        {-A_{12,r}}/{\lambda_{r}^{l+1}(A_{11})}
    \end{array}\right] \cdot\left(\sum_{k=r+1}^{p} s_{k} \lambda_{k}^{ l}(A) \beta^{(k)}\right)\right\|_{l_2} \\
    \leq & \sum_{l=0}^{\infty}\left\|\left[\begin{array}{c}
        {-A_{12,1}}/{\lambda_{1}^{l+1}(A_{11})} \\
        \vdots \\
        {-A_{12,r}}/{\lambda_{r}^{l+1}(A_{11})}
    \end{array}\right] \right\| \cdot \left\|\left(\sum_{k=r+1}^{p} s_{k} \lambda_{k}^{ l}(A) \beta^{(k)}\right)\right\|_{l_2} \\
    \leq & \sum_{l=0}^{\infty} \|-A_{12}\| \cdot 
    \left\|\left[\begin{array}{ccc}
        \frac{1}{\lambda_{1}^{l+1}(A_{11})} & & \\
        & \ddots &\\
        &  & \frac{1}{\lambda_{r}^{l+1}(A_{11})}
    \end{array}\right] \right\| \cdot 
    \left\|\left[\beta^{(r+1)} \cdots \beta^{\left(p\right)}\right] 
    \left[\begin{array}{c}
        s_{r+1}\lambda^l_{r+1}(A)\\
        \vdots \\
        s_{p}\lambda^l_{p}(A)
    \end{array}\right]\right\|_{l_2} \\
    \leq & \sum_{l=0}^{\infty} \frac{\left\|A_{12}\right\|}{\lambda_{r}^{l+1}(A_{11})} \cdot\left\|\left[\beta^{(r+1)} \cdots \beta^{\left(p\right)}\right]\right\|\cdot
    \left\|\left(s_{r+1}\lambda^l_{r+1}(A), \cdots s_{p}\lambda^l_{p}(A)\right)\T\right\|_{l_2}\\
    \leq & \sum_{l=0}^{\infty} \frac{\lambda^l_{r+1}(A)\left\|A_{12}\right\|}{\lambda_{r}^{l+1}(A_{11})} \cdot\
    \left\|s\right\|_{l_2}\\
    = & \frac{\left\|A_{12}\right\|}{\lambda_{r}(A_{11})-\lambda_{r+1}(A)},
\end{align*}
Then, 
\begin{align*}
    \|\sin \Theta(V, W)\|
    = &\|W\T V_\perp\|\\
    = &\left\|\begin{bmatrix}\alpha^{(r+1)}, \cdots \alpha^{(p)}\end{bmatrix}\right\|\\
    \leq & \frac{\|A_{12}\|}{\lambda_r(A_{11}) - \lambda_{k+1}(A)}\\ 
    =& \frac{\left\|W\T A W_{\perp}\right\|}{\lambda_{r}(W\T A W)-\lambda_{r+1}(A)}\\
    =& \frac{\left\|P_{W\T A W}W\T A W_{\perp}\right\|}{\lambda_{r}(W\T A W)-\lambda_{r+1}(A)}.
\end{align*}

Now we have proved Theorem \ref{Theorem_perturbation} for $A$ and $W$ satisfying (\ref{form_A}) and (\ref{form_W}). Then, for any general symmetric $A$ and $W$ satisfying (\ref{form_W}), let\[
B = \begin{bmatrix}
    \bar U &0 \\
    0&I 
\end{bmatrix} 
\begin{bmatrix}
    W\T\\
    W_\perp\T
\end{bmatrix}
A 
\begin{bmatrix}
    W, W_\perp
\end{bmatrix}
\begin{bmatrix}
    \bar U\T &0 \\
    0&I 
\end{bmatrix} 
=\begin{bmatrix}
    \bar \Sigma & \bar{U} \T W\T A W_\perp\\
    W_\perp\T A  W\bar{U}  &W_\perp \T A W_\perp
\end{bmatrix} ,
\]where $\bar{U} \bar{\Sigma} \bar{U}\T$ is the spectral decomposition of $W\T A W$. 

Then, since $\lambda_{r}(W\T B W) = \lambda_{r}(W\T A W)>\lambda_{r+1}(A) = \lambda_{r+1}(B)$, we have \[
\begin{aligned}
    \|\sin \Theta(V_B, W)\|_{F} 
    \leq& \frac{\left\|W\T B W_{\perp}\right\|_F}{\lambda_{r}(W\T B W)-\lambda_{r+1}(B)} \\
    =& \frac{\left\|\bar{U} \T W\T A W_\perp\right\|_F}{\lambda_{r}(\bar{\Sigma})-\lambda_{r+1}(A)}\\
    =& \frac{\left\|P_{W\T A W}W\T A W_\perp\right\|_F}{\lambda_{r}(W\T A W)-\lambda_{r+1}(A)}. 
\end{aligned}
\]
Also notice by comparing spectral decomposition of $B$ and $A$: \[
\begin{bmatrix}
    V_B, V_{B\perp} 
\end{bmatrix}\Sigma
\begin{bmatrix}
    V_B\T\\ V_{B\perp} \T
\end{bmatrix} = 
\begin{bmatrix}
    U_1\\ U_2
\end{bmatrix}
\begin{bmatrix}
    V_A, V_{A\perp} 
\end{bmatrix}\Sigma
\begin{bmatrix}
    V_A\T\\ V_{A\perp} \T
\end{bmatrix}
\begin{bmatrix}
    U_1\T, U_2\T
\end{bmatrix} , 
\]where $U_1 = [\bar{U}, 0]$ and $U_2 = [0, I]$. 

Hence \[
V_B = \begin{bmatrix}
    U_1 V_A\\ U_2 V_A
\end{bmatrix}
\]and
\[
V_B\T W_\perp = \begin{bmatrix}
    V_A\T U_1\T, V_A\T U_2\T
\end{bmatrix}
\begin{bmatrix}
    0\\ I
\end{bmatrix}
=V_A\T U_2\T = V_A\T W_\perp.
\]
Thus, 
\[
\|\sin \Theta(V_B, W)\|_{F} = \|V_B\T W_\perp\|_{F} =
\|\sin \Theta(V_A, W)\|_{F},
\]i.e., we have proved Theorem \ref{Theorem_perturbation} for general $A$. Finally, for general $W \in \mathbb{O}_{p,r}$, let $\widetilde{W} = [W, W_\perp]$ and notice that \[
\|\sin \Theta(V_A, W)\|_{F} 
= \|\sin \Theta(\widetilde{W}\T V_A, 
\widetilde{W}\T W)\|_{F} 
= \left\|\sin \Theta\left(\widetilde{W}\T V_A, 
\begin{bmatrix}
    I\\ 0
\end{bmatrix}\right)\right\|_{F} . 
\]
Let 
\[
B = \widetilde{W}\T A\widetilde{W} 
= \begin{bmatrix}
    W\T A W & W\T A W_\perp\\ 
    W_\perp A W\T & W_\perp\T  A W_\perp
\end{bmatrix}
= \widetilde{W}\T\begin{bmatrix}
    V_A, V_{A\perp} 
\end{bmatrix}\Sigma
\begin{bmatrix}
    V_A\T\\ V_{A\perp} \T
\end{bmatrix}\widetilde{W}, 
\]then $V_B = \widetilde{W}\T V_A$ and $\lambda_{r}(B_{11}) = \lambda_{r}(W\T A W)>\lambda_{r+1}(A) = \lambda_{r+1}(B)$. By previous result, \[
\|\sin \Theta(V_A, W)\|_{F} = \left\|\sin \Theta\left(V_B, 
\begin{bmatrix}
    I\\ 0
\end{bmatrix}\right)\right\|_{F} \leq \frac{\left\|P_{W\T A W}W\T A W_{\perp}\right\|_F}{\lambda_{r}(W\T A W)-\lambda_{r+1}(A)}. 
\]
Similarly, we can also generalize the upper bound of spectral norm for symmetric $A$ and $W \in \mathbb{O}_{p,r}$.

%%%%%%%%%%%%%%%%
\subsection{Proof of Corollary \ref{Corollary_perturbation_bound}}
%%%%%%%%%%%%%%%%

First, to prove \[
\|\sin \Theta(U, \hat{U})\| \leq \frac{\left\|{P}_{U} Z {P}_{U_{\perp}}\right\|}{\lambda_{r} (P_U \hat{X} P_U)-\lambda_{r+1}(\hat{X})}, 
\]let $Y = X + aI$ for sufficient large $a$, such that both $\hat{Y} = X +aI +Z$ is positive semi-definite. Notice the structure of eigenspace and eigen-gap do not change, i.e., \[
{Y} = \left[\begin{array}{ll}
    {U} & {U}_{\perp}
\end{array}\right] \cdot\left(\left[\begin{array}{cc}
    {\Sigma}_{1} & 0 \\
    0 & {\Sigma}_{2}
\end{array}\right]+aI\right) \cdot\left[\begin{array}{c}
    {U}^{\top} \\
    {U}_{\perp}^{\top}
\end{array}\right]
\]is the eigen-decomposition of ${Y}$, \[
\hat{Y} = \left[\begin{array}{ll}
    \hat{U} & \hat{U}_{\perp}
\end{array}\right] \cdot\left(\left[\begin{array}{cc}
    \hat{\Sigma}_{1} & 0 \\
    0 & \hat{\Sigma}_{2}
\end{array}\right]+aI\right) \cdot\left[\begin{array}{c}
    \hat{U}^{\top} \\
    \hat{U}_{\perp}^{\top}
\end{array}\right]
\]is the eigen-decomposition of $\hat{Y}$ and $\lambda_{r} (P_U \hat{Y} P_U) - \lambda_{r+1} (\hat{Y}) = \lambda_{r} (P_U \hat{X} P_U) - \lambda_{r+1} (\hat{X})$. So, we can apply Theorem \ref{Theorem_perturbation} on $\hat{Y}$, which yields \[
\|\sin \Theta(U, \hat{U})\| \leq \frac{\left\|{P}_{U} Z {P}_{U_{\perp}}\right\|}{\lambda_{r} (P_U \hat{Y} P_U)-\lambda_{r+1}(\hat{Y})} = \frac{\left\|{P}_{U} Z {P}_{U_{\perp}}\right\|}{\lambda_{r} (P_U \hat{X} P_U)-\lambda_{r+1}(\hat{X})}.		
\]

Since $\hat{X} = \hat{X}P_U + \hat{X}P_{U_\perp}$ and Lemma 2 in \cite{cai2018rate}, we have $\lambda_{r+1} (\hat{X}) \leq \sigma_{r+1} (\hat{X})\leq \sigma_1 (P_{U_\perp}\hat{X}) \leq \sigma_1 (P_{U_\perp}\hat{X}P_{U})+\sigma_1 (P_{U_\perp}\hat{X}P_{U_\perp}) = \|{P}_{U_{\perp}}\hat{X}{P}_{U_{\perp}}\|+\|{P}_{U} Z {P}_{U_{\perp}}\|$. Thus, the inequality 
\[
\|\sin \Theta(U, \hat{U})\| \leq \frac{\left\|{P}_{U} Z {P}_{U_{\perp}}\right\|}{\lambda_{r} (P_U \hat{X} P_U)-\|{P}_{U_{\perp}}\hat{X}{P}_{U_{\perp}}\|-\|{P}_{U} Z {P}_{U_{\perp}}\|} 
\]
follows.

%%%%%%%%%%%%%%%%%%%%%%%%
\section{Additional Theory for Order-$d$ \texttt{MOP-UP}}\label{section_general_tensor_theorems}
%%%%%%%%%%%%%%%%%%%%%%%%

The following theorem presents a mild identifiability condition for the order-$d$ spiked covariance model, expanding upon the findings of Theorem \ref{theorem_identifiability}.
\begin{Theorem}[Identifiability of order-$d$ spiked covariance model]\label{theorem_identifiability_tensor}
    Suppose $\Y = \sum_{k=1}^d \A_{k} \times_k U_k$, where $U_k\in \mathbb{O}_{p_r, r_k}$ is a deterministic matrix and $\A_k\in \mathbb{R}^{p_1\times \cdots \times p_{k-1}\times r_k\times p_{k+1}\times \cdots \times p_d}$ is a random tensor, $k=1, \cdots,d$. If for any $k = 1, \cdots, d$, any $v\in \mathbb{R}^{p_1\cdots p_{k-1} r_k p_{k+1}\cdots p_d}$ and any affine subspace $\mathcal{W} \subseteq \RR^{p_k}$, either $\PP\left(U_k\MM_k(\A_{k}) v \in \mathcal{W}| \A_h, h \neq k\right) = 0$ or $\Span(U_k)\in\mathcal{W}$. 
    Then, $U_k$ are identifiable in the sense that for any fixed $V_k \in \mathbb{O}_{p_k, r_k}, k = 1, \cdots d$, if $\|\sin \Theta (U_k, V_k)\| \neq 0$ for some $k$, then $\SSigma \times_{k = 1}^d P_{V_{k\perp}} \neq 0$, where $\SSigma$ is the covariance tensor of $\Y$. 
\end{Theorem}

The following proposition shows the \texttt{AP} algorithm (Algorithm \ref{algorithm_iterative_projection_tensor}) is essentially performing alternating minimization.
\begin{Proposition}[Generalization of Proposition \ref{proposition_minimizer}]\label{proposition_minimizer_tensor}
For any given $k \in \{1, \cdots d\}$, tensors $\X_i, i = 1, \cdots n$ and $V_h \in \mathbb{O}_{p_h, r_h}, h \neq k$,
we have
\[\begin{aligned}
        &\argmin_{V_k \in \mathbb{O}_{p_k, r_k}} \sum_{i = 1}^{n} \left\|(\X_i-\bar{\X})\times_{k=1}^d V_{k\perp}^\top\right\|_F^2\\
        =& \left\{\Eigen_{r_k}\left( \sum_{i = 1}^{n}\MM_k \left(  (\X_i-\bar{\X}) \times_{h \neq k} {V}_{h\perp}\T  \right) \MM_k \left(  (\X_i-\bar{\X}) \times_{h \neq k} {V}_{h\perp} \T  \right) \T \right) O: \forall O\in \mathbb{O}_{r_k}\right\}.
\end{aligned}\]
\end{Proposition}

Similarly, we can establish the linear convergence and statistical error properties for the \texttt{AP} algorithm (Algorithm \ref{algorithm_iterative_projection_tensor}).
\begin{Assumption}\label{assumption_conditionnumber_tensor}
Assume in decomposition (\ref{eq:decomposition-equivalence}), $\A_{k}$'s are independent and there is a constant $\mu$ such that
\[\PP \left\{\mu \leq \frac{\max_{k} \left\| \MM_h \left(\A_{k}\right) \right\|}{\sqrt{\prod_{k \neq h} p_k}}\right\}\leq \nu, \quad \text{for some small $\nu < 1$. }\]
Denote $\lambda$ as 
$$\lambda =\min_h \left(\lambda_{\min }\left(\frac{1}{\prod_{k \neq h} p_k}\mathbb{E} \MM_h \left(  \A_{h}\right) \left(\bigotimes_{j \neq h} P_{{U}_{j\perp}} \right)\MM_h \left(  \A_{h}\right) \T \right)\right).$$
We have $\frac{\mu^2}{{\lambda}}\leq C$ for some constant $C>0$.
\end{Assumption}
Here, $\mu^2/{\lambda}$ can be interpreted as a conditional number reflecting balance among singular values of $\A_{k}$. So Assumption \ref{assumption_conditionnumber_tensor} essentially means the condition number of the score matrices $\A_k$ is bounded. 

\begin{Assumption}\label{assumption_z_tensor}
$\Z$ has i.i.d. sub-Gaussian entries with sub-Gaussian norm $\tau$ and mean 0. 
\end{Assumption}

Define $u = \max_h\left\|\frac{\MM_h(\Z)}{\sqrt{\prod_{k\neq h}p_k}}\right\|_{\psi_2}$. We can now present the following theoretical guarantee for \texttt{AP}, which can be viewed as a generalization of Theorem \ref{theorem_convergence}.
\begin{Theorem}[Local Convergence and Statistical Error Bound]\label{theorem_convergence_tensor}
Let tensors $\{\X_i\}$ satisfy the decomposition (\ref{eq:decomposition-equivalence}).
Suppose the output of Algorithm \ref{algorithm_iterative_projection_tensor} is $\{\hat{U}_k^{(t)}\}_{k = 1}^{d}$ and define the error at $t$th iteration as
$$\operatorname{Error}^{(t)} = \max_{k \in \{1, \cdots d\}}\left\{\|\sin \Theta (U_k, \hat{U}_k^{(t)})\|\right\}. $$
Assume Assumptions \ref{assumption_conditionnumber_tensor} and \ref{assumption_z_tensor} hold and $d\geq 2$. For any given $c_1>0$, there exist constants $c_2$, $c_3$, $c_4<1$, $c_5$ (all independent of any variable in the following inequalities) such that if initialization error $\operatorname{Error}^{(0)} \leq c_3$ and $n$ satisfies:
\[
    n \geq c_2 r_{m} (p_m + p_{-m}) \max\left\{\frac{ u^4}{\mu^4}, \frac{\theta^2 u^2}{\mu^2}, \frac{\theta u}{\mu}, 1\right\},
\]
then we have that with probability at least  $1-e^{-c_1r_{-m} (p_{m} + p_{-m})} - \nu$, the estimation error in Algorithm \ref{algorithm_iterative_projection_tensor} converges linearly with rate $c_4$: 
\[
    \operatorname{Error}^{(t)} - \operatorname{Error} \leq c_4 \left(\operatorname{Error}^{(t-1)} - \operatorname{Error}\right),
\]
where $\operatorname{Error}$ is bounded by 
\begin{equation}\label{ineq:final-error}
\operatorname{Error} \leq c_5 \sqrt{\frac{\log p_{\max}}{n}} \max \left\{\theta {\frac{u}{\mu}}, {\frac{u^2}{\mu^2}}\right\}.
\end{equation}
Here, $r_{-h} = \prod_{k \neq h} r_k$, $p_{-h} = \prod_{k \neq h} p_k$, $\theta = \max\left\{ 1, \sqrt{\frac{p_h}{\prod_{k \neq h}p_k}}; h = 1, \cdots, d\right\}$, $m = \argmax_k\{r_{-k} (p_{k} + p_{-k})\}$ and $p_{\max} = \max_k\{p_k\}$. 
\end{Theorem}
\begin{Remark}
In the context of Theorem \ref{theorem_convergence_tensor}, $\theta$ represents the balance of orders of $\X$ -- as long as there is no one dimension greater than the product of the others, then $\theta = 1$. 

Now, assume the orders of $\X$ are balanced such that $\theta \leq C$ for some constant $C$. If all the entries of $\Z$ are i.i.d. sub-Gaussian distributed as $z$ is with sub-Gaussian norm $\tau$, then by Lemma \ref{lemma_subGaussian_norm}, we can conclude that $u = \left\|\frac{\MM_h(\Z)}{\sqrt{\prod_{k\neq h}p_k}}\right\|_{\psi_2} \lesssim \tau \left(1+\sqrt{\frac{p_h}{\prod_{k\neq h}p_k}}\right) \lesssim \tau$. The condition for sample size $n$ can be simplified to
\[
n \geq c_2 r_{-m} (p_{-m} + p_m) \max\left\{\frac{ \tau^4}{\mu^4}, 1\right\}, 
\]
and the error bound \eqref{ineq:final-error} is simplified to 
\be
\label{equation_error_bound_tensor}
\operatorname{Error} \leq c_5 \sqrt{\frac{\log p_{\max}}{n}} \max \left\{ {\frac{\tau}{\mu}}, {\frac{\tau^2}{\mu^2}}\right\}. 
\ee
\end{Remark}

%%%%%%%%%%%%%%%%%%%%%%%%
\subsection{Proof of Theorem \ref{theorem_equivalence_tensor}}
%%%%%%%%%%%%%%%%%%%%%%%%
\begin{proof}
    Without loss of generality, assume $\mathbb{E} \mathbf{X} = 0$ and $\sigma = 1$. And for tensors $\A$ and $\B$, $\A \otimes \B$ refers to their tensor product in this section. 
	
	Assume the decomposition equation\[
	\X = \mathbb{E}\X + \sum_{k=1}^d \A_k \times_k U_k + \Z 
	\] holds. Notice the facts: 
	\begin{enumerate}
		\item Consider a random tensor $\mathbf{X}$ as a multilinear transformation. Then for fixed $v$, 
		$$(\mathbb{E} \mathbf{X}) (v) = \mathbb{E} (\mathbf{X} v).$$
		\item Consider a random tensor $\mathbf{A}$ as a multilinear transformation: 
		$$
		\mathbb{A}: \RR^{\mathbf{p}} \longrightarrow \mathbb{R}.
		$$
		Then $\mathbf{A} \times_k U_k$ is the multilinear transformation after changing the base of $\mathbb{R}^{p_k}$ by linear transformation $U_k$ and hence, the order of tensor product and n-mode product with a matrix is changeable. And, $$\mathbb{E} (\mathbf{A}\times_k U_k) = (\mathbb{E} \mathbf{A})\times_k U_k. $$
	\end{enumerate}
	Hence, the covariance tensor $$\begin{aligned}
		\boldsymbol\Sigma 
		=& \mathbb{E} (\mathbf{X} \otimes \mathbf{X})\\
		=& \mathbb{E} \left(\sum_{k=1}^{d} \mathbf{A}_{k} \times_{k} U_{k}+\mathbf{Z}\right) \otimes \left(\sum_{k=1}^{d} \mathbf{A}_{k} \times_{k} U_{k}+\mathbf{Z}\right)\\
		=& \mathbb{E} \left(\sum_{k=1}^{d} \mathbf{A}_{k} \times_{k} U_{k}\right) \otimes \left(\sum_{k=1}^{d} \mathbf{A}_{k} \times_{k} U_{k}\right) + \mathbf{I_{\bp}}.
	\end{aligned}$$
	Denote $\boldsymbol{\Sigma}_{0} = \mathbb{E} \left(\sum_{k=1}^{d} \mathbf{A}_{k} \times_{k} U_{k}\right) \otimes \left(\sum_{k=1}^{d} \mathbf{A}_{k} \times_{k} U_{k}\right)$. Thus, 
	$$
	\begin{aligned}
		&\boldsymbol{\Sigma}_{0} \times_1 U_{1 \perp}\T \times_2 \cdots \times_d U_{d \perp}\T\\
		= &\mathbb{E} \left(\sum_{i,j=1}^{d} (\mathbf{A}_{i} \times_{i} U_{i}) \otimes (\mathbf{A}_{j} \times_{j} U_{j})\times_1 U_{1 \perp}\T \cdots \times_d U_{d \perp}\T\right) \\
		=& \mathbb{E} \left(\sum_{i,j=1}^{d} \left( (\mathbf{A}_{i} \times_{i} U_{i} \times_{i} U_{i \perp}\T) \otimes (\mathbf{A}_{j} \times_{j} U_{j})\times_1 U_{1 \perp}\T \cdots \times_{i-1} U_{i-1 \perp}\T \times_{i+1} U_{i+1 \perp}\T\cdots \times_d U_{d \perp}\T\right)\right) \\
		=& 0, 
	\end{aligned}$$which proved the sufficiency. 
 
    To prove the necessity, assume $\mathbf{X}$ has spiked covariance. Define $$
	\mathbf{A}_k = (\mathbf{X} - \mathbf{Z}) \times_1 (U_{1 \perp} U_{1 \perp}\T) \cdots \times_{k-1} (U_{k-1 \perp} U_{k-1 \perp}\T) \times_k U_k\T,
	$$where $\mathbf{Z}$ is some random tensor with $\mathbb{E}\mathbf{Z} = 0,\mathbb{E}(\mathbf{Z}\otimes\mathbf{Z}) = \mathbb{E}(\mathbf{X}\otimes\mathbf{Z}) = \mathbf{I}$. So $\mathbb{E}(\mathbf{A}_k \otimes \mathbf{Z}) = 0$. Here, to see the existence of such $\Z$, without loss of generality, assume $\mathbf{X} = x$ is a random vector. Then, the existence of $z$ is guaranteed by lemma \ref{Lemma_existence}.
	
	Now denote $\mathbf{Y} = \mathbf{X}- \mathbf{Z}$. Notice $\mathbb{R}^{p_i} = \Span(U_i) + \Span(U_{i \perp})$ implies that for any $v_i \in \mathbb{R}^{p_i}$ we have decomposition $v_i = u_i + u_i^\perp$ with $u_i \in \Span(U_i), u_i^\perp \in \Span(U_{i \perp})$. Thus,
	\begin{align*}
		&\left(\mathbf{Y} - \sum_{k=1}^{d} \mathbf{A}_{k} \times_{k} U_{k}\right)(v_1, \cdots v_d)\\
		=&\left(\mathbf{Y} - \sum_{k=1}^{d} \mathbf{A}_{k} \times_{k} U_{k}\right)(u_1, v_2,\cdots v_d)+\left(\mathbf{Y} - \sum_{k=1}^{d} \mathbf{A}_{k} \times_{k} U_{k}\right)(u_1^\perp, v_2,\cdots v_d)\\
		=&\left(\mathbf{Y} - \Y \times_1 P_{U_1} - \sum_{k=2}^{d} \mathbf{A}_{k} \times_{k} U_{k}\right)(u_1, v_2,\cdots v_d)+\left(\mathbf{Y} - \sum_{k=2}^{d} \mathbf{A}_{k} \times_{k} U_{k}\right)(u_1^\perp, v_2,\cdots v_d)\\
		=&\left(\mathbf{Y} - \sum_{k=2}^{d} \mathbf{A}_{k} \times_{k} U_{k}\right)(u_1^\perp, v_2,\cdots v_d)\\
		=& \cdots\\
		=&\mathbf{Y}(u_1^\perp, \cdots u_d^\perp). 
	\end{align*}
	Notice $\boldsymbol{\Sigma}_{0} \times_1 U_{1 \perp} \times_2 \cdots \times_d U_{d \perp}=0$ implies that for any $u_i^\perp \in \Span(U_{i \perp}), v_j \in \mathbb{R}^{p_i},$ we have $\boldsymbol{\Sigma}_{0} (u_1^\perp, \cdots u_d^\perp,v_{d+1}, \cdots v_{2d}) = 0$. Thus, 
	\begin{align*}
		&\left(\mathbb{E} \left(\mathbf{Y} - \sum_{k=1}^{d} \mathbf{A}_{k} \times_{k} U_{k}\right)\otimes \left(\mathbf{Y} - \sum_{k=1}^{d} \mathbf{A}_{k} \times_{k} U_{k}\right)\right)(v_1, \cdots v_{2d})\\
		=& \mathbb{E} \left( \left(\mathbf{Y} - \sum_{k=1}^{d} \mathbf{A}_{k} \times_{k} U_{k}\right)(v_1, \cdots v_{d})\right) \left( \left(\mathbf{Y} - \sum_{k=1}^{d} \mathbf{A}_{k} \times_{k} U_{k}\right)(v_{d+1}, \cdots v_{2d})\right)\\
		=& \mathbb{E} \left( \mathbf{Y}(u_1^\perp, \cdots u_d^\perp)\right) \left( \mathbf{Y} (u_{d+1}^\perp, \cdots u_{2d}^\perp)\right)\\
		=& \boldsymbol{\Sigma}_{0} (u_1^\perp, \cdots u_{2d}^\perp) \\
		=& 0,
	\end{align*} 
	i.e., the covariance tensor is 0, and hence, the decomposition equation holds a.s. and the Theorem is proved. 
\end{proof}

%%%%%%%%%%%%%%%%%%%%%%%%
\subsection{Proof of Theorem \ref{theorem_identifiability_tensor}}
%%%%%%%%%%%%%%%%%%%%%%%%
\begin{proof}
    Assume that for some $V_k$ we have $\Y\times_{k = 1}^d P_{V_{k\perp}} = 0.$ Notice that $\|\sin \Theta (U_k, V_k)\| = 0$ is equivalent to $P_{V_{k\perp}} U_k = 0$. Without loss of generality, assume $P_{V_{1\perp}} U_1 \neq 0$. We have 
    \begin{align*}
        0=&\MM_1\left(\Y\times_{h = 1}^d P_{V_{h\perp}}\right) \\
        =& \sum_{k=1}^d \MM_1\left(\A_{k} \times_k U_k \times_{h = 1}^d P_{V_{h\perp}}\right)\\
        =& P_{V_{1\perp}} U_1 \MM_1\left(\A_{1}\right) Q + P_{V_{1\perp}} \sum_{k=2}^d \MM_1\left(\A_{k}\right) Q_k
    \end{align*}
    where $Q = \bigotimes_{j = d}^2 P_{V_{j\perp}}$ and $Q_k = Q \bigotimes_{j = d}^{k+1} I_{p_j} \otimes U_k\T \bigotimes_{j = k-1}^{2} I_{p_j}$. Intuitively, to make the last line 0, we need its two terms to cancel out with each other. However, by the condition in the theorem, the probability for $P_{V_{1\perp}} U_1 \MM_1\left(\A_{1}\right) Q$ to cancel out with $P_{V_{1\perp}} \sum_{k=2}^d \MM_1\left(\A_{k}\right) Q_k$ for any given $\A_{k}, k\geq 2$ is 0. 

    To make the statement rigorous, notice that it follows$\Span\left(U_1 \MM_1\left(\A_{1}\right) Q + \sum_{k=2}^d \MM_1\left(\A_{k}\right) Q_k\right) \subseteq \ker(P_{V_{1\perp}})$, which implies for any $v$ such that $Qv\neq 0$, we have 
    \[
        U_1 \MM_1\left(\A_{1}\right) Q v + \sum_{k=2}^d \MM_1\left(\A_{k}\right) Q_k v \in \ker(P_{V_{1\perp}}),
    \]and hence
    \[
        U_1 \MM_1\left(\A_{1}\right) Q v \in \mathcal{A}, 
    \]where $\mathcal{A}$ represents the affine space $\{w = u - \sum_{k=2}^d \MM_1\left(\A_{k}\right) Q_k v: \forall u \in \ker(P_{V_{1\perp}})\}$.

    For given $\A_k, k \geq 2, $ if $\sum_{k=2}^d \MM_1\left(\A_{k}\right) Q_k v = 0$, we have $\mathcal{A} = \ker(P_{V_{1\perp}}) = \Span(V_1) \neq \Span(U_1)$, and thus $\Span(U_1) \not \subseteq \mathcal{A}$. 
    
    If $\sum_{k=2}^d \MM_1\left(\A_{k}\right) Q_k v \neq 0$, then $\mathcal{A}$ is a shifted $r_1$ dimensional space. 
    If the shift direction is in the subspace, i.e., $\sum_{k=2}^d \MM_1\left(\A_{k}\right) Q_k v \in \ker(P_{V_{1\perp}})$, then $\mathcal{A} = \ker(P_{V_{1\perp}})$, and hence $\Span(U_1) \not \subseteq \mathcal{A}$. 
    If the shift direction is not in the subspace, i.e., $\sum_{k=2}^d \MM_1\left(\A_{k}\right) Q_k v \notin \ker(P_{V_{1\perp}})$, then $\forall u$ with small enough $\|u\|$, we have $u\notin \mathcal{A}$. Thus, $\Span(U_1) \not \subseteq \mathcal{A}$.
    
    By above discuss, we always have $\Span(U_1) \not \subseteq \mathcal{A}$. 
    %On the other hand, if $U_1^{(-1)}(\mathcal{A}): = \{v:U_1 v \in \mathcal{A}\} = \RR^{P_1}$, then we must have $\Span(U_1) \subseteq \mathcal{A}$. Thus, $U_1^{(-1)}(\mathcal{A}) \neq \RR^{P_1}$. 
    Hence by the condition in the theorem, we have $\PP(U_1\MM_1(\A_1) Qv \in \mathcal{A}|\A_k, k \geq 2) = 0$, which concludes that 
    \[
        \PP\left(\Y\times_{k = 1}^d P_{V_{k\perp}} = 0\right) \leq 
        \EE \left( \PP \left(\left. U_1\MM_1(\A_1) Qv \in \mathcal{A}\right| \A_k, k \geq 2 \right)\right)  = 0. 
    \]
    Thus, by Theorem \ref{theorem_equivalence_tensor}, if the covariance tensor $\SSigma$ of $\Y$ satisfies $\SSigma \times_{k = 1}^d V_{k\perp} = 0$, then there exists $\{\B_i\}$ such that $\Y = \sum_{i = 1}^d \B_i \times_i P_{V_{k\perp}}$, and hence, $\PP\left(\Y\times_{k = 1}^d P_{V_{k\perp}} = 0\right) = 1$. Thus, $\SSigma \times_{k = 1}^d V_{k\perp} \neq 0$.
\end{proof}

%%%%%%%%%%%%%%%%%%%%%%%%
\subsection{Proof of Proposition \ref{proposition_minimizer_tensor}}
%%%%%%%%%%%%%%%%%%%%%%%%
\begin{proof}
    Without loss of generality, assume $\bar\X = 0$. Notice 
        \[
	\begin{aligned}
		\sum_{i = 1}^{n} \left\|\X_i\times_{k=1}^d U_{k\perp}^\top\right\|_F^2
            =& \sum_{i = 1}^{n} \left\| \MM_k \left(\X_i\times_{k=1}^d U_{k\perp}^\top\right)\right\|_F^2\\
            =& \sum_{i = 1}^{n} \tr\left(P_{U_{k\perp}}\MM_k \left(\X_i \times_{h \neq k} {U}_{h\perp}\T \right) \MM_k \left(\X_i \times_{h \neq k} {U}_{h\perp}\T \right)\T P_{U_{k\perp}}\right)\\
		=&\tr\left(P_{U_{k\perp}}\sum_{i = 1}^{n} \MM_k \left(\X_i \times_{h \neq k} {U}_{h\perp}\T \right) \MM_k \left(\X_i \times_{h \neq k} {U}_{h\perp}\T \right)\T P_{U_{k\perp}}\right)\\
            =& \tr\left(P_{U_{k\perp}} \sum_{i = 1}^{p_k}\lambda_ie_ie_i\T\right) =\sum_{i = 1}^{p_k} \lambda_i\tr(P_{U_{k\perp}} e_ie_i\T)\\
		=&\sum_{i = 1}^{p_k} \lambda_i e_i\T P_{U_{k\perp}} e_i\geq \sum_{i>r_k} \lambda_i,
	\end{aligned}
	\]
        where $e_i$ is the eigenvector of $\sum_{i = 1}^{n} \MM_k \left(\X_i \times_{h \neq k} {U}_{h\perp}\T \right) \MM_k \left(\X_i \times_{h \neq k} {U}_{h\perp}\T \right)\T$ and $\lambda_i$ is the corresponding eigenvalue satisfying $\lambda_i \geq \lambda_{i+1} \geq 0 $. The last inequality is due to the fact that $0 \leq e_i\T P_{U_{k\perp}} e_i \leq 1$ and $\sum_{i = 1}^{p_k}  e_i\T P_{U_{k\perp}} e_i = \tr(P_{U_{k\perp}} \sum_{i = 1}^{p_k} e_ie_i\T) = \tr(P_{U_{k\perp}}) = p_k - r_k$. The equality holds when $P_{U_{k\perp}} e_i = 0$ for $i = 1, \cdots r_k$, i.e. the minimizer should be $U_{k}$ such that $ \Span(U_k) = \Span\left(\Eigen_{r_k}\left(\frac{1}{n}\sum (X_i\T P_{U_\perp}X_i)\right)\right)$. 
\end{proof}

%%%%%%%%%%%%%%%%%%%%%%%%
\subsection{Proof of Theorem \ref{theorem_convergence_tensor}}
%%%%%%%%%%%%%%%%%%%%%%%%

Let's first introduce notations of set $\mathcal{A}$ of conditions on $\A_{i,k}$, $\Z_i$ and set $\mathcal{C}$ of condition on initialization. 
	\begin{Notation}\label{notation_W5678_tensor}
		Denote
		\begin{align*}
			T_h = & \left\{ W \in \RR^{\prod_{k \neq h}p_k \times \prod_{k \neq h}p_k}: \|W\|=1 , \tr(W) = 0, W = W\T, \rank(W) \leq 2^{d-1}\prod_{k \neq h}r_k\right\}, \\
			L_{h} = & \frac{1}{n\prod_{k \neq h}p_k}\sigma_{r_1}\left(\sum_{i = 1}^{n}\MM_h \left(  \A_{i,h}\right) \left(\bigotimes_{j \neq h} P_{{U}_{j\perp}} \right)\MM_h \left(  \A_{i,h}\right) \T   \right),\\
			M_{1,h} = &\frac{1}{n\prod_{k \neq h}p_k} \sup_{E\in T_h} \left\|\sum_{i = 1}^{n} \MM_h \left(  \A_{i,h} \right) E \MM_h \left( \sum_{k \neq h}\left(\A_{i,k} \times_k U_k\right) \right)\T \right\| , \\
			M_{2,h} = &\frac{1}{n\prod_{k \neq h}p_k}\max_{\substack{k_1, k_2\\k_1, k_2 \neq h; k_1 \neq k_2}}\sup_{E\in T_h} \left\|\sum_{i = 1}^{n} \MM_h \left( \A_{i,k_1} \times_{k_1} U_{k_1} \right) E \MM_h \left( \A_{i,k_2} \times_{k_2} U_{k_2} \right)\T \right\|  \\
			M_{3,h} = &\frac{1}{n\prod_{k \neq h}p_k}\sup_{E\in T_h}\left\|\sum_{i = 1}^{n} \MM_h \left(\Z_{i}\right) E \MM_h \left( \sum_{k = 1}^{d}\left(\A_{i,k} \times_k U_k\right) \right)\T   \right\|  , \\
			M_{4,h} = &\frac{1}{n\prod_{k \neq h}p_k}\sup_{E\in T_h} \left\|\sum_{i = 1}^{n} \MM_h \left(  \Z_{i}\right) E  \MM_h \left(\Z_{i}\right) \T \right\|, \\
			\Delta_h = & L_h - (d-1)(M_{1,h} + (d-1)^2M_{2,h} + M_{3,h} + M_{4,h}),\\
                \xi_h =& (d-1)(2+2d^2 +M_{1,1}+(d-1)^2M_{2,1}+ 3M_{3,1} +3M_{4,1}),\\
                \Gamma_h =& \Delta_h - \varepsilon_0 \xi_h, \\
			\kappa_h =& \frac{2}{n\prod_{k \neq h}p_k}\left\|\sum_{i = 1}^{n} \MM_h \left(\Z_{i}\right)\left(\bigotimes_{j \neq h} P_{{U}_{j\perp}} \right) \MM_h \left( \A_{i,h} \times_h U_h \right)\T   \right\|\\
			&+\frac{1}{n\prod_{k \neq h}p_k}\left\|\sum_{i = 1}^{n} \MM_h \left(  \Z_{i}\right) \left(\bigotimes_{j \neq h} P_{{U}_{j\perp}} \right)  \MM_h \left(\Z_{i}\right) \T - n\sigma^2\prod_{k \neq h}\left({p_k - r_k}\right) I \right\|,\\
			\mathcal{A}_1 = & \{\Delta_h > 0; h = 1, \cdots d\},\\
			\mathcal{A}_2 = & \left\{\frac{4\kappa_h}{L_h - \varepsilon_0\xi_h -  3 \kappa_h} \leq \frac{\Gamma_h}{L_h} \varepsilon_0;  h = 1, \cdots d\right\},\\
			\mathcal{A}_3 = & \left\{\max_{i,k} \left\| \MM_h \left(A_{i,k}\right) \right\| \leq \sqrt{\prod_{t \neq h}p_t}. \right\},\\
			\mathcal{A} = & \mathcal{A}_1 \cap \mathcal{A}_2 \cap\mathcal{A}_3
		\end{align*}
	\end{Notation}
	\begin{Notation}\label{notation_initialization2}
		Denote 
		\[
		\mathcal{B} = \{\Gamma_h > 0; h = 1, \cdots d\}. 
		\]
	\end{Notation}
	
	The strategy of this proof is to first establish a deterministic upper bound for estimation error given that $\A_{i,k}$, $\Z_i$ are nonrandom satisfying conditions $\{\A_{i,k}, \Z_i; i = 1,\cdots n, k = 1, \cdots d\} \subseteq \mathcal{A}\cap\mathcal{B}$ (in Section \ref{section_deterministic_bound_tensor_proof}), and then prove these conditions hold with high probability (in Section \ref{section_statistical_bound_tensor}). 
	
    \subsubsection{Deterministic Bound}\label{section_deterministic_bound_tensor_proof}
	We first introduce the following technical lemma that will be used in this section: 
	\begin{Lemma}\label{lemma_kronecker}
		Let $A_i$, $B_i$ be matrices with the same dimension, $\|A_i\|, \|B_i\| \leq 1$, and $\|A_i - B_i\| \leq \varepsilon$. Then \[
		\left\|\bigotimes_{i = 1}^d A_i - \bigotimes_{i = 1}^d B_i\right\| \leq d\varepsilon 
		\] 
	\end{Lemma}
	\begin{proof}
		Notice the following: \[\begin{aligned}
			\left\|\bigotimes_{i = 1}^k A_i \bigotimes_{i = k+1}^d B_i - \bigotimes_{i = 1}^{k-1} A_i \bigotimes_{i = k}^d B_i\right\| 
			=& \left\|\bigotimes_{i = 1}^{k-1} A_i \otimes(A_k - B_k) \bigotimes_{i = k+1}^d B_i\right\| \\
			=& \prod_{i = 1}^{k-1}\|A_i\|\|A_k - B_k\| \prod_{i = k+1}^{d}\|B_i\|\\
			\leq& \varepsilon 
		\end{aligned}
		\]So
		\begin{equation*}
		\begin{split}	
& \left\|\bigotimes_{i = 1}^d A_i - \bigotimes_{i = 1}^d B_i\right\| =  \left\|\sum_{k=1}^d\left(\bigotimes_{i = 1}^k A_i \bigotimes_{i = k+1}^d B_i - \bigotimes_{i = 1}^{k-1} A_i \bigotimes_{i = k}^d B_i\right)\right\|\\
\leq & \sum_{k=1}^d \left\|\bigotimes_{i = 1}^k A_i \bigotimes_{i = k+1}^d B_i - \bigotimes_{i = 1}^{k-1} A_i \bigotimes_{i = k}^d B_i\right\| \leq d\varepsilon.
\end{split}
\end{equation*}
	\end{proof}
	Define \[
	\S_i = \sum_{k=1}^d \A_{i,k} \times_k U_k = \X_i - \Z_i,
	\]
	\[
	H_k = \frac{1}{n\prod_{j\neq k}p_j}\sum_{i = 1}^{n}\MM_k \left(  \S_i \times_{j \neq k} \left( {U}_{j\perp} \right)\T  \right) \MM_k \left(  \S_i \times_{j \neq k} \left( {U}_{j\perp} \right)\T  \right) \T,
	\]
	\[
	\widehat H_k^{(t)} = \frac{1}{n\prod_{j\neq k}p_j}\sum_{i = 1}^{n}\MM_k \left(  \X_i \times_{j \neq k} \left( \widehat{U}_{j\perp}^{(t)} \right)\T  \right) \MM_k \left(  \X_i \times_{j \neq k} \left( \widehat{U}_{j\perp}^{(t)} \right)\T  \right) \T.
	\]
	\begin{Lemma}\label{parameters_2_tensor}
		In the context of Corollary \ref{Corollary_perturbation_bound}, let $\widehat{X} = \widehat{H}_1^{(t+1)} - \left(\prod_{k \neq 1}\frac{p_k - r_k}{p_k}\sigma^2\right)  I, {X} = {H}_1,$ and denote 
		$$\varepsilon_t = \max\left\{\|\sin \Theta (U_k, \widehat{U}_k^{(t)})\|, k = 1, \cdots n\right\}, \alpha = \lambda_{r_1} (P_{U_1} \widehat{X} P_{U_1}), \beta = \|{P}_{U_{1\perp}}\widehat{X}{P}_{U_{1\perp}}\|,z_{21} = \|{P}_{U_1} Z {P}_{U_{1\perp}}\|.$$ When $\A_{i,k}$, $\Z_i$ are nonrandom satisfying condition $\{\A_{i,k}, \Z_i; i = 1,\cdots n, k = 1, \cdots d\} \subseteq \mathcal{A}\cap\mathcal{B}$, we have: 
		$$
		\begin{aligned}
			\alpha \geq& L_1 - \varepsilon_t(d-1)(d^2 + 2M_{3,1} + M_{4,1}) - \kappa_1,\\
			\alpha \leq& L_1 + \varepsilon_t(d-1)(d^2 + 2M_{3,1} + M_{4,1}) + \kappa_1,\\
			\beta \leq& \varepsilon_t(d-1)(d^2+1 + M_{4,1}) + \kappa_1,\\
			z_{21} \leq& \varepsilon_t(d-1)(M_{1,1} + (d-1)^2M_{2,1}+ \varepsilon_t + M_{3,1} + M_{4,1}) + \kappa_1. 
		\end{aligned}
		$$
		where $M_{i,1}, \kappa_1, L_1$ are defined in Notation \ref{notation_W5678}. 
	\end{Lemma}
	
	\begin{proof}
		Firstly notice, by Lemma \ref{lemma_kronecker}, we have
		\be \label{inequality_norm_of_kronecker}
		\left\|\frac{\bigotimes_{j \neq 1} P_{{U}_{j\perp}} - \bigotimes_{j \neq 1} P_{\widehat{U}_{j\perp}^{(t)}} }{(d-1)\varepsilon_t} \right\| \leq 1. 
		\ee
            Also, $\rank(\bigotimes_{j \neq 1} P_{{U}_{j\perp}} - \bigotimes_{j \neq 1} P_{\widehat{U}_{j\perp}^{(t)}} ) = \rank(\bigotimes_{j \neq 1} (P_{{U}_{j\perp}} - P_{\widehat{U}_{j\perp}^{(t)}})) = \rank(\bigotimes_{j \neq 1} (P_{{U}_{j}} - P_{\widehat{U}_{j}^{(t)}})) \leq 2^{d-1} \prod_{k \neq 1} r_k$
		Hence, by Lemma \ref{lemma_Weyl_eigenvalues}, 
		\begin{align*}
			     &n\prod_{k \neq 1}p_k \alpha\\
			=&n\prod_{k \neq 1}p_k \lambda_{r_1}\left(U_1\T \widehat{H}_1^{(t+1)} U_1 -  \sigma^2\prod_{k \neq 1}\frac{p_k - r_k}{p_k}\sigma^2 \cdot U_1\T I U_1\right) \\
			\overset{(\ref{inequality_eigenvalue_sum})}{\leq} & \lambda_{r_1}\left(\sum_{i = 1}^{n}\MM_1 \left(  \A_{i,1} \times_{j \neq 1} \left( \widehat{U}_{j\perp}^{(t)} \right)\T  \right) \MM_1 \left(  \A_{i,1} \times_{j \neq 1} \left( \widehat{U}_{j\perp}^{(t)} \right)\T  \right) \T\right) \\
			&+ 2\left\|\sum_{i = 1}^{n} \MM_1 \left(  \A_{i,1} \times_{j \neq 1} \left( \widehat{U}_{j\perp}^{(t)} \right)\T  \right) \MM_1 \left(  \sum_{k \neq 1}\left(\A_{i,k} \times_k U_k\right) \times_{j \neq 1} \left( \widehat{U}_{j\perp}^{(t)} \right)\T  \right)\T U_1\right\|\\
			& + \left\|U_1\T\sum_{i = 1}^{n} \MM_1 \left(  \sum_{k \neq 1}\left(\A_{i,k} \times_k U_k\right) \times_{j \neq 1} \left( \widehat{U}_{j\perp}^{(t)} \right)\T  \right) \MM_1 \left(  \sum_{k \neq 1}\left(\A_{i,k} \times_k U_k\right) \times_{j \neq 1} \left( \widehat{U}_{j\perp}^{(t)} \right)\T  \right)\T U_1\right\| \\
			& + 2\left\|U_1\T \sum_{i = 1}^{n} \MM_1 \left(  \Z_{i} \times_{j \neq 1} \left( \widehat{U}_{j\perp}^{(t)} \right)\T  \right) \MM_1 \left(  \sum_{k = 1}^{d}\left(\A_{i,k} \times_k U_k\right) \times_{j \neq 1} \left( \widehat{U}_{j\perp}^{(t)} \right)\T  \right) \T U_1\right\| \\
			& + \left\|U_1\T\sum_{i = 1}^{n} \MM_1 \left(  \Z_{i} \times_{j \neq 1} \left( \widehat{U}_{j\perp}^{(t)} \right)\T  \right) \MM_1 \left(  \Z_{i} \times_{j \neq 1} \left( \widehat{U}_{j\perp}^{(t)} \right)\T  \right) \T U_1 - n\sigma^2\prod_{k \neq 1}\left({p_k - r_k}\right) I U_1\right\|\\
			= & \sigma_{r_1}\left(\sum_{i = 1}^{n}\MM_1 \left(  \A_{i,1}\right) \left(\bigotimes_{j \neq 1} P_{\widehat{U}_{j\perp}^{(t)}} \right)\MM_1 \left(  \A_{i,1}\right) \T   \right) 
                + 2\left\|\sum_{i = 1}^{n} \MM_1 \left(  \A_{i,1} \right) \left(\bigotimes_{j \neq 1} P_{\widehat{U}_{j\perp}^{(t)}} \right) \MM_1 \left( \sum_{k \neq 1}\left(\A_{i,k} \times_k U_k\right) \right)\T \right\|\\
			& + \left\|\sum_{i = 1}^{n} \MM_1 \left( \sum_{k \neq 1}\left(\A_{i,k} \times_k U_k\right) \right) \left(\bigotimes_{j \neq 1} P_{\widehat{U}_{j\perp}^{(t)}} \right) \MM_1 \left( \sum_{k \neq 1}\left(\A_{i,k} \times_k U_k\right) \right)\T \right\| \\
			& + 2\left\|\sum_{i = 1}^{n} \MM_1 \left(\Z_{i}\right) \left(\bigotimes_{j \neq 1} P_{\widehat{U}_{j\perp}^{(t)}} \right) \MM_1 \left( \sum_{k = 1}^{d}\left(\A_{i,k} \times_k U_k\right) \right)\T   \right\| +  \left\|\sum_{i = 1}^{n} \MM_1 \left(  \Z_{i}\right) \left(\bigotimes_{j \neq 1} P_{\widehat{U}_{j\perp}^{(t)}} \right)  \MM_1 \left(\Z_{i}\right) \T - n\sigma^2\prod_{k \neq 1}\left({p_k - r_k}\right) I\right\|\\	
			\leq & \sigma_{r_1}\left(\sum_{i = 1}^{n}\MM_1 \left(  \A_{i,1}\right) \left(\bigotimes_{j \neq 1} P_{{U}_{j\perp}} \right)\MM_1 \left(  \A_{i,1}\right) \T   \right) + (d-1)\varepsilon_t\left\| \sum_{i = 1}^{n}\MM_1 \left(  \A_{i,1}\right)  \frac{\bigotimes_{j \neq 1} P_{{U}_{j\perp}} - \bigotimes_{j \neq 1} P_{\widehat{U}_{j\perp}^{(t)}} }{(d-1)\varepsilon_t}  \MM_1 \left(  \A_{i,1}\right) \T \right\|  \\
			&+ 2\left\|\sum_{i = 1}^{n} \MM_1 \left(  \A_{i,1} \right) \left(\bigotimes_{j \neq 1} P_{{U}_{j\perp}} \right) \MM_1 \left( \sum_{k \neq 1}\left(\A_{i,k} \times_k U_k\right) \right)\T \right\|\\
			&+ 2(d-1)\varepsilon_t\left\|\sum_{i = 1}^{n} \MM_1 \left(  \A_{i,1} \right) \frac{\bigotimes_{j \neq 1} P_{{U}_{j\perp}} - \bigotimes_{j \neq 1} P_{\widehat{U}_{j\perp}^{(t)}} }{(d-1)\varepsilon_t} \MM_1 \left( \sum_{k \neq 1}\left(\A_{i,k} \times_k U_k\right) \right)\T \right\|  \\
			& + \left\|\sum_{i = 1}^{n} \MM_1 \left( \sum_{k \neq 1}\left(\A_{i,k} \times_k U_k\right) \right) \left(\bigotimes_{j \neq 1} P_{{U}_{j\perp}} \right) \MM_1 \left( \sum_{k \neq 1}\left(\A_{i,k} \times_k U_k\right) \right)\T \right\|\\
			& + (d-1)\varepsilon_t \left\|\sum_{i = 1}^{n} \MM_1 \left( \sum_{k \neq 1}\left(\A_{i,k} \times_k U_k\right) \right) \frac{\bigotimes_{j \neq 1} P_{{U}_{j\perp}} - \bigotimes_{j \neq 1} P_{\widehat{U}_{j\perp}^{(t)}} }{(d-1)\varepsilon_t} \MM_1 \left( \sum_{k \neq 1}\left(\A_{i,k} \times_k U_k\right) \right)\T \right\| \\
			& + 2\left\|\sum_{i = 1}^{n} \MM_1 \left(\Z_{i}\right)\left(\bigotimes_{j \neq 1} P_{{U}_{j\perp}} \right) \MM_1 \left( \sum_{k = 1}^{d}\left(\A_{i,k} \times_k U_k\right) \right)\T   \right\|\\
			&+ 2(d-1)\varepsilon_t \left\|\sum_{i = 1}^{n} \MM_1 \left(\Z_{i}\right) \frac{\bigotimes_{j \neq 1} P_{{U}_{j\perp}} - \bigotimes_{j \neq 1} P_{\widehat{U}_{j\perp}^{(t)}} }{(d-1)\varepsilon_t} \MM_1 \left( \sum_{k = 1}^{d}\left(\A_{i,k} \times_k U_k\right) \right)\T   \right\|   \\
			&+\left\|\sum_{i = 1}^{n} \MM_1 \left(  \Z_{i}\right) \left(\bigotimes_{j \neq 1} P_{{U}_{j\perp}} \right)  \MM_1 \left(\Z_{i}\right) \T - n\sigma^2\prod_{k \neq 1}\left({p_k - r_k}\right) I \right\|+ (d-1)\varepsilon_t \left\|\sum_{i = 1}^{n} \MM_1 \left(  \Z_{i}\right) \frac{\bigotimes_{j \neq 1} P_{{U}_{j\perp}} - \bigotimes_{j \neq 1} P_{\widehat{U}_{j\perp}^{(t)}} }{(d-1)\varepsilon_t}  \MM_1 \left(\Z_{i}\right) \T \right\|
                \end{align*}
                \begin{align*}
			= & \sigma_{r_1}\left(\sum_{i = 1}^{n}\MM_1 \left(  \A_{i,1}\right) \left(\bigotimes_{j \neq 1} P_{{U}_{j\perp}} \right)\MM_1 \left(  \A_{i,1}\right) \T   \right) + (d-1)\varepsilon_t\left\| \sum_{i = 1}^{n}\MM_1 \left(  \A_{i,1}\right)  \frac{\bigotimes_{j \neq 1} P_{{U}_{j\perp}} - \bigotimes_{j \neq 1} P_{\widehat{U}_{j\perp}^{(t)}} }{(d-1)\varepsilon_t}  \MM_1 \left(  \A_{i,1}\right) \T \right\|  \\
			&+ 2(d-1)\varepsilon_t\left\|\sum_{i = 1}^{n} \MM_1 \left(  \A_{i,1} \right) \frac{\bigotimes_{j \neq 1} P_{{U}_{j\perp}} - \bigotimes_{j \neq 1} P_{\widehat{U}_{j\perp}^{(t)}} }{(d-1)\varepsilon_t} \MM_1 \left( \sum_{k \neq 1}\left(\A_{i,k} \times_k U_k\right) \right)\T \right\|  \\
			& + (d-1)\varepsilon_t \left\|\sum_{i = 1}^{n} \MM_1 \left( \sum_{k \neq 1}\left(\A_{i,k} \times_k U_k\right) \right) \frac{\bigotimes_{j \neq 1} P_{{U}_{j\perp}} - \bigotimes_{j \neq 1} P_{\widehat{U}_{j\perp}^{(t)}} }{(d-1)\varepsilon_t} \MM_1 \left( \sum_{k \neq 1}\left(\A_{i,k} \times_k U_k\right) \right)\T \right\| \\
			& + 2\left\|\sum_{i = 1}^{n} \MM_1 \left(\Z_{i}\right)\left(\bigotimes_{j \neq 1} P_{{U}_{j\perp}} \right) \MM_1 \left( \A_{i,1} \times_1 U_1 \right)\T   \right\|\\
			&+ 2(d-1)\varepsilon_t \left\|\sum_{i = 1}^{n} \MM_1 \left(\Z_{i}\right) \frac{\bigotimes_{j \neq 1} P_{{U}_{j\perp}} - \bigotimes_{j \neq 1} P_{\widehat{U}_{j\perp}^{(t)}} }{(d-1)\varepsilon_t} \MM_1 \left( \sum_{k = 1}^{d}\left(\A_{i,k} \times_k U_k\right) \right)\T   \right\|   \\
			&+\left\|\sum_{i = 1}^{n} \MM_1 \left(  \Z_{i}\right) \left(\bigotimes_{j \neq 1} P_{{U}_{j\perp}} \right)  \MM_1 \left(\Z_{i}\right) \T - n\sigma^2\prod_{k \neq 1}\left({p_k - r_k}\right) I \right\|\\
			&+ (d-1)\varepsilon_t \left\|\sum_{i = 1}^{n} \MM_1 \left(  \Z_{i}\right) \frac{\bigotimes_{j \neq 1} P_{{U}_{j\perp}} - \bigotimes_{j \neq 1} P_{\widehat{U}_{j\perp}^{(t)}} }{(d-1)\varepsilon_t}  \MM_1 \left(\Z_{i}\right) \T \right\| \\
			{\leq}&n\prod_{k \neq 1}p_k\left( L_1 + \varepsilon_t(d-1)((d-1)^2+2(d-1)+1 + 2M_{3,1} + M_{4,1}) + \kappa_1\right)\\
			= &n\prod_{k \neq 1}p_k\left( L_1 + \varepsilon_t(d-1)(d^2 + 2M_{3,1} + M_{4,1}) + \kappa_1\right), 
		\end{align*}
		where the last inequality holds by (\ref{inequality_norm_of_kronecker}) and $\{\A_{i,k}, \Z_i; i = 1,\cdots n, k = 1, \cdots d\} \subseteq \mathcal{A}\cap\mathcal{B}$. On the other hand, we can similarly derive a lower bound: 
		\begin{align*}
			&n\prod_{k \neq 1}p_k \alpha\\
			\overset{(\ref{inequality_eigenvalue_sum})}{\geq} & \sigma_{r_1}\left(\sum_{i = 1}^{n}\MM_1 \left(  \A_{i,1} \times_{j \neq 1} \left( \widehat{U}_{j\perp}^{(t)} \right)\T  \right) \MM_1 \left(  \A_{i,1} \times_{j \neq 1} \left( \widehat{U}_{j\perp}^{(t)} \right)\T  \right) \T\right) \\
			&- 2\left\|\sum_{i = 1}^{n} \MM_1 \left(  \A_{i,1} \times_{j \neq 1} \left( \widehat{U}_{j\perp}^{(t)} \right)\T  \right) \MM_1 \left(  \sum_{k \neq 1}\left(\A_{i,k} \times_k U_k\right) \times_{j \neq 1} \left( \widehat{U}_{j\perp}^{(t)} \right)\T  \right)\T U_1\right\|\\
			& - \left\|U_1\T\sum_{i = 1}^{n} \MM_1 \left(  \sum_{k \neq 1}\left(\A_{i,k} \times_k U_k\right) \times_{j \neq 1} \left( \widehat{U}_{j\perp}^{(t)} \right)\T  \right) \MM_1 \left(  \sum_{k \neq 1}\left(\A_{i,k} \times_k U_k\right) \times_{j \neq 1} \left( \widehat{U}_{j\perp}^{(t)} \right)\T  \right)\T U_1\right\| \\
			& - 2\left\|U_1\T \sum_{i = 1}^{n} \MM_1 \left(  \Z_{i} \times_{j \neq 1} \left( \widehat{U}_{j\perp}^{(t)} \right)\T  \right) \MM_1 \left(  \sum_{k = 1}^{d}\left(\A_{i,k} \times_k U_k\right) \times_{j \neq 1} \left( \widehat{U}_{j\perp}^{(t)} \right)\T  \right) \T U_1\right\| \\
			& - \left\|U_1\T\sum_{i = 1}^{n} \MM_1 \left(  \Z_{i} \times_{j \neq 1} \left( \widehat{U}_{j\perp}^{(t)} \right)\T  \right) \MM_1 \left(  \Z_{i} \times_{j \neq 1} \left( \widehat{U}_{j\perp}^{(t)} \right)\T  \right) \T U_1 - n\sigma^2\prod_{k \neq 1}\left({p_k - r_k}\right) I U_1\right\|\\
			\geq & \sigma_{r_1}\left(\sum_{i = 1}^{n}\MM_1 \left(  \A_{i,1}\right) \left(\bigotimes_{j \neq 1} P_{{U}_{j\perp}} \right)\MM_1 \left(  \A_{i,1}\right) \T   \right)\\
			&- (d-1)\varepsilon_t\left\| \sum_{i = 1}^{n}\MM_1 \left(  \A_{i,1}\right)  \frac{\bigotimes_{j \neq 1} P_{{U}_{j\perp}} - \bigotimes_{j \neq 1} P_{\widehat{U}_{j\perp}^{(t)}} }{(d-1)\varepsilon_t}  \MM_1 \left(  \A_{i,1}\right) \T \right\|  \\
			&- 2(d-1)\varepsilon_t\left\|\sum_{i = 1}^{n} \MM_1 \left(  \A_{i,1} \right) \frac{\bigotimes_{j \neq 1} P_{{U}_{j\perp}} - \bigotimes_{j \neq 1} P_{\widehat{U}_{j\perp}^{(t)}} }{(d-1)\varepsilon_t} \MM_1 \left( \sum_{k \neq 1}\left(\A_{i,k} \times_k U_k\right) \right)\T \right\|  \\
			&- (d-1)\varepsilon_t \left\|\sum_{i = 1}^{n} \MM_1 \left( \sum_{k \neq 1}\left(\A_{i,k} \times_k U_k\right) \right) \frac{\bigotimes_{j \neq 1} P_{{U}_{j\perp}} - \bigotimes_{j \neq 1} P_{\widehat{U}_{j\perp}^{(t)}} }{(d-1)\varepsilon_t} \MM_1 \left( \sum_{k \neq 1}\left(\A_{i,k} \times_k U_k\right) \right)\T \right\| \\
			&- 2\left\|\sum_{i = 1}^{n} \MM_1 \left(\Z_{i}\right)\left(\bigotimes_{j \neq 1} P_{{U}_{j\perp}} \right) \MM_1 \left( \A_{i,1} \times_1 U_1 \right)\T   \right\|\\
			&- 2(d-1)\varepsilon_t \left\|\sum_{i = 1}^{n} \MM_1 \left(\Z_{i}\right) \frac{\bigotimes_{j \neq 1} P_{{U}_{j\perp}} - \bigotimes_{j \neq 1} P_{\widehat{U}_{j\perp}^{(t)}} }{(d-1)\varepsilon_t} \MM_1 \left( \sum_{k = 1}^{d}\left(\A_{i,k} \times_k U_k\right) \right)\T   \right\|   \\
			&- \left\|\sum_{i = 1}^{n} \MM_1 \left(  \Z_{i}\right) \left(\bigotimes_{j \neq 1} P_{{U}_{j\perp}} \right)  \MM_1 \left(\Z_{i}\right) \T - n\sigma^2\prod_{k \neq 1}\left({p_k - r_k}\right) I \right\|\\
			&- (d-1)\varepsilon_t \left\|\sum_{i = 1}^{n} \MM_1 \left(  \Z_{i}\right) \frac{\bigotimes_{j \neq 1} P_{{U}_{j\perp}} - \bigotimes_{j \neq 1} P_{\widehat{U}_{j\perp}^{(t)}} }{(d-1)\varepsilon_t}  \MM_1 \left(\Z_{i}\right) \T \right\| \\
			&\geq {n\prod_{k \neq 1}p_k}\left( L_1 - \varepsilon_t(d-1)(d^2 + 2M_{3,1} + M_{4,1}) - \kappa_1\right). 
		\end{align*}
		Similarly, we can give a upper bound of $\beta$: 
		\begin{align*}
			&n\prod_{k \neq 1}p_k\beta\\
			=& n\prod_{k \neq 1}p_k\left\|U_{1\perp}\T \widehat{H}_1^{(t+1)} U_{1\perp}- \prod_{k \neq 1}\frac{p_k - r_k}{p_k}\sigma^2 \cdot I\right\|
            \end{align*}
		\begin{align*}
                \leq & 
			\left\|U_{1\perp}\T\sum_{i = 1}^{n} \MM_1 \left(  \sum_{k \neq 1}\left(\A_{i,k} \times_k U_k\right) \times_{j \neq 1} \left( \widehat{U}_{j\perp}^{(t)} \right)\T  \right) \MM_1 \left(  \sum_{k \neq 1}\left(\A_{i,k} \times_k U_k\right) \times_{j \neq 1} \left( \widehat{U}_{j\perp}^{(t)} \right)\T  \right)\T U_{1\perp}\right\| \\
			&+2\left\|U_{1\perp}\T \sum_{i = 1}^{n} \MM_1 \left(  \Z_{i} \times_{j \neq 1} \left( \widehat{U}_{j\perp}^{(t)} \right)\T  \right) \MM_1 \left(  \sum_{k = 1}^d  \left(\A_{i,k} \times_k U_k\right) \times_{j \neq 1} \left( \widehat{U}_{j\perp}^{(t)} \right)\T  \right) \T U_{1\perp}\right\| \\ 
			&+ \left\|U_{1\perp}\T\sum_{i = 1}^{n} \MM_1 \left(  \Z_{i} \times_{j \neq 1} \left( \widehat{U}_{j\perp}^{(t)} \right)\T  \right) \MM_1 \left(  \Z_{i} \times_{j \neq 1} \left( \widehat{U}_{j\perp}^{(t)} \right)\T  \right) \T U_{1\perp} - n\sigma^2\prod_{k \neq 1}\left({p_k - r_k}\right) I\right\|\\
			\leq& 
			(d-1)\varepsilon_t\left\|\sum_{i = 1}^n \MM_1 \left( \sum_{k \neq 1}\left(\A_{i,k} \times_k U_k\right) \right) \frac{\bigotimes_{j \neq 1} P_{{U}_{j\perp}} - \bigotimes_{j \neq 1} P_{\widehat{U}_{j\perp}^{(t)}} }{(d-1)\varepsilon_t} \MM_1 \left( \sum_{k \neq 1}\left(\A_{i,k} \times_k U_k\right) \right)\T\right\| \\
			& + 2(d-1)\varepsilon_t \left\|\sum_{i = 1}^n \MM_1 \left( \sum_{k = 1}^d \left(\A_{i,k} \times_k U_k\right) \right) \frac{\bigotimes_{j \neq 1} P_{{U}_{j\perp}} - \bigotimes_{j \neq 1} P_{\widehat{U}_{j\perp}^{(t)}} }{(d-1)\varepsilon_t} \MM_1 \left(  \Z_{i}\right)\T \right\| \\
			& + \left\|\sum_{i = 1}^{n} \MM_1 \left(  \Z_{i}\right) \left(\bigotimes_{j \neq 1} P_{{U}_{j\perp}} \right)  \MM_1 \left(\Z_{i}\right) \T - n\sigma^2\prod_{k \neq 1}\left({p_k - r_k}\right) I\right\|\\
			& + (d-1)\varepsilon_t \left\|\sum_{i = 1}^{n} \MM_1 \left(  \Z_{i}\right) \frac{\bigotimes_{j \neq 1} P_{{U}_{j\perp}} - \bigotimes_{j \neq 1} P_{\widehat{U}_{j\perp}^{(t)}} }{(d-1)\varepsilon_t}  \MM_1 \left(\Z_{i}\right) \T \right\| \\
			&\leq {n\prod_{k \neq 1}p_k}\left(\varepsilon_t(d-1)((d-1)^2 + 2d + M_{4,1}) + \kappa_1\right)\\
			&= {n\prod_{k \neq 1}p_k}\left(\varepsilon_t(d-1)(d^2+1 + M_{4,1}) + \kappa_1\right)
		\end{align*}
		To upper bound $z_{21}$, firstly notice
		\begin{align*}
			&\left\|\sum_{i = 1}^{n} \MM_1 \left(  \sum_{k \neq 1}\A_{i,k} \times_k \left(\widehat{U}_{k\perp}^{(t)\top} U_k  \right) \times_{j \neq 1,k} \widehat{U}_{j\perp}^{(t)\top}  \right) \MM_1 \left(  \sum_{k \neq 1}\A_{i,k} \times_k \left(\widehat{U}_{k\perp}^{(t)\top} U_k  \right) \times_{j \neq 1,k} \widehat{U}_{j\perp}^{(t)\top}  \right)\T \right\|\\
			\leq & 
			\left\|\sum_{i = 1}^{n} \sum_{k \neq 1} \MM_1 \left( \A_{i,k} \times_k \left(\widehat{U}_{k\perp}^{(t)\top} U_k  \right) \times_{j \neq 1,k} \widehat{U}_{j\perp}^{(t)\top}  \right) \MM_1 \left( \A_{i,k} \times_k \left(\widehat{U}_{k\perp}^{(t)\top} U_k  \right) \times_{j \neq 1,k} \widehat{U}_{j\perp}^{(t)\top}  \right)\T \right\|\\
			&+ \left\|\sum_{i = 1}^{n} \sum_{\substack{k_1, k_2 \neq 1 \\ k_1 \neq k_2}} \MM_1 \left( \A_{i,k_1} \times_{k_1} U_{k_1} \times_{j \neq 1} \widehat{U}_{{j}\perp}^{(t)\top}  \right) \MM_1 \left( \A_{i,k_2} \times_{k_2} U_{k_2} \times_{j \neq 1} \widehat{U}_{{j}\perp}^{(t)\top}  \right)\T \right\|\\
			= & 
			\left\|\sum_{i = 1}^{n} \sum_{k \neq 1} \MM_1 \left( \A_{i,k} \right) \left(\left(\bigotimes_{j \neq 1, j \leq k-1} P_{\widehat{U}_{j\perp}^{(t)}}\right) \otimes \left(U_k\T \widehat{U}_{k\perp}^{(t)}\right)^2\otimes \left( \bigotimes_{j \neq 1, j \geq k+1} P_{\widehat{U}_{j\perp}^{(t)}} \right)\right)\MM_1 \left( \A_{i,k}   \right)\T \right\|\\
			&+ \left\|\sum_{i = 1}^{n} \sum_{\substack{k_1, k_2 \neq 1 \\ k_1 \neq k_2}} \MM_1 \left( \A_{i,k_1} \times_{k_1} U_{k_1} \times_{j \neq 1} \widehat{U}_{{j}\perp}^{(t)\top} \right) \left(\bigotimes_{j \neq 1}P_{\widehat{U}_{j\perp}^{(t)}}\right) \MM_1 \left( \A_{i,k_2} \times_{k_2} U_{k_2} \times_{j \neq 1} \widehat{U}_{{j}\perp}^{(t)\top}  \right)\T \right\|\\
			\leq &
			\sum_{k \neq 1} \left\|\sum_{i = 1}^{n} \MM_1 \left( \A_{i,k} \right) \left(\left(\bigotimes_{j \neq 1, j \leq k-1} P_{\widehat{U}_{j\perp}^{(t)}}\right) \otimes \left(U_k\T \widehat{U}_{k\perp}^{(t)}\right)^2\otimes \left( \bigotimes_{j \neq 1, j \geq k+1} P_{\widehat{U}_{j\perp}^{(t)}} \right)\right)\MM_1 \left( \A_{i,k}   \right)\T \right\|\\
			&+ \left\|\sum_{i = 1}^{n} \sum_{\substack{k_1, k_2 \neq 1 \\ k_1 \neq k_2}} \MM_1 \left( \A_{i,k_1} \times_{k_1} U_{k_1} \right) \left(\bigotimes_{j \neq 1}P_{\widehat{U}_{j\perp}^{(t)}}\right) \MM_1 \left( \A_{i,k_2} \times_{k_2} U_{k_2}\right)\T \right\|.
		\end{align*}
		So, 
		\begin{align*}	
			& n\prod_{k \neq 1}p_k z_{21} = n\prod_{k \neq 1}p_k \left\|P_{U_{1\perp}} \widehat{H}_1^{(t+1)} P_{U_1}\right\| \\
			\leq & 
			\left\|\sum_{i = 1}^{n} \MM_1 \left(  \sum_{k \neq 1}\left(\A_{i,k} \times_k U_k\right) \times_{j \neq 1} \left( \widehat{U}_{j\perp}^{(t)} \right)\T  \right) \MM_1 \left(  \A_{i,1}\times_1 U_1 \times_{j \neq 1} \left( \widehat{U}_{j\perp}^{(t)} \right)\T  \right)\T \right\|\\
			&+ \left\|\sum_{i = 1}^{n} \MM_1 \left(  \sum_{k \neq 1}\A_{i,k} \times_k \left(\widehat{U}_{k\perp}^{(t)\top} U_k  \right) \times_{j \neq 1,k} \widehat{U}_{j\perp}^{(t)\top}  \right) \MM_1 \left(  \sum_{k \neq 1}\A_{i,k} \times_k \left(\widehat{U}_{k\perp}^{(t)\top} U_k  \right) \times_{j \neq 1,k} \widehat{U}_{j\perp}^{(t)\top}  \right)\T \right\| \\
			&+ \left\| \sum_{i = 1}^{n}P_{U_{1\perp}} \MM_1 \left(  \sum_{k = 1}^d \left(\A_{i,k} \times_k U_k\right) \times_{j \neq 1} \left( \widehat{U}_{j\perp}^{(t)} \right)\T  \right)  \MM_1 \left(  \Z_{i} \times_{j \neq 1} \left( \widehat{U}_{j\perp}^{(t)} \right)\T  \right)\T \right\|\\
			&+ \left\| \sum_{i = 1}^{n} \MM_1 \left(  \Z_{i} \times_{j \neq 1} \left( \widehat{U}_{j\perp}^{(t)} \right)\T  \right) \MM_1 \left(  \sum_{k = 1}^d \left(\A_{i,k} \times_k U_k\right) \times_{j \neq 1} \left( \widehat{U}_{j\perp}^{(t)} \right)\T  \right) \T \right\|\\
			&+ \left\|\sum_{i = 1}^{n} \MM_1 \left(  \Z_{i} \times_{j \neq 1} \left( \widehat{U}_{j\perp}^{(t)} \right)\T  \right) \MM_1 \left(  \Z_{i} \times_{j \neq 1} \left( \widehat{U}_{j\perp}^{(t)} \right)\T  \right) \T\right\|\\
			\leq & 
			(d-1)\varepsilon_t\left\|\sum_{i = 1}^{n} \MM_1 \left(  \A_{i,1} \right) \frac{\bigotimes_{j \neq 1} P_{{U}_{j\perp}} - \bigotimes_{j \neq 1} P_{\widehat{U}_{j\perp}^{(t)}} }{(d-1)\varepsilon_t} \MM_1 \left( \sum_{k \neq 1}\left(\A_{i,k} \times_k U_k\right) \right)\T \right\|\\
			&+ \varepsilon_t^2 \sum_{k \neq 1} \left\|\sum_{i = 1}^{n} \MM_1 \left( \A_{i,k} \right) \frac{\left(\left(\bigotimes_{j \neq 1, j \leq k-1} P_{\widehat{U}_{j\perp}^{(t)}}\right) \otimes \left(U_k\T \widehat{U}_{k\perp}^{(t)}\right)^2\otimes \left( \bigotimes_{j \neq 1, j \geq k+1} P_{\widehat{U}_{j\perp}^{(t)}} \right)\right)}{\varepsilon_t^2}\MM_1 \left( \A_{i,k}   \right)\T \right\|\\
			&+ (d-1)\varepsilon_t\left\|\sum_{i = 1}^{n} \sum_{\substack{k_1, k_2 \neq 1 \\ k_1 \neq k_2}} \MM_1 \left( \A_{i,k_1} \times_{k_1} U_{k_1} \right) \frac{\bigotimes_{j \neq 1} P_{{U}_{j\perp}} - \bigotimes_{j \neq 1} P_{\widehat{U}_{j\perp}^{(t)}} }{(d-1)\varepsilon_t} \MM_1 \left( \A_{i,k_2} \times_{k_2} U_{k_2} \right)\T \right\|\\
			&+ \left\|\sum_{i = 1}^{n} \MM_1 \left(\Z_{i}\right)\left(\bigotimes_{j \neq 1} P_{{U}_{j\perp}} \right) \MM_1 \left( \A_{i,1} \times_1 U_1 \right)\T   \right\|\\
			&+ 2(d-1)\varepsilon_t \left\|\sum_{i = 1}^{n} \MM_1 \left(\Z_{i}\right) \frac{\bigotimes_{j \neq 1} P_{{U}_{j\perp}} - \bigotimes_{j \neq 1} P_{\widehat{U}_{j\perp}^{(t)}} }{(d-1)\varepsilon_t} \MM_1 \left( \sum_{k = 1}^{d}\left(\A_{i,k} \times_k U_k\right) \right)\T   \right\|   \\
			& + \left\|\sum_{i = 1}^{n} \MM_1 \left(  \Z_{i}\right) \left(\bigotimes_{j \neq 1} P_{{U}_{j\perp}} \right)  \MM_1 \left(\Z_{i}\right) \T - n\sigma^2\prod_{k \neq 1}\left({p_k - r_k}\right) I\right\|\\
			& + (d-1)\varepsilon_t \left\|\sum_{i = 1}^{n} \MM_1 \left(  \Z_{i}\right) \frac{\bigotimes_{j \neq 1} P_{{U}_{j\perp}} - \bigotimes_{j \neq 1} P_{\widehat{U}_{j\perp}^{(t)}} }{(d-1)\varepsilon_t}  \MM_1 \left(\Z_{i}\right) \T \right\| \\
			&\leq {n\prod_{k \neq 1}p_k}\left(\varepsilon_t(d-1)(M_{1,1} + (d-1)^2M_{2,1}+ \varepsilon_t + M_{3,1} + M_{4,1}) + \kappa_1\right). 
		\end{align*}
	\end{proof}
	We are ready to prove the following: 
	\begin{Theorem}\label{theorem_convergence_deterministic_tensor}
		In Algorithm \ref{algorithm_iterative_projection}, let $\X_i =\sum_{k=1}^d \A_{i,k} \times_k U_k + \Z_i \in \mathbb{R}^{p_1\times \cdots \times p_d}, i = 1, \cdots n$. 
		Denote the estimation error $\varepsilon_t = \max\left\{\|\sin \Theta (U_k, \widehat{U}_k^{(t)})\|, k = 1, \cdots n\right\}$. 
		When $\A_{i,k}$, $\Z_i$ are nonrandom satisfying condition $\{\A_{i,k}, \Z_i; i = 1,\cdots n, k = 1, \cdots d\} \subseteq \mathcal{A}\cap\mathcal{B}$, there is a constant $\chi = \max_h \frac{L_h - \Gamma_h}{L_h}<1$, which does not depend on $t$, such that for $t=0,1,\ldots, m$, 
		\begin{equation}
			\varepsilon_{t+1} \leq \chi \varepsilon_t + K_1\leq \varepsilon_0 , \notag
		\end{equation}
		where $$K_1 = \max_h\frac{4\kappa_h}{L_h - \xi_h -  3 \kappa_h}. $$
		Consequently, 
		\begin{align}
			\varepsilon_t \leq \chi^t \varepsilon_0 + K_2,\notag 
		\end{align}where $K_2 = K_1 \frac{1-\chi^t}{1-\chi}$.

	\end{Theorem}
	
	\begin{proof}
		We will prove this by induction. Assume the statement holds for $m<t$. Consider $m = t$, where we are calculating $\widehat U^{(t+1)}_h$ for $h = 1$. 
		
		By Corollary \ref{Corollary_perturbation_bound}:
		$$
		\|\sin \Theta(U_1, \widehat{U}_1^{(t+1)})\| \leq \frac{z_{21}}{\alpha-\beta-z_{21}}. 
		$$
		Denote$$
		\begin{aligned}
			\alpha_1 =& L_1 - \varepsilon_t(d-1)(d^2 + 2M_{3,1} + M_{4,1}),\\
			\beta_1 =& \varepsilon_t(d-1)(d^2+1 + M_{4,1}),\\
			z_{1} =& \varepsilon_t(d-1)(M_{1,1} + (d-1)^2M_{2,1}+ \varepsilon_t + M_{3,1} + M_{4,1}). 
		\end{aligned}
		$$
		By Lemma \ref{parameters_2_tensor}, 
		\begin{align}
			\frac{z_{21}}{\alpha-\beta-z_{21}}
			& \leq \frac{z_1 + \kappa_1}{\alpha_1 - \beta_1 - z_1 - 3\kappa_1} \notag\\
			& = \frac{z_1}{\alpha_1 - \beta_1 - z_1}+ \frac{\kappa_1}{\alpha_1 - \beta_1 - z_1-3\kappa_1} + \frac{z_1}{\alpha_1 - \beta_1 - z_1}\frac{3\kappa_1}{\alpha_1 - \beta_1 - z_1-3\kappa_1}. \label{inequality_zab_tensor}
		\end{align}
		To bound the first term on right-hand side of (\ref{inequality_zab_tensor}), notice that the function $f(y) = (x-y)/x$ is monotone decreasing for $y<x$ with any given $x$ and that by $\{\A_{i,k}, \Z_i; i = 1,\cdots n, k = 1, \cdots d\} \subseteq \mathcal{A}\cap\mathcal{B}$, we have
		$$
		\begin{aligned}
			\alpha_1 - \beta_1 - z_1 - \frac{z_1}{\varepsilon_t }
			=&  L_1 - (d-1)(M_{1,1} + (d-1)^2M_{2,1} + M_{3,1} + M_{4,1})\\
			&- \varepsilon_t (d-1)(2+2d^2 +M_{1,1}+(d-1)^2M_{2,1}+ 3M_{3,1} +3M_{4,1} + )\\
			\geq& \Gamma_1 > 0. 
		\end{aligned}
		$$So, \be
		\label{inequality_term1_1_tensor}
		\frac{z_1/\varepsilon_t}{\alpha_1 - \beta_1 - z_1} \leq \frac{\alpha_1 - \beta_1 - z_1 - \Gamma_1}{\alpha_1 - \beta_1 - z_1}.
		\ee
		Further notice that the function $g(x) = (x-y)/x$ is monotone increasing on $x>y$ for fixed $y$ and $\alpha_1 - \beta_1 - z_1 \leq L_1$. So, 
		\be
		\label{inequality_term1_2_tensor}
		\frac{\alpha_1 - \beta_1 - z_1 - \Gamma_1}{\alpha_1 - \beta_1 - z_1} \leq \frac{L_1 - \Gamma_1}{L_1}. 
		\ee
		Thus, combining (\ref{inequality_term1_1_tensor}) and (\ref{inequality_term1_2_tensor}), we have
		\be
		%\label{inequality_term1}
		\frac{z_1}{\alpha_1 - \beta_1 - z_1} \leq \frac{L_1 - \Gamma_1}{L_1}\varepsilon_t. 
		\ee
		To bound the remaining two terms of (\ref{inequality_zab_tensor}), by (\ref{inequality_term1}) and $\{\A_{i,k}, \Z_i; i = 1,\cdots n, k = 1, \cdots d\} \subseteq \mathcal{A}\cap\mathcal{B}$, we have \[
		\frac{\kappa_1}{\alpha_1 - \beta_1 - z_1-3\kappa_1} + \frac{z_1}{\alpha_1 - \beta_1 - z_1}\frac{3\kappa_1}{\alpha_1 - \beta_1 - z_1-3\kappa_1} \leq \frac{4\kappa_1}{\alpha_1 - \beta_1 - z_1-3\kappa_1}, 
		\]
		and
		\[
		\begin{aligned}
			\alpha_1 - \beta_1 - z_1 
			\geq  L_1-  \varepsilon_0 \xi_1,
		\end{aligned}
		\]
		which yield
		\be
		\label{inequality_term2_tensor}
			\frac{\kappa_1}{\alpha_1 - \beta_1 - z_1-3\kappa_1} + \frac{z_1}{\alpha_1 - \beta_1 - z_1}\frac{3\kappa_1}{\alpha_1 - \beta_1 - z_1-3\kappa_1} \leq \frac{4\kappa_1}{L_1 - \varepsilon_0\xi_1 -  3 \kappa_1}. 
		\ee
		Combining (\ref{inequality_zab_tensor}), (\ref{inequality_term1}) and (\ref{inequality_term2_tensor}), we finally have \[
		\|\sin \Theta(U_1, \widehat{U}_1^{(t+1)})\| \leq \frac{z_{21}}{\alpha-\beta-z_{21}} \leq \frac{L_1 - \Gamma_1}{L_1} \varepsilon_t + \frac{4\kappa_1}{L_1 - \varepsilon_0\xi_1 -  3 \kappa_1} \leq \chi \varepsilon_t + K_1. 
		\]
		By $\{\A_{i,k}, \Z_i; i = 1,\cdots n, k = 1, \cdots d\} \subseteq \mathcal{A}\cap\mathcal{B}$, we further have\[\begin{aligned}
			\frac{L_1 - \Gamma_1}{L_1} \varepsilon_t + \frac{4\kappa_1}{L_1 - \varepsilon_0\xi_1 -  3 \kappa_1}
			&\leq \frac{L_1 - \Gamma_1}{L_1} \varepsilon_0 + \frac{\Gamma_1}{L_1} \varepsilon_0 = \varepsilon_0. 
		\end{aligned}
		\]Thus, we have proved\[
		\|\sin \Theta(U_1, \widehat{U}_1^{(t+1)})\| \leq \chi \varepsilon_t + K_1 \leq \varepsilon_0. 
		\]Similarly, we can prove $\|\sin \Theta(U_h, \widehat{U}_h^{(t+1)})\| \leq \chi \varepsilon_t + K_1 \leq \varepsilon_0$ for other $h$. So the statement holds by induction.
	\end{proof}
    \subsubsection{Statistical Bound}\label{section_statistical_bound_tensor}
	In this section, we are going to argue that when $\A_{i,k}$ and $\Z_i$ are random tensors satisfying Assumption \ref{assumption_conditionnumber_tensor} and \ref{assumption_z_tensor} and with proper initialization, the probability of $\mathcal{A}_1 \cap \mathcal{A}_2 \cap \mathcal{B}$ is high and the estimation error converges to 0 in probability.  
	
	Recall Assumption \ref{assumption_conditionnumber_tensor}: 
	\begin{manualassumption}{\ref{assumption_conditionnumber_tensor}}
		Assume in decomposition (\ref{eq:decomposition-equivalence}), $\A_{k}$'s are independent and there is a constant $\mu$ such that
            \[\PP \left\{\mu \leq \frac{\max_{k} \left\| \MM_h \left(\A_{k}\right) \right\|}{\sqrt{\prod_{k \neq h} p_k}}\right\}\leq \nu, \quad \text{for some small $\nu < 1$. }\]
            Denote $\lambda$ as 
            $$\lambda =\min_h \left(\lambda_{\min }\left(\frac{1}{\prod_{k \neq h} p_k}\mathbb{E} \MM_h \left(  \A_{h}\right) \left(\bigotimes_{j \neq h} P_{{U}_{j\perp}} \right)\MM_h \left(  \A_{h}\right) \T \right)\right).$$
            We have $\frac{\mu^2}{{\lambda}}\leq C$ for some constant $C>0$.
	\end{manualassumption}
	
We additionally assume $\mu = 1$ and $\nu = 0$ for now, i.e., $\PP(\mathcal{A}_3) = 1$. The following lemma bounds the sub-Gaussian norm of $\|Z_i\|$. Denote the variance and fourth moment of each entry of $Z$ as $\sigma^2$ and $\zeta^4$. 
 
	Let's first bound the term $L$ in Notation \ref{notation_W5678_tensor} by the well-known Matrix Chernoff inequality (Lemma \ref{lemma_Chernoff}). In our setting, since $\lambda \leq \lambda_{\min }\left(\frac{1}{\prod_{k \neq h} p_k}\mathbb{E} \MM_h \left(  \A_{1,h}\right) \left(\bigotimes_{j \neq h} P_{{U}_{j\perp}} \right)\MM_h \left(  \A_{1,h}\right) \T \right)$, apply this on $L_h$: 
	\begin{align*}
		\PP\{L_h \leq (1-\delta) \lambda\}
		\leq&
		r_h \cdot \left[\frac{\mathrm{e}^{-\delta}}{(1-\delta)^{1-\delta}}\right]^{n\lambda}. 
	\end{align*}
	Taking $\delta = 1/2$, since $e^{-0.5}/0.5^{0.5} \leq 0.86$, we have 
	\begin{Corollary}\label{corollary_L_tensor}
		\[
		\PP\{L_h \leq \lambda/2\}
		\leq
		r_h \cdot  \exp\left\{\log(0.86)n\lambda\right\}.
		\]
	\end{Corollary}
	
	To bound the terms $M$ in Notation \ref{notation_W5678_tensor}, we are going to use the strategy called `union bound'. To that end, let's first estimate the covering number. 
	Recall $$T_1 = \left\{ W \in \RR^{\prod_{k \neq 1}p_k \times \prod_{k \neq 1}p_k}: \|W\|=1 , \tr(W) = 0, W = W\T, \rank(W) \leq \prod_{k \neq 1}r_k \right\}.$$ By Lemma \ref{lemma_covering_number}, we have
	\begin{Corollary}\label{corollary_epsilon_net_tensor}
		There exists subset $\bar{T}_1$ of ${T}_1$ such that for some absolute constant $C_0$, $$|\bar{T}_1|\leq \exp(  2^{d-1} \prod_{k \neq 1}r_k (p_1 + \prod_{k \neq 1}p_k)\log (C_0/\varepsilon)),$$
		\[
		M_{1,1} \leq \frac{1}{n\prod_{k \neq 1}p_k} \sup_{E\in \bar T_1} \left\|\sum_{i = 1}^{n} \MM_1 \left(  \A_{i,1}\right) E \MM_1 \left( \sum_{k \neq 1}\left(\A_{i,k} \times_k U_k\right) \right)\T \right\| + \varepsilon, 
		\] and similar bounds hold for other $M_{i,1}$ for $i= 2,3,4$ in Notation \ref{notation_W5678}. 
	\end{Corollary}
	Then, we use the Bernstein bound (Lemma \ref{lemma_bernstein}) and union bound (Corollary \ref{corollary_epsilon_net_tensor}), which yield the following corollaries for different $M$ in Notation \ref{notation_W5678_tensor}. 
	\begin{Corollary}\label{corollary_M1_tensor}
		For given $d$, $c_1$ and $c_2$, there exists constant $c_3$ such that if $$n \geq  c_3\left(\frac{1}{\lambda}\right)^2(\prod_{k \neq 1} r_k) (p_1 + \prod_{k \neq 1} p_k ),$$ then with probability at least  $1-e^{-(c_2)\prod_{k \neq 1}r_k (p_1 + \prod_{k \neq 1}p_k)}$, 
		we have
		\[
		M_{1,1} \leq c_1\lambda. 
		\]
	\end{Corollary}
	
	\begin{proof}
		Let $Z_i = \frac{1}{\prod_{k \neq 1}p_k}\MM_1 \left(  \A_{i,1} \right) E \MM_1 \left( \sum_{k \neq 1}\left(\A_{i,k} \times_k U_k\right) \right)\T$, then $\|Z_i\| \leq (d-1)$ and hence $\exp((\log 2) \|Z_i\|/(d-1)) \leq 2$. Also, $\|Z_i Z_i^{\top}\| = \|Z_i\T Z_i\| \leq (d-1)^2$. So in Lemma \ref{lemma_bernstein}, for any given $\varepsilon>0$, constant $C$ and $E \in T_1$ , let $t = C(\prod_{k \neq 1} r_k) (p_1 + \prod_{k \neq 1} p_k )$, $\sigma_Z = (d-1)$, $\alpha = 1$ and $U_Z^{(\alpha)} = (d-1)/\log2$, which yields that there exists constant $C_2$ such that with probability at least  $1-e^{-C\prod_{k \neq 1}r_k (p_1 + \prod_{k \neq 1}p_k)}$, the following holds: 
		\[\begin{aligned}
			&\frac{1}{n\prod_{k \neq 1}p_k} \left\|\sum_{i = 1}^{n} \MM_1 \left(  \A_{i,1}\right) E \MM_1 \left( \sum_{k \neq 1}\left(\A_{i,k} \times_k U_k\right) \right)\T \right\| \\
			\leq& C_2 \max \left\{(d-1) \sqrt{\frac{(\prod_{k \neq 1} r_k) (p_1 + \prod_{k \neq 1} p_k )}{n}}, (d-1) \frac{(\prod_{k \neq 1} r_k) (p_1 + \prod_{k \neq 1} p_k )}{n}\right\}. 
		\end{aligned}
		\]Then by union bound from Corollary \ref{corollary_epsilon_net_tensor}, it yields that with probability at least  $$1-e^{-(C-\log(C_0/\varepsilon))\prod_{k \neq 1}r_k (p_1 + \prod_{k \neq 1}p_k)},$$ we have 
		\[
		M_{1,1} \leq C_2 \max \left\{(d-1) \sqrt{\frac{(\prod_{k \neq 1} r_k) (p_1 + \prod_{k \neq 1} p_k )}{n}}, (d-1) \frac{(\prod_{k \neq 1} r_k) (p_1 + \prod_{k \neq 1} p_k )}{n}\right\} + \varepsilon. 
		\]
		So for given $d$ and constants $C_4$ and $C_5$, let $\varepsilon$ be a constant multiplier of $\lambda$ and hence, if 
		$$n \geq  C_3\left(\frac{1}{\lambda}\right)^2(\prod_{k \neq 1} r_k) (p_1 + \prod_{k \neq 1} p_k )$$ 
		for some large enough $C_3$, then with probability at least  $1-e^{-(C_4)\prod_{k \neq 1}r_k (p_1 + \prod_{k \neq 1}p_k)}$, we have
		\[
		M_{1,1} \leq C_5\lambda. 
		\]
	\end{proof}
	We can similarly prove the following: 
	\begin{Corollary}\label{corollary_M2_tensor}
		For given $d$, $c_1$ and $c_2$, there exists constant $c_3$ such that if $$n \geq  c_3\left(\frac{1}{\lambda}\right)^2(\prod_{k \neq 1} r_k) (p_1 + \prod_{k \neq 1} p_k ),$$ then with probability at least  $1-e^{-(c_2)\prod_{k \neq 1}r_k (p_1 + \prod_{k \neq 1}p_k)}$, we have
		\[
		M_{2,1} \leq c_1\lambda. 
		\]
	\end{Corollary}	
	And we also have: 
	\begin{Corollary}\label{corollary_M3_tensor}
		For given $d$, $c_1$ and $c_2$, there exists constant $c_3$ such that if 
		$$n \geq c_3 (\prod_{k \neq 1} r_k) (p_1 + \prod_{k \neq 1} p_k ) \max\left\{\frac{\theta^2u^2}{\lambda^2}, \frac{\theta u}{\lambda}\right\},$$ 
		where $\theta = \max\left\{ \sqrt{\frac{p_1}{\prod_{k \neq 1}p_k}}, 1 \right\}$, then with probability at least  $1-e^{-(c_2)\prod_{k \neq 1}r_k (p_1 + \prod_{k \neq 1}p_k)}$, we have
		\[
		M_{3,1} \leq c_1\lambda. 
		\]
	\end{Corollary}
	\begin{proof}
		Let $Z_i = \frac{1}{\prod_{k \neq 1}p_k}\MM_1 \left(\Z_{i}\right) E \MM_1 \left( \sum_{k = 1}^{d}\left(\A_{i,k} \times_k U_k\right) \right)\T$, then $\|Z_i\| \leq (d-1)\frac{\|\MM_1 \left(\Z_{i}\right)\|}{\sqrt{\prod_{k \neq 1}p_k}}$ and \[
		\sqrt{\log 2}\cdot \left\| \frac{\|\MM_1 \left(\Z_{i}\right)\|}{\sqrt{\prod_{k \neq 1}p_k}} \right\|_{\psi_1} \leq \left\| \frac{\|\MM_1 \left(\Z_{i}\right)\|}{\sqrt{\prod_{k \neq 1}p_k}} \right\|_{\psi_2}. 
		\]
		Hence, 
		$\mathbb{E} \exp \left(\frac{\sqrt{\log 2}\|Z_i\|}{u(d-1)}\right) \leq \mathbb{E} \exp \left(\frac{\sqrt{\log 2}\|\MM_1 \left(\Z_{i}\right)\|}{u\sqrt{\prod_{k \neq 1}p_k}}\right) \leq 2$. Also, for some absolute constant $C_6>1$, we have
		$$
		\begin{aligned}
			& \left\|\EE Z_i Z_i^{\top}\right\|\\
			=& \frac{1}{\prod_{k \neq 1}p_k^2} \left\|\EE \left(\EE \left(\left.  \MM_1 \left(\Z_{i}\right) E \MM_1 \left( \sum_{k = 1}^{d}\left(\A_{i,k} \times_k U_k\right) \right)\T \MM_1 \left( \sum_{k = 1}^{d}\left(\A_{i,k} \times_k U_k\right) \right) E \MM_1 \left(\Z_{i}\right) \T \right| \A_{i,k} \right)\right)\right\|\\
			=& \frac{1}{\prod_{k \neq 1}p_k^2} \left\|\EE \left( \sigma^2 \tr\left( E \MM_1 \left( \sum_{k = 1}^{d}\left(\A_{i,k} \times_k U_k\right) \right)\T \MM_1 \left( \sum_{k = 1}^{d}\left(\A_{i,k} \times_k U_k\right) \right) E \right) I \right) \right\|\\
			\leq& \frac{1}{\prod_{k \neq 1}p_k}\EE \left( \sigma^2 \left\| \MM_1 \left( \sum_{k = 1}^{d}\left(\A_{i,k} \times_k U_k\right) \right) \right\|^2 \right) \\
			\leq& d^2\sigma^2\\
			\leq& C_6^2d^2u^2, 
		\end{aligned}$$and 
		$$
		\begin{aligned}
			& \left\|\EE Z_i^{\top}Z_i \right\|\\
			=& \frac{1}{\prod_{k \neq 1}p_k^2} \left\|\EE \left(\EE \left(\left. \MM_1 \left( \sum_{k = 1}^{d}\left(\A_{i,k} \times_k U_k\right) \right) E \MM_1 \left(\Z_{i}\right) \T \MM_1 \left(\Z_{i}\right) E \MM_1 \left( \sum_{k = 1}^{d}\left(\A_{i,k} \times_k U_k\right) \right)\T  \right| \A_{i,k} \right)\right)\right\|\\
			=& \frac{1}{\prod_{k \neq 1}p_k^2} \left\|\EE \left( \sigma^2 p_1  \MM_1 \left( \sum_{k = 1}^{d}\left(\A_{i,k} \times_k U_k\right) \right) E^2 \MM_1 \left( \sum_{k = 1}^{d}\left(\A_{i,k} \times_k U_k\right) \right)\T \right) \right\|\\
			\leq& \frac{1}{\prod_{k \neq 1}p_k}\EE \left( \sigma^2 p_1 \left\| \MM_1 \left( \sum_{k = 1}^{d}\left(\A_{i,k} \times_k U_k\right) \right) \right\|^2 \right) \\
			\leq& \frac{d^2 p_1}{\prod_{k \neq 1}p_k}\sigma^2\\
			\leq& C_6^2\frac{d^2 p_1}{\prod_{k \neq 1}p_k} u^2. 
		\end{aligned}$$
		So in Lemma \ref{lemma_bernstein}, for any given $\varepsilon>0$, constant $C$ and $E \in T_1$ , let $t = C(\prod_{k \neq 1} r_k) (p_1 + \prod_{k \neq 1} p_k )$, $\sigma_Z = C_6 du \theta$, $\alpha = 1$ and $U_Z^{(\alpha)} = C_6ud\theta/\sqrt{\log 2}$, where $\theta = \max\left\{ \sqrt{\frac{p_1}{\prod_{k \neq 1}p_k}}, 1 \right\}$. It yields that, there exists constant $C_2$ such that, with probability at least  $1-e^{-C\prod_{k \neq 1}r_k (p_1 + \prod_{k \neq 1}p_k)}$, the following holds: 
		\[\begin{aligned}
			&\frac{1}{n\prod_{k \neq 1}p_k} \left\|\sum_{i = 1}^{n} \MM_1 \left(\Z_{i}\right) E \MM_1 \left( \sum_{k = 1}^{d}\left(\A_{i,k} \times_k U_k\right) \right)\T   \right\| \\
			\leq& C_2 \max \left\{du\theta \sqrt{\frac{(\prod_{k \neq 1} r_k) (p_1 + \prod_{k \neq 1} p_k )}{n}}, du\theta \frac{(\prod_{k \neq 1} r_k) (p_1 + \prod_{k \neq 1} p_k )}{n}\right\},
		\end{aligned}
		\]where $\theta = \max\left\{ \sqrt{\frac{p_1}{\prod_{k \neq 1}p_k}}, 1 \right\}$. Then by union bound from Corollary \ref{corollary_epsilon_net_tensor}, it yields that with probability at least  $1-e^{-(C-\log(C_0/\varepsilon))\prod_{k \neq 1}r_k (p_1 + \prod_{k \neq 1}p_k)}$: 
		\[
		M_{3,1} \leq C_2 \max \left\{du\theta \sqrt{\frac{(\prod_{k \neq 1} r_k) (p_1 + \prod_{k \neq 1} p_k )}{n}}, du\theta \frac{(\prod_{k \neq 1} r_k) (p_1 + \prod_{k \neq 1} p_k )}{n}\right\} + \varepsilon.
		\]
		So for given $d$ and constants $C_4$ and $C_5$, let $\varepsilon$ be a constant multiplier of $\lambda$ and hence, if 
		$$n \geq C_3 (\prod_{k \neq 1} r_k) (p_1 + \prod_{k \neq 1} p_k ) \max\left\{\frac{\theta^2u^2}{\lambda^2}, \frac{\theta u}{\lambda}\right\}$$
		for some large enough $C_3$, then with probability at least  $1-e^{-(C_4)\prod_{k \neq 1}r_k (p_1 + \prod_{k \neq 1}p_k)}$, we have\[
		M_{3,1} \leq C_5\lambda. 
		\]
	\end{proof}
	
	\begin{Corollary}\label{corollary_M4_tensor}
		For given $d$, $c_1$ and $c_2$, there exists constant $c_3$ such that if $$n \geq c_3 (\prod_{k \neq 1} r_k) (p_1 + \prod_{k \neq 1} p_k ) \max\left\{\frac{u^4}{\lambda^2}, \frac{u^2}{\lambda}\right\},$$  then with probability at least  $1-e^{-c_2\prod_{k \neq 1}r_k (p_1 + \prod_{k \neq 1}p_k)}$, we have
		\[
		M_{4,1} \leq c_1\lambda. 
		\]
	\end{Corollary}
	
	\begin{proof}
		Let $Z_i = \frac{1}{\prod_{k \neq 1}p_k}\MM_1 \left(\Z_{i}\right) E \MM_1 \left(\Z_{i}\right)\T$, then $\|Z_i\| \leq \frac{\|\MM_1 \left(\Z_{i}\right)\|^2}{{\prod_{k \neq 1}p_k}}$ and hence $\mathbb{E} \exp \left(\|Z_i\|/u^2\right) \leq \mathbb{E} \exp \left(\frac{\|\MM_1 \left(\Z_{i}\right)\|^2}{u^2{\prod_{k \neq 1}p_k}}\right) \leq 2$. Also, for some absolute constant $C_6 > 1$,  
		$$
		\begin{aligned}
			& \left\|\EE Z_i Z_i^{\top}\right\|\\
			=& \frac{1}{\prod_{k \neq 1}p_k^2} \left\|\EE  \left.  \MM_1 \left(\Z_{i}\right) E \MM_1 \left(\Z_{i}\right)\T \MM_1 \left(\Z_{i}\right) E \MM_1 \left(\Z_{i}\right) \T  \A_{i,k} \right)\right\|\\
			\overset{\text{Lemma \ref{lemma_Z}}}{\leq}& \frac{1}{\prod_{k \neq 1}p_k}\zeta^4 tr(E^2)\\
			\leq& \zeta^4 \|E\|^2\\
			\leq& \zeta^4\\
			\leq& C_6^2 u^4.
		\end{aligned}$$
		So in Lemma \ref{lemma_bernstein}, for any given $\varepsilon>0$, constant $C$ and $E \in T_1$ , let $t = C(\prod_{k \neq 1} r_k) (p_1 + \prod_{k \neq 1} p_k )$, $\sigma_Z = C_6 u^2$, $\alpha = 1$ and $U_Z^{(\alpha)} = 2C_6u^2$, which yields that there exists constant $C_2$ such that with probability at least  $1-e^{-C\prod_{k \neq 1}r_k (p_1 + \prod_{k \neq 1}p_k)}$, the following holds: 
		\[\begin{aligned}
			&\frac{1}{n\prod_{k \neq 1}p_k} \left\|\sum_{i = 1}^{n} \MM_1 \left(\Z_{i}\right) E \MM_1 \left(\Z_{i}\right)\T   \right\| \\
			\leq& C_2 \max\left\{u^2 \sqrt{\frac{(\prod_{k \neq 1} r_k) (p_1 + \prod_{k \neq 1} p_k )}{n}}, u^2 \frac{(\prod_{k \neq 1} r_k) (p_1 + \prod_{k \neq 1} p_k )}{n}\right\}.
		\end{aligned}
		\]
		Then by union bound from Corollary \ref{corollary_epsilon_net_tensor}, it yields that with probability at least  $$1-e^{-(C-\log(C_0/\varepsilon))\prod_{k \neq 1}r_k (p_1 + \prod_{k \neq 1}p_k)},$$ we have 
		\[
		M_{4,1} \leq C_2 \max\left\{u^2 \sqrt{\frac{(\prod_{k \neq 1} r_k) (p_1 + \prod_{k \neq 1} p_k )}{n}}, u^2 \frac{(\prod_{k \neq 1} r_k) (p_1 + \prod_{k \neq 1} p_k )}{n}\right\} + \varepsilon. 
		\]
		So for given $d$ and constants $C_4$ and $C_5$, let $\varepsilon$ be a constant multiplier of $\lambda$ and hence, if 
		$$n \geq C_3(\prod_{k \neq 1} r_k) (p_1 + \prod_{k \neq 1} p_k ) \max\left\{\frac{u^4}{\lambda^2}, \frac{u^2}{\lambda}\right\}$$ 
		for some large enough $C_3$, then with probability at least  $1-e^{-(C_4)\prod_{k \neq 1}r_k (p_1 + \prod_{k \neq 1}p_k)}$, we have
		\[
		M_{4,1} \leq C_5\lambda. 
		\]
	\end{proof}
	
	\begin{Corollary}\label{corollary_kappa1_tensor}
		For given $d$, $c_1$ and $c_2$, there exists constant $c_3$ such that if $$n \geq c_3 \log p_1 \max\left\{\frac{\theta^2 u^2}{\lambda^6}, \frac{\theta u}{\lambda^3}\right\},$$where $\theta = \max\left\{ \sqrt{\frac{p_1}{\prod_{k \neq 1}p_k}}, 1 \right\}$, then with probability at least  $1-e^{-c_2\log p_1}$, we have
		\[
		\frac{1}{n\prod_{k \neq 1}p_k} \left\|\sum_{i = 1}^{n} \MM_1 \left(\Z_{i}\right) \left(\bigotimes_{j \neq 1} P_{{U}_{j\perp}} \right) \MM_1 \left( \A_{i,1} \times_1 U_1 \right)\T   \right\| \leq c_1\lambda^3. 
		\]
	\end{Corollary}
	\begin{proof}
		Let $Z_i = \frac{1}{\prod_{k \neq 1}p_k}\MM_1 \left(\Z_{i}\right)\left(\bigotimes_{j \neq 1} P_{{U}_{j\perp}} \right) \MM_1 \left( \A_{i,1} \times_1 U_1 \right)\T$, then $\|Z_i\| \leq \frac{\|\MM_1 \left(\Z_{i}\right)\|}{\sqrt{\prod_{k \neq 1}p_k}}$ and \[
		\sqrt{\log 2} \cdot \left\| \frac{\|\MM_1 \left(\Z_{i}\right)\|}{\sqrt{\prod_{k \neq 1}p_k}} \right\|_{\psi_1} \leq \left\| \frac{\|\MM_1 \left(\Z_{i}\right)\|}{\sqrt{\prod_{k \neq 1}p_k}} \right\|_{\psi_2}.
		\]
		Hence, 
		$\mathbb{E} \exp \left(\frac{\|Z_i\|\sqrt{\log 2}}{u}\right) \leq \mathbb{E} \exp \left(\frac{\|\MM_1 \left(\Z_{i}\right)\|\sqrt{\log 2}}{u\sqrt{\prod_{k \neq 1}p_k}}\right) \leq 2$. Also, there exists an absolute constant $C_6>1$ such that
		$$
		\begin{aligned}
			& \left\|\EE Z_i Z_i^{\top}\right\|\\
			=& \frac{1}{\prod_{k \neq 1}p_k^2} \left\|\EE \left(\EE \left(\left.  \MM_1 \left(\Z_{i}\right)\left(\bigotimes_{j \neq 1} P_{{U}_{j\perp}} \right) \MM_1 \left( \A_{i,1} \times_1 U_1 \right)\T \right.\right.\right. \right.\\
   & \qquad \qquad \qquad \qquad \left.\left.\left.\left. \MM_1 \left( \A_{i,1} \times_1 U_1 \right) \left(\bigotimes_{j \neq 1} P_{{U}_{j\perp}} \right) \MM_1 \left(\Z_{i}\right) \T \right| \A_{i,k} \right)\right)\right\|\\
			=& \frac{1}{\prod_{k \neq 1}p_k^2} \left\|\EE \left( \sigma^2 \tr\left( \left(\bigotimes_{j \neq 1} P_{{U}_{j\perp}} \right) \MM_1 \left( \A_{i,1} \times_1 U_1 \right)\T \MM_1 \left( \A_{i,1} \times_1 U_1 \right) \left(\bigotimes_{j \neq 1} P_{{U}_{j\perp}} \right) \right) I \right) \right\|\\
			\leq& \frac{1}{\prod_{k \neq 1}p_k}\EE \left( \sigma^2 \left\| \MM_1 \left( \A_{i,1} \times_1 U_1 \right) \right\|^2 \right) \leq \sigma^2\leq C_6^2 u^2,
		\end{aligned}$$and 
		$$
		\begin{aligned}
			& \left\|\EE Z_i^{\top}Z_i \right\|\\
			=& \frac{1}{\prod_{k \neq 1}p_k^2} \left\|\EE \left(\EE \left(\left. \MM_1 \left( \A_{i,1} \times_1 U_1 \right) \left(\bigotimes_{j \neq 1} P_{{U}_{j\perp}} \right)\T \MM_1 \left(\Z_{i}\right)\T \right.\right.\right.\right.\\
                & \qquad \qquad \qquad \qquad \left.\left.\left.\left. \MM_1 \left(\Z_{i}\right)\left(\bigotimes_{j \neq 1} P_{{U}_{j\perp}} \right) \MM_1 \left( \A_{i,1} \times_1 U_1 \right)\T  \right| \A_{i,k} \right)\right)\right\|\\
			=& \frac{1}{\prod_{k \neq 1}p_k^2} \left\|\EE \left( \sigma^2 p_1  \MM_1 \left( \A_{i,1} \times_1 U_1 \right) \left(\bigotimes_{j \neq 1} P_{{U}_{j\perp}} \right)^2 \MM_1 \left( \A_{i,1} \times_1 U_1 \right)\T \right) \right\|\\
			\leq& \frac{1}{\prod_{k \neq 1}p_k}\EE \left( \sigma^2 p_1 \left\| \MM_1 \left( \A_{i,1} \times_1 U_1 \right) \right\|^2 \right) \\
			\leq& \frac{p_1}{\prod_{k \neq 1}p_k}\sigma^2\\
			\leq& C_6^2\frac{p_1}{\prod_{k \neq 1}p_k}u^2.
		\end{aligned}$$
		So in Lemma \ref{lemma_bernstein}, for any given $\varepsilon>0$, constant $C$ and $E \in T_1$ , let $t = C\log p_1$, $\sigma_Z = C_6\theta u$, $\alpha = 1$ and $U_Z^{(\alpha)} = C_6 \theta u/\sqrt{\log 2}$, where $\theta = \max\left\{ \sqrt{\frac{p_1}{\prod_{k \neq 1}p_k}}, 1 \right\}$. It yields that there exists constant $C_2$ such that with probability at least  $1-e^{-C\log p_1}$, the following holds: 
		\begin{align}
			&\frac{1}{n\prod_{k \neq 1}p_k} \left\|\sum_{i = 1}^{n} \MM_1 \left(\Z_{i}\right) \left(\bigotimes_{j \neq 1} P_{{U}_{j\perp}} \right) \MM_1 \left( \A_{i,1} \times_1 U_1 \right)\T   \right\| \notag \\
			\leq& C_2 \max \left\{u\theta \sqrt{\frac{\log p_1}{n}}, u\theta \frac{\log p_1}{n}\right\} \label{inequality_kappa_1_tensor},
		\end{align}
		where $\theta = \max\left\{ \sqrt{\frac{p_1}{\prod_{k \neq 1}p_k}}, 1 \right\}$.	So, for given $d$ and constants $C_4$ and $C_5$ if 
		$$n \geq C_3\log p_1 \max\left\{\frac{\theta^2 u^2}{\lambda^6}, \frac{\theta u}{\lambda^3}\right\}$$ 
		for some large enough $C_3$, then with probability at least  $1-e^{-C_4\log p_1}$, we have
		\[
		\frac{1}{n\prod_{k \neq 1}p_k} \left\|\sum_{i = 1}^{n} \MM_1 \left(\Z_{i}\right) \left(\bigotimes_{j \neq 1} P_{{U}_{j\perp}} \right) \MM_1 \left( \A_{i,1} \times_1 U_1 \right)\T   \right\| \leq C_5\lambda^3. 
		\]
	\end{proof}
	\begin{Corollary}\label{corollary_kappa2_tensor}
		For given $d$, $c_1$ and $c_2$, there exists constant $c_3$ such that if $$n \geq c_3 \log p_1 \max\left\{\frac{u^4}{\lambda^6}, \frac{u^2}{\lambda^3}\right\}, $$ then with probability at least  $1-e^{-c_2\log p_1}$, we have
		\[
		\frac{1}{n\prod_{k \neq h}p_k} \left\|\sum_{i = 1}^{n} \MM_1 \left(\Z_{i}\right) \left(\bigotimes_{j \neq 1} P_{{U}_{j\perp}} \right) \MM_1 \left(\Z_{i}\right)\T- n\sigma^2\prod_{k \neq 1}\left({p_k - r_k}\right) I  \right\| \leq c_1\lambda^3. 
		\]
	\end{Corollary}
	\begin{proof}
		Let $Z_i = \frac{1}{\prod_{k \neq 1}p_k}\left(\MM_1 \left(\Z_{i}\right)\left(\bigotimes_{j \neq 1} P_{{U}_{j\perp}} \right) \MM_1 \left(\Z_{i}\right)\T - \sigma^2\prod_{k \neq 1}\left({p_k - r_k}\right) I \right)$, then $\|Z_i\| \leq \max\left\{\frac{\|\MM_1 \left(\Z_{i}\right)\|^2}{\prod_{k \neq 1}p_k}, \sigma^2r\right\}$, where $r = \frac{\prod_{k \neq 1}\left({p_k - r_k}\right)}{\prod_{k \neq 1}{p_k}}$. \\
		
		Hence, 
		$\mathbb{E} \exp \left(\frac{\|Z_i\|}{\tau}\right) \leq \mathbb{E} \exp \left(\max\left\{\frac{\|\MM_1 \left(\Z_{i}\right)\|^2}{\tau\prod_{k \neq 1}p_k}, \frac{\sigma^2r}{\tau}\right\}\right) \leq 2$, 
		where $\tau = \max\left\{u^2, \frac{\sigma^2 r}{\log 2}\right\} \leq u^2 / \log2$. Also, 
		$$
		\begin{aligned}
			& \EE Z_i Z_i^{\top}\\
			=&\frac{1}{\prod_{k \neq 1}p_k^2}\left(\EE \MM_1 \left(\Z_{i}\right)\left(\bigotimes_{j \neq 1} P_{{U}_{j\perp}} \right) \MM_1 \left(\Z_{i}\right)\T\MM_1 \left(\Z_{i}\right)\left(\bigotimes_{j \neq 1} P_{{U}_{j\perp}} \right) \MM_1 \left(\Z_{i}\right)\T\right)\\
			&-\frac{2\sigma^2\prod_{k \neq 1}\left({p_k - r_k}\right)}{\prod_{k \neq 1}p_k^2}\left(\EE \MM_1 \left(\Z_{i}\right)\left(\bigotimes_{j \neq 1} P_{{U}_{j\perp}} \right) \MM_1 \left(\Z_{i}\right)\T \right) + \frac{\sigma^4\prod_{k \neq 1}\left({p_k - r_k}\right)^2}{\prod_{k \neq 1}p_k^2}I\\
			\overset{\text{Lemma \ref{lemma_Z}}}{\preccurlyeq} & \frac{\zeta^4}{\prod_{k \neq 1}p_k}\tr \left(\bigotimes_{j \neq 1} P_{{U}_{j\perp}}^2 \right)I-\frac{2\sigma^4\prod_{k \neq 1}\left({p_k - r_k}\right)}{\prod_{k \neq 1}p_k^2} \tr\left(\bigotimes_{j \neq 1} P_{{U}_{j\perp}}\right)I + \frac{\sigma^4\prod_{k \neq 1}\left({p_k - r_k}\right)^2}{\prod_{k \neq 1}p_k^2}I\\
			\preccurlyeq & \frac{\prod_{k \neq 1}{p_k - r_k}}{\prod_{k \neq 1}p_k}\zeta^4I.
		\end{aligned}$$
		Thus, for some absolute constant $C_6> 1$,
		\[
		\left\|\EE Z_i Z_i^{\top}\right\| \leq \zeta^4 \leq C_6^2 u^4. 
		\]
		So in Lemma \ref{lemma_bernstein}, for any given $\varepsilon>0$, constant $C$ and $E \in T_1$ , let $t = C\log p_1$, $\sigma_Z =C_6 u^2$, $\alpha = 1$ and $U_Z^{(\alpha)} =C_6 u^2 / \log2$, which yields that there exists constant $C_2$ such that with probability at least  $1-e^{-C\log p_1}$, the following holds: 
		\begin{align}
			&\frac{1}{n\prod_{k \neq h}p_k} \left\|\sum_{i = 1}^{n} \MM_1 \left(\Z_{i}\right) \left(\bigotimes_{j \neq 1} P_{{U}_{j\perp}} \right) \MM_1 \left(\Z_{i}\right)\T - n\sigma^2\prod_{k \neq 1}\left({p_k - r_k}\right) I \right\| \notag\\
			\leq& C_2 \max \left\{u^2 \sqrt{\frac{\log p_1}{n}}, u^2\frac{\log p_1}{n}\right\} \label{inequality_kappa_2_tensor}.
		\end{align}
		So, for given $d$ and constants $C_4$ and $C_5$ if 
		$$n \geq C_3\log p_1 \max\left\{\frac{u^4}{\lambda^6}, \frac{u^2}{\lambda^3}\right\}$$ 
		for some large enough $C_3$, then with probability at least  $1-e^{-C_4\log p_1}$, we have
		\[
		\frac{1}{n\prod_{k \neq h}p_k} \left\|\sum_{i = 1}^{n} \MM_1 \left(\Z_{i}\right) \left(\bigotimes_{j \neq 1} P_{{U}_{j\perp}} \right) \MM_1 \left(\Z_{i}\right)\T- n\sigma^2\prod_{k \neq 1}\left({p_k - r_k}\right) I  \right\| \leq C_5\lambda^3. 
		\]
	\end{proof}
	
	Corollaries \ref{corollary_L_tensor}, \ref{corollary_M1_tensor}, \ref{corollary_M2_tensor}, \ref{corollary_M3_tensor} and \ref{corollary_M4_tensor} imply that with proper choice of $c_1$, we have
	$$\Delta_1 \geq L_1 - C_d(M_{1,h} + M_{2,h} + M_{3,h} + M_{4,h})\gtrsim \lambda,$$
	and 
	$$\xi_1 \leq C_d^\prime + C_d(M_{1,h} + M_{2,h} + M_{3,h} + M_{4,h}) \lesssim 1$$
	with high probability, where $C_d$ and $C_d^\prime$ are constants only related to $d$. If $d$ is fixed and we set $\varepsilon_0$ as a small constant multiplier of $\lambda$, we have 
	$$\Gamma_1 = \Delta_1 - \epsilon_0 \xi_1 \gtrsim \lambda. $$ 
	So, if further pick $\kappa_1\lesssim\lambda^3 \leq \lambda$ small enough, by Corollaries \ref{corollary_kappa1_tensor} and \ref{corollary_kappa2_tensor}, we have 
	$$\frac{4\kappa_1}{L_1 - \varepsilon_0 \xi_1 - 3\kappa_1} \lesssim \lambda^2 \lesssim \frac{\Gamma_1\varepsilon_0}{L_1}$$with high probability. 
	Additionally, we can bound the error $K_2$ in Theorem \ref{theorem_convergence_deterministic_tensor} by (\ref{inequality_kappa_1_tensor}) and (\ref{inequality_kappa_2_tensor}). 
	
	The argument above can be summarized as the following lemma.
	\begin{Lemma}
		Given tensor order $d\geq 2$, constants $c_2>0$ and assuming Assumption \ref{assumption_conditionnumber}, \ref{assumption_z} hold with $\mu = 1$ in Assumption \ref{assumption_conditionnumber}, then there exist constants $c_1$, $c_4$, $c_5<1$, $c_6$ (does not depend on any variable appeared in the following equations) such that if $\varepsilon_0 \leq c_4 \lambda$, \[
		n \geq c_1 \log p_1 \max\left\{\frac{u^4}{\lambda^6}, \frac{u^2}{\lambda^3}, \frac{\theta u^2}{\lambda^6}, \frac{\theta u}{\lambda^3}\right\},
		\]
		and
		\[
		n \geq c_1 (\prod_{k \neq 1} r_k) (p_1 + \prod_{k \neq 1} p_k ) \max\left\{\frac{1}{\lambda^2}, \frac{u^4}{\lambda^2}, \frac{u^2}{\lambda}, \frac{\theta^2 u^2}{\lambda^2}, \frac{\theta u}{\lambda}\right\}, 
		\]where $\theta = \max\left\{ 1, \sqrt{\frac{p_h}{\prod_{k \neq h}p_k}}; h = 1, \cdots, d\right\}$, 
		then with probability at least  $1- e^{-c_2\prod_{k \neq 1}r_k (p_1 + \prod_{k \neq 1}p_k)}$, $\{\A_{i,k}, \Z_i; i = 1,\cdots n, k = 1, \cdots d\} \subseteq \mathcal{A}\cap\mathcal{B}$. Additionally, in Theorem \ref{theorem_convergence_deterministic_tensor}, we have $\chi \leq c_5$ and $$K_2\leq c_6 \lambda \max \left\{u\theta \sqrt{\frac{\log p_1}{n}}, u\theta \frac{\log p_1}{n}, u^2 \sqrt{\frac{\log p_1}{n}}, u^2\frac{\log p_1}{n}\right\}.$$. 
	\end{Lemma}
	Now let's consider general $\mu$. We can let $\Y_i = \X_i / \mu(\{\X_i\})$ and then $\mu(\{\Y_i\} ) = 1$ and $\lambda(\{\Y_i\}) = \lambda(\{\X_i\}) / \mu(\{\X_i\})^2 \geq 1/C$, where $\mu(\{\X_i\})$ refers the $\mu$ of $\X_i$ in Assumption \ref{assumption_conditionnumber}, $\mu(\{\Y_i\})$ refers the $\mu$ of $\Y_i$ and $\lambda(\{\Y_i\})$ are defined similarly. Then we can apply the above lemma and Theorem \ref{theorem_convergence_deterministic_tensor} to $\Y_i$. So finally for general $\nu$, by taking out the probability of $\mu \leq \frac{\max_{k} \left\| \MM_h \left(\A_{k}\right) \right\|}{\sqrt{\prod_{k \neq h} p_k}}$ we get:

	\begin{manualtheorem}{\ref{theorem_convergence_tensor}}
		Given constant $c_1>0$, tensors $\{\X_i\}$ satisfying decomposition (\ref{eq:decomposition-equivalence}) and Assumption \ref{assumption_conditionnumber_tensor} and \ref{assumption_z_tensor} with order $d\geq 2$, then when applying Algorithm \ref{algorithm_iterative_projection_tensor}, there exists constant $c_2$, $c_3$, $c_4<1$, $c_5$ (does not depend on any variable appeared in the following equations) such that if $\varepsilon_0 \leq c_3$ and $n$ satisfies the following for all $h$: 
		\[
		n \geq c_2 r_{-h} (p_h + p_{-h}) \max\left\{\frac{ u^4}{\mu^4}, \frac{\theta^2 u^2}{\mu^2}, \frac{\theta u}{\mu}, 1\right\}, 
		\]
		then with probability at least  $1-e^{-c_1r_{-m} (p_{m} + p_{-m})} - \nu$, the estimation error $\max_{k \in \{1, \cdots d\}}\left\{\|\sin \Theta (U_k, \hat{U}_k^{(t)})\|\right\}$ in Algorithm \ref{algorithm_iterative_projection_tensor} converges linearly with $\chi \leq c_4$: 
		\[
		\max_{k \in \{1, \cdots d\}}\left\{\|\sin \Theta (U_k, \hat{U}_k^{(t)})\|\right\} - \operatorname{Error} \leq \chi \left(\max_{k \in \{1, \cdots d\}}\left\{\|\sin \Theta (U_k, \hat{U}_k^{(t-1)})\|\right\} - \operatorname{Error}\right) ,
		\]
		and the final error is bounded by 
		$$\operatorname{Error} \leq c_5 \sqrt{\frac{\log p_{\max}}{n}} \max \left\{\theta {\frac{u}{\mu}}, {\frac{u^2}{\mu^2}}\right\},$$ 
		where $r_{-h} = \prod_{k \neq h} r_k$, $p_{-h} = \prod_{k \neq h} p_k$, $\theta = \max\left\{ 1, \sqrt{\frac{p_h}{\prod_{k \neq h}p_k}}; h = 1, \cdots, d\right\}$, $m = \argmax\{r_{-k} (p_{k} + p_{-k})\}$ and $p_{\max} = \max_k\{p_k\}$. 
	\end{manualtheorem}

\end{sloppypar}
\end{document}